\documentclass[11pt]{article}
\usepackage[T1]{fontenc}
\usepackage{fullpage}

 \usepackage{array,tabularx}

\usepackage{booktabs}
\usepackage{amsfonts}
\usepackage{amssymb}
\usepackage{tikz}
\usetikzlibrary{arrows.meta,calc,decorations.pathreplacing,fit,positioning,shapes.geometric}
\usepackage{amstext}
\usepackage{amsmath}
\usepackage{mathtools}
\usepackage{xspace}
\usepackage{amsthm}
\usepackage{graphicx}
\usepackage{subcaption}
\usepackage[framemethod=TikZ]{mdframed}

\usepackage{xcolor}
\usepackage{graphics}
\usepackage{colordvi}
\usepackage{color}
\usepackage{paralist}
\usepackage{bm}
\usepackage{mathpazo}
\usepackage{float}

\usepackage[ruled,vlined]{algorithm2e}
\usepackage{xurl}
\usepackage[hidelinks]{hyperref}
\usepackage{thmtools}
\usepackage{thm-restate}
 
\newcommand{\OPT}{\mbox{\sf OPT}}

\makeatletter
\@ifundefined{newcounteralias}{}{\renewcommand\thmt@autorefsetup{\@xa\def\csname\thmt@envname autorefname\@xa\endcsname\@xa{\thmt@thmname}}}
\makeatother
\usepackage[capitalize,noabbrev]{cleveref}

\newcommand{\am}[1]{{\color{black}#1}}

\newtheorem{theorem}{Theorem}[section]

\newtheorem{lemma}[theorem]{Lemma}
\newtheorem{observation}[theorem]{Observation}
\Crefname{observation}{Observation}{Observations}

\newtheorem{corollary}[theorem]{Corollary}
\newtheorem{claim}[theorem]{Claim}
\Crefname{claim}{Claim}{Claims}
\newtheorem{proposition}[theorem]{Proposition}

\newtheorem{definition}[theorem]{Definition}

\mdfdefinestyle{ProblemFrame}{
    linecolor=black!65, linewidth=.6pt, roundcorner=6pt,
    innertopmargin=8pt, innerbottommargin=8pt,
    innerleftmargin=10pt, innerrightmargin=10pt,
    skipabove=.5\baselineskip, skipbelow=.5\baselineskip,
    nobreak=true
}

\newcommand{\opt}{{\sf OPT}\xspace}

\newenvironment{proofof}[1]
  {\begin{proof}[Proof of #1]}
  {\end{proof}}

\newenvironment{claimproof}[1][Proof of claim]
  {\begin{proof}[#1]}
  {\end{proof}}

\newcommand{\Oh}{O}

\newcommand{\Aa}{\mathcal{A}}

\newcommand{\N}{\mathbb{N}}

\newcommand{\R}{\mathbb{R}}
\newcommand{\Bb}{\mathcal{B}}

\newcommand{\Tt}{\mathcal{T}}

\newcommand{\poly}{\mathrm{poly}}

\renewcommand{\leq}{\leqslant}
\renewcommand{\geq}{\geqslant}
\renewcommand{\le}{\leqslant}

\renewcommand{\subset}{\subseteq}
\renewcommand{\supset}{\supseteq}

\newcommand*{\eg}{e.g.\@\xspace}
\newcommand*{\ie}{i.e.\@\xspace}

\newcommand{\cpi}{c_\pi}
\newcommand{\LN}{\texttt{l\_ngb}}
\newcommand{\RN}{\texttt{r\_ngb}}
\newcommand{\LB}{\texttt{l\_bdr}}
\newcommand{\RB}{\texttt{r\_bdr}}
\newcommand{\rect}[2]{\ensuremath{R[{#1}, {#2}]}}

\newcommand{\proofcase}[1]{\par\smallskip\noindent\textit{#1}\enspace}

\providecommand{\MCFigurePalette}{4}
\newcommand{\MCSetFigurePalette}[5]{\definecolor{mcink}{HTML}{#1}\definecolor{mcgray}{HTML}{#2}\definecolor{mcblue}{HTML}{#3}\definecolor{mcred}{HTML}{#4}\definecolor{mcgreen}{HTML}{#5}}
\ifcase\MCFigurePalette
  \PackageError{mcfigures}{Palette must be 1, 2, 3, or 4}{}
\or
  \MCSetFigurePalette{1A1A1A}{AAAAAA}{245EA6}{C04535}{23817A}
\or
  \MCSetFigurePalette{20252B}{A5ABB3}{2166AC}{B56A16}{187D65}
\or
  \MCSetFigurePalette{25222A}{AAA4AF}{6F4C9B}{BE5925}{007C83}
\or
  \MCSetFigurePalette{171717}{B9B9B9}{3C3C3C}{666666}{858585}
\else
  \PackageError{mcfigures}{Palette must be 1, 2, 3, or 4}{}
\fi

\tikzset{
  mcfig/.style={font=\footnotesize,text=mcink,draw=mcink,
    line cap=butt,line join=miter},
  mcgrid/.style={draw=mcgray,line width=.25pt},
  mcaxis/.style={draw=mcink,line width=.55pt},
  mcbox/.style={draw=mcgray,line width=1pt},
  mcblock/.style={draw=mcink,fill=white,line width=.65pt,inner sep=4pt},
  mcpoint/.style={circle,fill=mcink,draw=none,inner sep=1.8pt},
  mcsmallpoint/.style={mcpoint,inner sep=1.45pt},
  mcselected/.style={mcpoint,fill=mcred,inner sep=1.95pt},
  mcadded/.style={circle,fill=white,draw=mcred,line width=.85pt,inner sep=1.7pt},
  mcbluepoint/.style={mcpoint,fill=mcblue},
  mcgreenpoint/.style={mcpoint,fill=mcgreen},
  mcgraypoint/.style={mcpoint,fill=mcgray},
  mcnode/.style={circle,draw=mcink,fill=white,line width=.65pt,
    minimum size=4.6mm,inner sep=1.5pt},
  mcpath/.style={draw=mcblue,line width=.9pt},
  mcgreenpath/.style={draw=mcgreen,line width=.9pt},
  mcredpath/.style={draw=mcred,line width=.9pt},
  mcguide/.style={draw=mcgray,dash pattern=on 2.5pt off 2pt,line width=.5pt},
  mcarrow/.style={-{Stealth[length=4.5pt,width=4pt]},draw=mcgray,line width=.8pt},
  mcbluearrow/.style={mcarrow,draw=mcblue},
  mcgreenarrow/.style={mcarrow,draw=mcgreen},
  mcredarrow/.style={mcarrow,draw=mcred},
  mclabel/.style={font=\scriptsize,inner sep=2pt,text=mcink},
  mcpanel/.style={font=\footnotesize,anchor=north,inner sep=2pt,text=mcink},
mcmainfig/.style={mcfig,scale=1.14,font=\small,
    mclabel/.append style={font=\footnotesize,inner sep=2.5pt},
    mcpanel/.append style={font=\small},
    mcnode/.append style={minimum size=5.2mm},
    mcpoint/.append style={inner sep=2pt},
    mcsmallpoint/.append style={inner sep=1.6pt},
    mcselected/.append style={inner sep=2.2pt},
    mcadded/.append style={inner sep=1.9pt}}
}

\title{Optimal Pattern-Avoidance Bounds for Binary Search Trees}

\author{
Anita D\"urr\thanks{ETH Z\"urich, Z\"urich, Switzerland}\and
Arturo Merino\thanks{Department of Computer Science, University of Chile,
Santiago, Chile. Supported by ANID FONDECYT Iniciación No. 11251528.}
\and Karol W\k{e}grzycki\thanks{Max Planck Institute for Informatics,
Saarbr\"ucken, Germany. Supported by the Deutsche Forschungsgemeinschaft
(DFG, German Research Foundation), grant number 559177164.}
\and Sorrachai Yingchareonthawornchai\thanks{Institute for Theoretical Studies, ETH Z\"urich, Z\"urich, Switzerland. Supported by Dr.~Max R\"ossler, the Walter Haefner Foundation, and the ETH Z\"urich Foundation.}}
\date{}

\begin{document}
\pagenumbering{gobble}
\maketitle

\begin{abstract}

Self-adjusting binary search trees (BSTs) perform better than balanced BSTs if the access sequence has additional structure. For example, splay trees serve a sequential access sequence in linear time.  Pattern avoidance captures regularity in an access sequence and unifies several classic examples of efficient BST access, such as deque and traversal sequences. A central question is how well a BST algorithm can exploit a pattern-avoiding access sequence. 

We present an online BST algorithm that serves an access sequence avoiding a permutation pattern $\pi$ in linear time, matching the information-theoretic lower bound. The algorithm does not need to know that pattern $\pi$ a priori. Our results answer an open problem asked by [Kozma 2016, Problem 44, strong variant]. Previously, the best known offline bound was optimal only for constant-size patterns [Berendsohn,  Kozma, and  Opler, STOC'24], and the best known online bound was superlinear [Chalermsook,  Pettie,  Yingchareonthawornchai, SODA'24] even for a constant-size pattern. As a corollary, our BST algorithm yields an optimal sorting
algorithm for a pattern-avoiding sequence that supports online insertions. Previously, an optimal sorting algorithm was achieved using a mergesort-like approach [Opler, FOCS'24] in a batch setting. 

A key technique is to apply recursive split decomposition [Chalermsook, Chuzhoy, Saranurak, ToC'23] jointly in time and space, along with the Marcus-Tardos proof of the Stanley-Wilf conjecture.  We show that the framework can be implemented online using $k$-finger BSTs, which is of independent interest. Finally, this framework also yields a constant-factor-approximation FPT algorithm in the offline setting.

 \end{abstract}
\thispagestyle{empty}

\clearpage
\pagenumbering{arabic}

\section{Introduction}
A self-adjusting binary search tree (BST) is an online algorithm that must handle the input request and readjust its structure without knowing future access requests.   It can outperform balanced BSTs such as AVL and red-black trees~\cite{adelson1962algorithm,Bay72} when the input sequence exhibits structure, patterns, or regularity. Pattern avoidance is a classical notion in enumerative and extremal
combinatorics, with algorithmic roots in Knuth's analysis of stack
sorting~\cite{Knuth68}.
In the BST setting, the study of structured access sequences dates
back to the introduction of splay trees, with classic examples
including sequential access and preorder
traversals~\cite{sleator1985self,Tarjan85}.
These
examples have been generalized to a common framework based on permutation pattern avoidance~\cite{chalermsook2015pattern}.  For a permutation pattern $\pi$, a sequence over $[n]$ is \emph{$\pi$-avoiding} if no subsequence has the same relative order as $\pi$.  For example, the sequence $(1,2,\ldots, n)$ avoids $(2,1)$.  
The central question is how well an online BST can exploit this ``easy'' structure hidden in the sequence.

The growth rate of the avoidance class gives a natural information-theoretic benchmark. Let $\Gamma_n(\pi)$ denote the set of
$\pi$-avoiding permutations of $[n]$. A standard encoding argument\footnote{Fix the initial
tree. A BST execution of cost $C$ can be encoded using $O(C)$ bits.
The accessed permutation can be recovered by replaying the execution.
Consequently, executions of cost at most $C$ can represent at most
$2^{O(C)}$ different permutations.}
implies that every BST algorithm incurs cost
$\Omega(n+\log|\Gamma_n(\pi)|)$ on some input in
$\Gamma_n(\pi)$, even if the algorithm is
offline~\cite{berendsohn2024optimization}. Let $
    s_\pi=\lim_{n\to\infty}|\Gamma_n(\pi)|^{1/n}$ be the Stanley--Wilf limit, which is finite by Klazar's
reduction~\cite{Klazar00} to the F\"uredi--Hajnal
conjecture~\cite{FurediH92} and its subsequent proof by Marcus
and Tardos~\cite{marcus2004excluded}.  This yields
a worst-case lower bound of $\Omega(n+n\log s_{\pi})$ for each fixed $\pi$ and sufficiently large $n$, and raises the question of whether an online BST matches this
information-theoretic lower bound:
\begin{quote}
Is there an online BST algorithm on $n$ keys that serves every
$\pi$-avoiding access sequence of length $m\geq n$ with total cost
$O(m \cdot (1+\log s_\pi))$?
\end{quote}

This question was asked explicitly by Kozma~\cite[Problem~44]{Kozma16}.  Special cases of this question have been studied since the introduction of splay trees and the dynamic optimality 
conjecture~\cite{sleator1985self}.  Table~\ref{tab:bst-pattern-bounds} summarizes prior upper bounds
for permutation inputs. For general pattern avoidance,
Berendsohn, Kozma, and Opler~\cite{berendsohn2024optimization}
established the offline bound $O(n\,s_\pi^{O(1)}) = O_{\pi}(n)$.
In contrast, the best-known online bound remained superlinear:
Chalermsook, Pettie, and
Yingchareonthawornchai~\cite{chalermsook2024sorting}
proved an online BST cost bound of
$n\,2^{O(|\pi|^2)+(1+o(1))\alpha(n)}$,
where $\alpha$ is the inverse-Ackermann function. The pattern length $|\pi| := k$ and the Stanley--Wilf limit $s_\pi$
measure different aspects of the permutation structure.
Although $\Omega(k^2)\leq s_\pi\leq 2^{O(k)}$, the value of
$s_\pi$ depends on the pattern itself, not merely its
length~\cite{fox2013stanley}.

\paragraph{Broader Context: Amortized analysis and pattern avoidance.}  Resolving this question is an important step toward a broader challenge in the beyond-worst-case analysis of online algorithms and data structures: identifying input structure that permits stronger guarantees and designing algorithms that exploit it
without knowing future requests. Potential functions are a central technique in amortized analysis, but designing a suitable potential often requires substantial problem-specific insight.  The state-of-the-art understanding of potential function design is much more ad hoc and mostly tailored to specific cases. Indeed, no systematic method is known for designing a potential function. The longstanding difficulty of resolving this problem highlights the limitations of existing techniques for analyzing how online data structures adapt to input structure.

Extremal combinatorics methods such as forbidden submatrix theory have been used successfully in amortized analysis as an alternative to potential function design~\cite{pettie2010applications,chalermsook2015pattern,chalermsook2023improved,berendsohn2024optimization,opler2024sorting}.  
The best-known online upper bound is obtained by analyzing the cost of Greedy~\cite{demaine2009geometry}, an online BST that is $O(1)$-competitive with the natural offline greedy algorithm~\cite{lucas1988canonical,munro2000competitiveness}. A binary matrix $M$ \emph{contains} a binary matrix $P$ if $P$
can be obtained from $M$ by deleting rows and columns, preserving
their order, and changing some remaining one-entries to zero.
Otherwise, $M$ \emph{avoids} $P$.  

Chalermsook, Goswami, Kozma, Mehlhorn, and Saranurak~\cite{chalermsook2015pattern} showed that
the total cost of Greedy on a $\pi$-avoiding permutation
of length $n$ is bounded by
$O(\operatorname{ex}(n,P_\pi\otimes H))$, where $P_\pi$
is the permutation matrix of $\pi$, $\otimes$ denotes
the Kronecker product, and
\[
H=\begin{pmatrix}
0 & 1 & 0\\
1 & 0 & 1
\end{pmatrix}.
\]
Here, $\operatorname{ex}(n,Q)$ denotes the maximum number
of ones in an $n\times n$ binary matrix avoiding $Q$.
A connection to Davenport--Schinzel sequences yields
a near-linear upper bound~\cite{FurediH92,pettie2015three}.
Recently, Chalermsook, Pettie, and
Yingchareonthawornchai~\cite{chalermsook2024sorting}
established nearly tight bounds on this extremal function,
improving the upper bound to \[\operatorname{ex}(n,P_\pi\otimes H) \leq n\,2^{O(|\pi|^2)+(1+o(1))\alpha(n)}.\]

For the best-known offline upper bound, twin-width provides a structural framework for efficient algorithms on permutations, often through a bounded-width decomposition~\cite{GM14}. For BST access, Berendsohn, Kozma, and
Opler~\cite{berendsohn2024optimization} showed that a
permutation of length $n$ and twin-width $d$ admits an
offline BST execution of cost $O(nd^2)$. Since every $\pi$-avoiding permutation has twin-width
polynomially bounded in $s_\pi$~\cite{GM14,Cibulka09},
this yields a bound of
 $O(n\,s_\pi^{O(1)})=O_\pi(n)$.

\paragraph{Barrier for Improvements.} Despite these advances, obtaining the optimal online bound requires overcoming limitations of both approaches. For the forbidden-matrix analysis of Greedy,
known lower bounds show that the superlinear
inverse-Ackermann dependence is unavoidable in the relevant
extremal function~\cite{chalermsook2024sorting}.
Thus, refining this extremal estimate alone cannot yield
linear cost, although Greedy itself may still achieve
the optimal bound.
On the other hand, the twin-width approach relies on a decomposition
of the entire input and yields an offline construction
with polynomial dependence on $s_\pi$.
Reaching the desired bound therefore requires exploiting
this structure without future knowledge while reducing
the parameter dependence from polynomial to logarithmic.

\begin{table}[!th]
\centering
\small
\setlength{\tabcolsep}{5pt}
\renewcommand{\arraystretch}{1.22}
\begin{tabularx}{\linewidth}{
  @{}
  >{\raggedright\arraybackslash}p{2.3cm}
  >{\raggedright\arraybackslash}p{2.3cm}
  >{\raggedright\arraybackslash}X
  >{\raggedright\arraybackslash}p{3.8cm}
  @{}
}
\toprule
\textbf{Sequence}
& \textbf{Pattern(s)}
& \textbf{Online upper bound}
& \textbf{Remarks (offline)} \\
\midrule

Sequential
& $21$
& $O(n)$ (Splay, \textsc{Greedy})
  ~\cite{Tarjan85,chalermsook2015pattern}.
&  \\
\addlinespace

Deque  
& $213,231$
& $O(n)$~\cite{chalermsook2015pattern}.
&  \\
\addlinespace

Preorder  
& $231$
& $O(n)$ (\textsc{Greedy})
  ~\cite{chalermsook2015pattern}.
&  \\
\addlinespace

Postorder  
& $312$
& $O(n)$ (\textsc{Greedy})
  ~\cite{chalermsook2023improved}.
&  \\
\addlinespace

Separable
& $2413,3142$
& $O(n)$ (\textsc{Greedy})
  ~\cite{chalermsook2015pattern,goyal2019better}.
&  \\
\addlinespace

$k$-Separable
& All simple permutations of length $>k$
& $O(n\log k)$ (\textsc{Greedy})
  ~\cite{chalermsook2018multi,goyal2019better}.
&    $O(n\log k)$
  ~\cite{chalermsook2015pattern}. \\
\addlinespace

$k$-increasing
& $k(k-1)\cdots1$
& $O(n\min\{k^2,k\alpha(n)\})$ (\textsc{Greedy})
  ~\cite{chalermsook2023improved}.
&    $O(nk\log k)$~\cite{chalermsook2018multi};
   $O(nk\alpha(n))$~\cite{chalermsook2023improved}. \\
\addlinespace

General pattern
& $\pi$
& $n\,2^{O(|\pi|^2)+(1+o(1))\alpha(n)}$ (\textsc{Greedy})
  ~\cite{chalermsook2024sorting}.
&    
  $O(n\,\operatorname{poly}(s_\pi))$
  \cite{berendsohn2024optimization}. \\
\midrule

\textbf{General pattern}
& $\pi$
& $\boldsymbol{O(n (1+\log s_\pi))}$ \textbf{(This work).}
& \textbf{Optimal} \\

\bottomrule
\end{tabularx}

\caption{Online upper bounds for pattern-avoiding BST access
permutations of $[n]$. All bounds are total costs, with $m=n$.
Input-independent $O(n)$ preprocessing is allowed. A permutation $\pi$ of $[n]$ is \emph{simple} if no block of
consecutive positions of size between $2$ and $n-1$ has values
forming a set of consecutive integers.}
\label{tab:bst-pattern-bounds}

\end{table}

\paragraph{Our Results.}

In this paper, we close this gap.  

\begin{restatable}{theorem}{onlineBST} \label{thm:main}
\label{thm:online-bst-bound}
\label{thm:online-bst-main}
There is an online BST algorithm on $n$ keys that serves a
$\pi$-avoiding access sequence of length $m\geq n$ in 
$O(m(1+ \log s_\pi))$ cost and time. Furthermore, the algorithm is pattern-agnostic, which means it does not take $\pi$ as input. 
\end{restatable} 

The cost of a BST algorithm is the total number of pointer movements and rotations it performs while serving the access sequence, counting each access as at least one unit of cost. 
Here, time complexity refers to the time the algorithm spends, including outside the BST model, to run the BST execution. Our result implies optimal bounds for all previously studied patterns shown in \Cref{tab:bst-pattern-bounds} and answers an open problem asked by Kozma~\cite[Problem~44]{Kozma16}: whether avoiding any fixed pattern guarantees linear online cost in an online BST.

In terms of techniques, our approach builds on \emph{key-split
decomposition}~\cite{ChalermsookCS20}: partition the keys into consecutive ranges, recursively solve the access sequence within each range and the compressed sequence recording which range each request visits, and combine the resulting solutions. Chalermsook, Chuzhoy, and  Saranurak~\cite[Section~1.2]{ChalermsookCS20} suggested that decomposing in both space and time could lead
to improved algorithms. Our work realizes this direction for pattern-avoiding sequences:
we combine partitions of the key space and the access sequence
with the Marcus--Tardos theorem~\cite{marcus2004excluded} to derive a recurrence that
yields the optimal bound. 

A key technical challenge is implementing online \emph{time-split decomposition}. A time-split decomposition partitions the access sequence into consecutive intervals, solves each interval locally, and combines these solutions through a smaller recursive instance.  Chalermsook, Chuzhoy, and Saranurak~\cite{ChalermsookCS20}
showed how to implement online key-split decomposition which is in the key
space.
To handle time intervals, our second ingredient is
$k$-finger BSTs (see~\cite{chalermsook2018multi} and \cref{sec:preliminaries_kfingerBST}): we recursively
serve the first visit to each key range within an interval
and keep fingers on those keys to serve repeated visits.

Furthermore, our bound yields a guarantee in terms of twin-width.
The connection between twin-width and pattern avoidance, together
with Cibulka's bound~\cite{Cibulka09}, implies that every
permutation of twin-width at most $d$ avoids a grid pattern $\pi$
with Stanley--Wilf limit $s_\pi=2^{O(d)}$.
Thus, $\log s_\pi=O(d)$, yielding the following
corollary.

\begin{corollary}
For $d \geq 1$, the same BST algorithm serves any
permutation access sequence of length $n$ and twin-width
at most $d$ with cost $O(n d)$, without knowing $d$ a priori.
\end{corollary}

Our corollary improves the dependence from $O(nd^2)$~\cite{berendsohn2024optimization} to $O(nd)$,
without requiring prior knowledge of $d$ or a decomposition
as part of the input. Thus, twin-width bounds the cost of BST access even when
a certificate of this structure is not supplied. Moreover, this bound is achieved online, without knowledge
of future accesses.

\paragraph{Consequences for the dynamic optimality conjecture.}

The dynamic optimality conjecture postulates the existence of
an online BST whose cost is within a universal constant factor
of the offline optimum for every sufficiently long access
sequence~\cite{sleator1985self}.
Let $\OPT_{bst}(X)$ denote the minimum cost of serving $X$ by an offline BST algorithm that knows the entire sequence in advance.

\begin{corollary} We obtain two new bounds for the offline optimum. 
\begin{itemize}
    \item For every $\pi$-avoiding sequence $X$ of length
$m \geq n$ on $n$ keys,  $OPT_{bst}(X) = O(m\log(1+s_{\pi}))$.
    \item If $X$ is a permutation of length $n$ with twin-width $d \geq 1$, then  $OPT_{bst}(X) = O(n d)$. 
\end{itemize}
\end{corollary}

For every fixed pattern $\pi$, \Cref{thm:main} gives an
$O_\pi(1)$-competitive online BST on $\pi$-avoiding sequences,
since $\OPT_{bst}(X)\geq \Omega(|X|)$.
Crucially, a single algorithm achieves this guarantee without
knowing $\pi$, with the optimal dependence on $s_\pi$.
The previously known offline linear bound already implied that
any dynamically optimal BST must serve every fixed
pattern-avoidance class in linear cost~\cite{berendsohn2024optimization}.
Our result sharpens this implication to
$O(m(1+\log s_\pi))$.
Thus, proving dynamic optimality for splay trees or Greedy
would establish this optimal pattern-dependent bound for
that algorithm as well.

\paragraph{Consequence for sorting.} In fact, we prove a stronger statement than \Cref{thm:main}:
our construction is \emph{insertion-compatible}, a notion
introduced by Kozma and Saranurak~\cite{kozma2019smooth} (see \cref{sec:BST}).
Informally, a key $x$ is never touched before it is first
requested, so its presence is not needed to serve earlier
requests. Any insertion-compatible BST can be transformed into a BST algorithm operating in the \emph{insert-only mode} with $O(1)$ overhead. The insert-only mode refers to a sequence of $n$ keys being inserted into an initially empty BST.  As a corollary, we obtain a BST-based sorting algorithm by running the BST algorithm in \Cref{thm:main} in insert-only mode on the input sequence, then reading off the sorted list by in-order traversal in an additional $O(n)$ time.

\begin{corollary}[Optimal sorting by online BST insertion]
\label{cor:opt tree sort}
There is an online, pattern-agnostic BST insertion strategy
that, starting from an empty tree, processes any $\pi$-avoiding
sequence $X$ of length $n$ in input order with time $O(n(1+\log s_\pi))$. Each insertion depends only on the input received so far, and the strategy does not take $\pi$ as input. Repeated keys are represented by multiplicities at their
corresponding nodes. A final in-order traversal outputs $X$ in sorted order in $O(n)$ additional time. If $X$ is a permutation of twin-width at most $d\geq1$,
the total cost and time are $O(nd)$, without prior knowledge
of $d$ or a twin-width decomposition.
\end{corollary}

Previously, 
Opler~\cite{opler2024sorting} established the same asymptotic
bounds for general comparison-based sorting in the batch
setting, without prior knowledge of the avoided pattern. \Cref{cor:opt tree sort} shows that these bounds remain achievable through online insertion into an initially empty BST, without knowing the avoided pattern or future requests. Our result shows that online BST insertion can exploit global regularity through local tree operations as keys arrive, matching
the information-theoretic sorting bound for every fixed
pattern-avoidance class.

\paragraph{Consequence for offline BST algorithm.}
We also obtain an offline approximation in fixed parameter time parameterized by $\pi$.

\begin{theorem}\label{thm:fpt-approx} 
  There is an algorithm that computes a BST execution serving a $\pi$-avoiding permutation $X$ with cost at most $O(\mathrm{OPT}(X))$ in time $2^{\poly(s_\pi)} \cdot n$.
\end{theorem}

For general access sequences, Chalermsook, Chuzhoy, and
Saranurak~\cite{ChalermsookCS20} obtained a constant-factor approximation to the
optimal offline BST cost in subexponential time in $n$.  For permutations
supplied with a twin-width decomposition of width $d$, the construction of
Berendsohn et al.~\cite{berendsohn2024optimization} gives a polynomial-time
$O(d^2)$-approximation.  In contrast, \Cref{thm:fpt-approx} gives a truly
$\Oh(1)$-approximation in fixed parameter time (by either twin-width or pattern
length) without any knowledge of $\pi$.

\subsection{Convention}\label{sec:convention}
We have so far expressed our bounds in terms of the Stanley--Wilf
limit $s_\pi$. For the proofs and analysis, it is more convenient
to use the \emph{F\"uredi--Hajnal limit} $c_\pi$.
Let $\operatorname{ex}_\pi(r)$ denote the maximum number of ones
in an $r\times r$ binary matrix avoiding the permutation matrix
$P_\pi$. The Marcus--Tardos theorem~\cite{marcus2004excluded},
together with superadditivity, guarantees that the limit
\[
    c_\pi := \lim_{r\to\infty}
    \frac{\operatorname{ex}_\pi(r)}{r}
\]
exists and is finite. Thus, $c_\pi$ depends only on $\pi$.

Cibulka~\cite{Cibulka09} proved that these two parameters are
polynomially related: $
    c_\pi=O(s_\pi^{4.5})$
    and $ s_\pi=O(c_\pi^2)$
with universal implicit constants. Since $c_\pi\geq2$ for
patterns of length at least two, these inequalities imply
$1+\log s_\pi=\Theta(\log c_\pi),$
again with universal implicit constants. Consequently,
replacing $1+\log s_\pi$ by $\log c_\pi$ changes our bounds
only by constant factors. We use $c_\pi$ throughout the
rest of the paper.

\subsection{Further Related Work}
 \paragraph{Dynamic Optimality.}
Research on dynamic optimality has developed both competitive
BST algorithms and lower bounds on the offline optimum.
Wilber~\cite{wilber1989lower} introduced the alternation and
funnel bounds, which remain central to this line of work.
Using the alternation bound, Demaine, Harmon, Iacono, and
P{\u{a}}tra{\c{s}}cu~\cite{demaine2007dynamic} introduced
Tango trees, the first $O(\log\log n)$-competitive online BSTs.
Wang, Derryberry, and Sleator~\cite{wang2006competitive}
subsequently obtained the same competitive ratio with
multi-splay trees, which also achieve linear cost for
sequential access.
Lecomte and Weinstein~\cite{lecomte2020settling} proved that
the funnel bound dominates the alternation bound and exhibited
a tight $\Theta(\log\log n)$ separation between them.
Jiang, Lecomte, Weinstein, and
Yingchareonthawornchai~\cite{jiang2024hardness} established
direct-sum theorems for both bounds and showed that Tango
trees are optimal among algorithms whose analyses charge
their cost to the alternation bound.

Another approach is to develop general frameworks for adaptive
BST guarantees. Chalermsook, Gupta, Jiamjitrak, Pareek, and
Yingchareonthawornchai~\cite{chalermsook2024group} introduced
the \emph{group access bounds}, which generalize the classical
access lemma through a hierarchy of groups of keys.
They proved that every group access tree admits an online
BST simulation with constant-factor overhead.
Their framework yields an $O(\sqrt{\log n})$-competitive
BST and versions of the $k$-finger and unified bounds with
additive overhead.

Recent work has also established sublogarithmic competitive
ratios for the two main candidates for dynamic optimality.
Chmel et al.~\cite{chmel2026splay} proved that splay trees are
$O(\log\log n\cdot(\log\log\log n)^2)$-competitive by comparing
their cost directly with the offline optimum.
Guo, Pettie, Skora, and Wan~\cite{guo2026greedy} proved that
\textsc{Greedy} is $2^{O(\sqrt{\log\log n})}$-competitive,
giving its first $o(\log n)$ competitive ratio.
Their analysis relates costs at different scales and uses
Wilber's interleave lower bound.
\paragraph{Sorting pattern-avoiding permutations.}
The comparison complexity of sorting a restricted family of
permutations was studied by Fredman~\cite{fredman1976information}.
For any known family $\Gamma$ of permutations of length $n$,
his result gives a comparison decision tree of depth
$O(n+\log|\Gamma|)$. Applied to $\Gamma_n(\pi)$, this yields
the bound $O(n(1+\log s_\pi))$ on the number of comparisons.
However, this decision-tree guarantee does not by itself
provide an algorithm with the same running time or remove
the need to know the input family.
Arthur~\cite{arthur2007fast} developed early algorithms
specifically for sorting pattern-avoiding permutations.
His framework gives linear-time sorting for several patterns
of length three and four, including $123$ and $1234$, and
$O(n\log\log\log n)$ time for several other patterns of length
four, including $1324$. These algorithms exploit the structure
of a prescribed avoided pattern. Kozma and Saranurak~\cite{kozma2019smooth} introduced
\emph{smooth heaps} through a dual view of heaps and BSTs
that exchanges the roles of time and key space.
In sorting mode, the input keys are inserted into the heap
and then extracted in increasing order by repeated
delete-min operations. They showed that the cost of
smooth-heap sorting on a permutation $X$ matches, up to
constant factors, the cost of \textsc{GreedyFuture} on
the inverse permutation $X^{-1}$.
Since $X$ avoids $\pi$ if and only if $X^{-1}$ avoids
$\pi^{-1}$, this correspondence transfers pattern-avoidance
bounds for BSTs to heap-based sorting.
Smooth heaps therefore provide a simple sorting algorithm
that adapts to pattern avoidance without knowing the
avoided pattern.
In particular, the analysis of Chalermsook, Pettie, and
Yingchareonthawornchai~\cite{chalermsook2024sorting} yields
the running-time bound
$n\,2^{O(|\pi|^2)+(1+o(1))\alpha(n)}$
for smooth-heap sorting of $\pi$-avoiding permutations.

\subsection{Technical overview}
\label{sec:technical-overview}

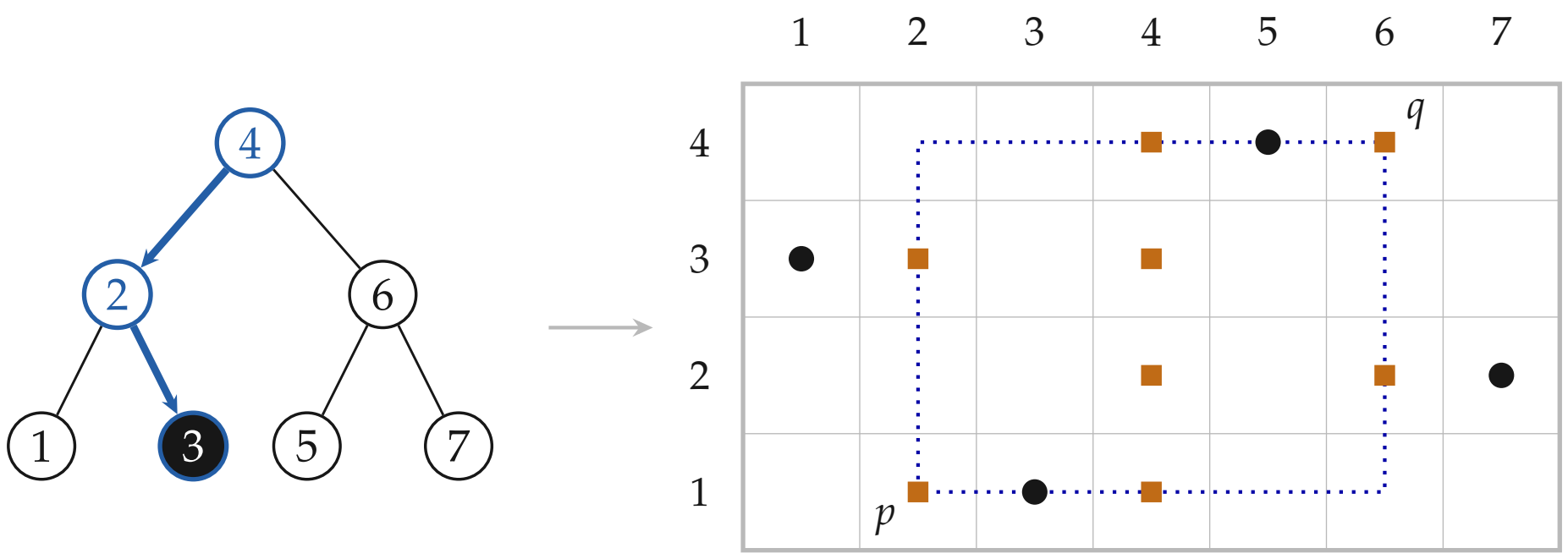
\begin{figure}[H]
\centering
\begin{tikzpicture}[mcmainfig,x=.65cm,y=.65cm,
  visited/.style={mcnode,draw=bstpath,text=bstpath,line width=1pt},
  requested/.style={mcnode,draw=bstpath,fill=mcink,text=white,line width=1pt,font=\small\bfseries},
  searchpath/.style={draw=bstpath,line width=1.65pt,-{Stealth[length=4pt,width=3.7pt]}},
  tracepoint/.style={rectangle,draw=bstadded,fill=bstadded,minimum size=4.2pt,inner sep=0pt}]
  \definecolor{bstpath}{HTML}{245EA6}
  \definecolor{bstadded}{HTML}{C06B16}

\node[visited] (n4) at (3,4.3) {$4$};
  \node[visited] (n2) at (1.6,2.7) {$2$};
  \node[mcnode] (n6) at (4.4,2.7) {$6$};
  \node[mcnode] (n1) at (.8,1.1) {$1$};
  \node[requested] (n3) at (2.4,1.1) {$3$};
  \node[mcnode] (n5) at (3.6,1.1) {$5$};
  \node[mcnode] (n7) at (5.2,1.1) {$7$};
  \draw[mcaxis] (n4)--(n6) (n2)--(n1) (n6)--(n5) (n6)--(n7);
  \draw[searchpath] (n4)--(n2);
  \draw[searchpath] (n2)--(n3);

  \draw[mcarrow] (6.15,2.35)--(7.25,2.35);
  \begin{scope}[shift={(8.2,0)},x=.80cm,y=.80cm]
\draw[mcbox] (0,0) rectangle (7,4);
    \foreach \x in {1,...,6} \draw[mcgrid] (\x,0)--(\x,4);
    \foreach \t in {1,...,3} \draw[mcgrid] (0,\t)--(7,\t);
    \foreach \x in {1,...,7} \node[mclabel] at ({\x-.5},4.45) {$\x$};
    \foreach \t in {1,...,4} \node[mclabel,anchor=east] at (-.18,{\t-.5}) {$\t$};

    \draw[draw=blue!65!black,dotted,line width=.8pt] (1.5,.5) rectangle (5.5,3.5);

\foreach \t/\x in {1/2,1/4,2/4,2/6,3/2,3/4,4/4,4/6}
      \node[tracepoint] at ({\x-.5},{\t-.5}) {};
    \foreach \t/\x in {1/3,2/7,3/1,4/5}
      \node[mcpoint] at ({\x-.5},{\t-.5}) {};
    \node[mclabel,inner sep=.6pt,anchor=north east,xshift=-4pt,yshift=-2pt] at (1.5,.5) {$p$};
    \node[mclabel,inner sep=.6pt,anchor=south west,xshift=4pt,yshift=2pt] at (5.5,3.5) {$q$};
\end{scope}
\end{tikzpicture}
\caption{A BST serving $X=(3,7,1,5)$ without rotations and its geometric trace. The blue path $4\to2\to3$ gives the bottom row. Black circles mark requests and orange squares mark other touched keys. The dotted rectangle spanned by $p$ and $q$ contains a third point at key $4$, illustrating arboral satisfaction.}
\label{fig:bst_to_manhattan}
\end{figure}

We use the geometric view of BSTs~\cite{demaine2009geometry} (see~\cref{sec:geometricBST}).  Represent
$X=(x_1,\ldots,x_m)\in[n]^m$ by $P_X=\{(t,x_t):t\in[m]\}$ (row $t$, column
$x_t$).  Following the notation of Demaine et al.~\cite{demaine2009geometry}, a
solution $S_X\supseteq P_X$ is \emph{arborally satisfied} if every rectangle
spanned by two of its points in different rows and columns contains a third
point of $S_X$ (see~\Cref{fig:bst_to_manhattan}). Its size measures BST cost
up to constant factors. We write $\OPT(X)=\OPT(P_X)$ for the minimum number
of added points.

Our starting point is a rectangular consequence of the Marcus–Tardos theorem~\cite{marcus2004excluded}: an \(m\times n\) binary matrix avoiding \(P_\pi\) has at most \(c_\pi\max\{m,n\}\) ones.  Recall that a binary matrix $M$ \emph{contains} a binary matrix $P$ if we
can select rows and columns of $M$, keeping their original
order, to obtain a submatrix of the same size as $P$ that has
a one wherever $P$ has a one. Otherwise, $M$ \emph{avoids} $P$.

The next ingredient we will use is the framework of Chalermsook, Chuzhoy, and
Saranurak~\cite{ChalermsookCS20}. To give an $\Oh(\log\log{n})$-competitive
algorithm for BSTs, they partition the columns into consecutive super-columns of size
$\Theta(\sqrt{n})$, solve each super-column and the instance obtained by contracting
the super-columns, and combine the solutions using \emph{glue} points (points on
the boundaries that connect solutions). To bound the competitive ratio, they show that the
local and contracted optima sum to at most the original optimum, and there are
$\Oh(\OPT(X))$ glue points. 
We extend their framework with the following two ideas:
\begin{compactdesc}
    \item[\textbf{Idea 1:}] \textbf{Block contraction.}
          We contract both super-rows and super-columns of width
          $\Theta(c_\pi)$ (compared with the super-column width
          $\Theta(\sqrt{n})$ in~\cite{ChalermsookCS20}). For pattern-avoiding inputs,
          Marcus--Tardos limits how many blocks are occupied, giving a
          small recursive instance $Z$ with one representative per block.
          Solve the super-rows and super-columns locally and glue their solutions, with gluing
          inspired by Berendsohn et al.~\cite{berendsohn2024optimization}.

    \item[\textbf{Idea 2:}] \textbf{$k$-finger simulation.} In the online algorithm,
      future requests are unknown, so we cannot solve an entire super-row
      in advance. Instead, we serve its super-column requests with an $\Oh(c_\pi\log c_\pi)$-finger BST and each super-row visits at most $\Oh(c_\pi\log c_\pi)$ distinct super-columns. 
      The simulation of Chalermsook et al.~\cite[Theorem~2]{chalermsook2018multi} turns this strategy
      into a standard BST with $\Oh(\log c_\pi)$ overhead.
      We apply it to arbitrary pattern-avoiding inputs.
  \end{compactdesc}
Next, we explain both ideas in more detail.

\begin{figure}[h]
\centering
\begin{subfigure}[b]{.32\textwidth}
\centering
\begin{tikzpicture}[mcfig,x=.40cm,y=.32cm]
  \useasboundingbox (-2.2,-.2) rectangle (8.2,10.2);

\begin{scope}[shift={(0,1.9)}]
    \draw[mcbox] (0,0) rectangle (6,6);
\foreach \x in {3,5} \draw[mcgrid] (\x,0)--(\x,6);
    \foreach \x in {1,2,4} \draw[mcguide,draw=mcink] (\x,0)--(\x,6);
\foreach \y in {2,5} \draw[mcgrid] (0,\y)--(6,\y);
    \foreach \y in {1,3,4} \draw[mcguide,draw=mcink] (0,\y)--(6,\y);
    \foreach \x/\y in {1/1,3/2,5/3,2/4,4/5,6/6}{
      \node[mcpoint] at ({\x-.5},{\y-.5}) {};
    }
  \end{scope}
\node[mclabel] at (3,9.3) {$Z$};
\end{tikzpicture}
\caption{Recurse on $Z$}
\label{fig:technical-overview-recursive}
\end{subfigure}
\hfill
\begin{subfigure}[b]{.32\textwidth}
\centering
\begin{tikzpicture}[mcfig,x=.40cm,y=.32cm]
  \useasboundingbox (-2,-.2) rectangle (8.4,10.2);
\foreach \i in {0,...,3}{
    \begin{scope}[shift={(0,{2.6*\i})}]
      \draw[mcbox] (0,0) rectangle (8,2);
      \foreach \x in {1,3,5,7} \draw[mcgrid] (\x,0)--(\x,2);
      \foreach \x in {2,4,6} \draw[mcguide,draw=mcink] (\x,0)--(\x,2);
      \draw[mcgrid] (0,1)--(8,1);
      \node[mclabel,anchor=east] at (-.15,1) {$R_{\the\numexpr\i+1\relax}$};
    \end{scope}
  }
  \foreach \x/\y in {1/1,2/2,5/3,7/4,3/5,4/6,6/7,8/8}{
    \node[mcpoint] at ({\x-.5},{\y-.5+.6*floor((\y-1)/2)}) {};
  }
  \foreach \x/\y in {1/2,5/4,3/6,6/8}{
    \node[mcadded] at ({\x-.5},{\y-.5+.6*floor((\y-1)/2)}) {};
  }
\end{tikzpicture}
\caption{Solve each $R_i$}
\label{fig:technical-overview-rows}
\end{subfigure}
\hfill
\begin{subfigure}[b]{.32\textwidth}
\centering
\begin{tikzpicture}[mcfig,x=.40cm,y=.32cm]
  \useasboundingbox (-.3,-.2) rectangle (10.1,10.2);

\begin{scope}[shift={(0,.9)}]
    \foreach \j in {0,...,3}{
      \begin{scope}[shift={({2.6*\j},0)}]
        \draw[mcbox] (0,0) rectangle (2,8);
        \foreach \y in {1,3,5,7} \draw[mcgrid] (0,\y)--(2,\y);
        \foreach \y in {2,4,6} \draw[mcguide,draw=mcink] (0,\y)--(2,\y);
        \draw[mcgrid] (1,0)--(1,8);
        \node[mclabel,anchor=south,yshift=3pt] at (1,8) {$C_{\the\numexpr\j+1\relax}$};
      \end{scope}
    }
    \foreach \x/\y in {1/1,2/2,5/3,7/4,3/5,4/6,6/7,8/8}{
      \node[mcpoint] at ({\x-.5+.6*floor((\x-1)/2)},{\y-.5}) {};
    }
    \foreach \x/\y in {1/2,3/6,5/7,7/8}{
      \node[mcadded] at ({\x-.5+.6*floor((\x-1)/2)},{\y-.5}) {};
    }
  \end{scope}
\end{tikzpicture}
\caption{Solve each $C_j$}
\label{fig:technical-overview-columns}
\end{subfigure}
\caption{Requests are black and local additions white (super-column widths and super-row heights are $2$). We omit glue from this picture.}
\label{fig:technical-overview-contraction}
\end{figure}
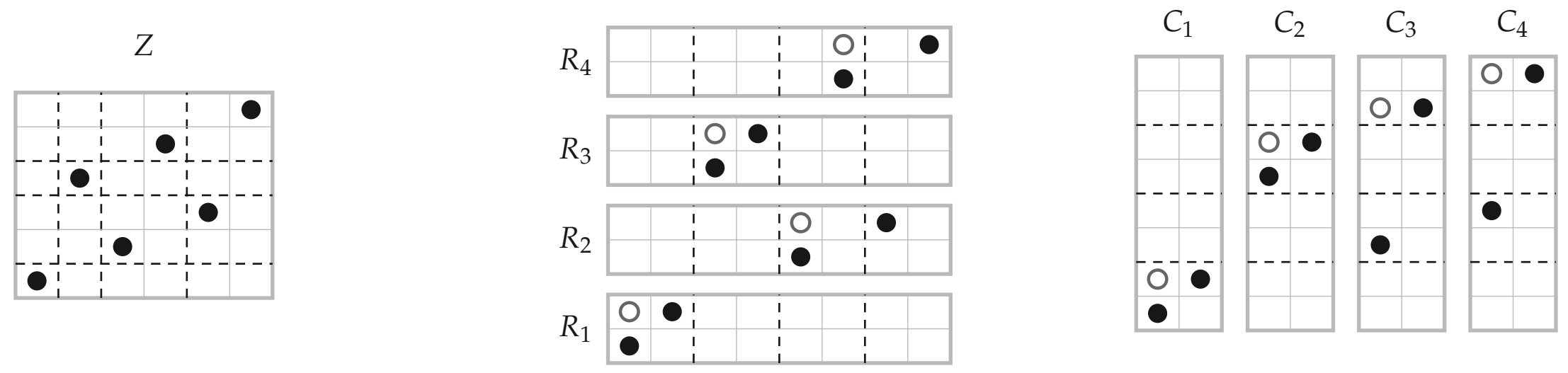

\paragraph{Idea 1: Block contraction.}
We present a new block contraction procedure that achieves a tight \emph{offline} bound. 
Consider a $\pi$-avoiding permutation $X$ of $[n]$ and let
$d=\lceil32c_\pi\rceil$.  We partition the columns into super-columns $C_j$ of
$d$ columns and partition the rows into super-rows $R_i$ of $d$ rows. Using a balanced
BST we can solve each super-column and super-row locally, using $\Oh(n\log d)$ points in total.  
By keeping an original request per occupied block, and renumbering the selected rows and
keys accordingly, we obtain a permutation $Z$. 
The key observation is that $Z$ is $\pi$-avoiding and has size $|Z| \leq 2c_\pi n/d$.
To see this, consider the $n/d\times n/d$ matrix $A$ whose $(i,j)$ entry is $1$ if the block $R_i\times C_j$ contains
a request and $0$ otherwise. 
By construction, $A$ avoids $\pi$, so Marcus--Tardos gives $|Z|\leq2c_\pi n/d\leq n/16$.

The idea is therefore to solve $Z$ recursively, and solve each super-row and each super-column locally. See~\Cref{fig:technical-overview-contraction}.
To aggregate the obtained solutions, we first \am{combine the solution for $Z$ with the local super-row solutions while keeping super-columns contracted, using glue points on the top and bottom boundaries of super-rows. We then lift the resulting solution to the original grid using copies on the left and right boundaries of super-columns and adding the local super-column solutions.}

For the analysis, the maximum total solution size $N_\pi(n)$ over $\pi$-avoiding permutations
with at most $n$ points satisfies
\[
    N_\pi(n)\leq4N_\pi(\lfloor n/16\rfloor)+\Oh(n\log c_\pi).
\]
This recurrence solves to $N_\pi(n)=\Oh(n\log c_\pi)$. See~\Cref{sec:offline_warmup} for details.

\paragraph{Idea 2: $k$-finger simulation}
As mentioned before, the above block contraction idea cannot directly be implemented in the online setting. Indeed, in the online model points on each row arrive one by one, from bottom to top, and the algorithm can only put points on the current row. Furthermore, this decision is irrevocable. 
In particular, the algorithm cannot wait to see the entire super-row to solve it locally and then add points, since this requires adding points on previous rows.\footnote{A similar issue arises for adding the four corner points when lifting the solution of $Z$. We omit this issue in this overview.}

To address this issue, we define a different block-contraction procedure. Super-columns $C_j$ are defined as above. Contracting them yields a point set $Y$, where $y_t=\lceil x_t/d\rceil$.
Super-rows, on the other hand, are defined using a greedy algorithm on $Y$ (inspired by
Opler~\cite{opler2024sorting}) such that a super-row contains at most $d$ distinct super-columns.
By keeping the first occurrence of each super-column within each super-row, we obtain a point set $Z \subset Y$.

Similarly to the offline analysis, we can show that $Z$ is $\pi$-avoiding and $|Z|\leq n/16$, so that $Z$ can be solved recursively. Now, instead of solving each super-row locally, we need to design a new online procedure to solve $Y$. We do this by exploiting the $k$-finger BST model (see~\cref{sec:preliminaries_kfingerBST}). In the $k$-finger BST model, there are $k$ pointers, called fingers, that keep their position on the tree between requests, as opposed to a single pointer starting at the root at each request in the standard BST model.
For each request of $Z$, we can thus use one working finger to serve the key, according to the solution of $Z$ recursively computed, and additionally place a cache finger on the served key. Since $Y$ contains subsequent requests of the keys in $Z$ within each super-row, subsequent repeated requests can be served using the placed finger in a single access operation. Finally, on the last row of each super-row, we need to bring all the fingers back to the root to prepare for the execution of the next super-row. Surprisingly, we show that this can be done with only a constant-factor overhead, independent of the number of fingers $k$.
Finally, super-columns can be solved locally using balanced BSTs. To aggregate the solutions of distinct super-columns, we first add neighboring points to the solution of $Y$, before lifting this solution to the original grid.  We leave the details to \Cref{sec:online_glueYtoX}.

\paragraph{A constant-factor approximation.}

Now, we briefly sketch an $\Oh(1)$-approximation in FPT time parameterized by $c_\pi$ (\ie \cref{thm:fpt-approx}). We use the offline blocks and
double the value of $d$ until our online algorithm 
finds a solution $S_Z$ of at most $n/10$
points for their reduced permutation $Z$.  The main idea is to solve $X$ within
each $C_j$ and $Y$ within each $R_i$ exactly in exponential time in $c_\pi$
(rather than use a BST algorithm that guarantees a solution of size
$\Oh(n \log{d})$).  Their optimum solutions sum to at most $\OPT(P_X)$, and $S_Z$ allows
us to connect the offline boundary copies cheaply.
See~\Cref{sec:fpt_approximation} for details.

\section{Preliminaries}
\label{sec:preliminaries}

For a positive integer $n$, let $[n]\coloneqq\{1,\ldots,n\}$.

\subsection{Forbidden submatrices}\label{sec:forbidden_matrix}
 
A binary matrix $A$ \emph{contains} a binary matrix $P$ if $P$
can be obtained from $A$ by deleting rows and columns and
changing some remaining one-entries to zero.
Otherwise, $A$ \emph{avoids} $P$.
For a permutation $\pi$ of $[k]$, its permutation matrix
$P_\pi$ has a one in position $(i,\pi(i))$ for every $i\in[k]$
and zeros elsewhere. Throughout, we assume $k\geq2$.

Let $\operatorname{ex}_\pi(n,m)$ denote the maximum number
of ones in an $m\times n$ binary matrix avoiding $P_\pi$,
and write
$\operatorname{ex}_\pi(n):=\operatorname{ex}_\pi(n,n)$.
The Marcus--Tardos theorem~\cite{marcus2004excluded} states
that $\operatorname{ex}_\pi(n)=\Oh_\pi(n)$.
Together with superadditivity, this implies that the
\emph{F\"uredi--Hajnal limit}
\[
    c_\pi
    :=\lim_{r\to\infty}\frac{\operatorname{ex}_\pi(r)}{r}
    =\sup_{r\geq1}\frac{\operatorname{ex}_\pi(r)}{r}
\]
exists and is finite~\cite{Cibulka09}.
In particular, $\operatorname{ex}_\pi(n)\leq c_\pi n$
for every $n$, and $c_\pi\geq2$.
We use the following rectangular formulation.

\begin{theorem}[Marcus--Tardos;
    {\cite[Lemma~2.1]{opler2024sorting}}]
\label{lem:rectangular-furedi-hajnal}
For every permutation $\pi$ and positive integers $n,m$,
\[
    \operatorname{ex}_\pi(n,m)
    \leq c_\pi\max\{n,m\}.
\]
Equivalently, every $m\times n$ binary matrix with more than
$c_\pi\max\{n,m\}$ ones contains $P_\pi$.
\end{theorem}

Cibulka~\cite{Cibulka09} proved that $s_\pi$ and $c_\pi$
are polynomially related. Since $c_\pi\geq2$, this gives
\[
    1+\log s_\pi=\Theta(\log c_\pi),
\]
with universal implicit constants.
Thus, expressing our bounds in terms of $\log c_\pi$
instead of $1+\log s_\pi$ changes them only by constant
factors.

This convention also preserves the information-theoretic
benchmark for BSTs.
An execution of cost $C$, starting from a fixed initial tree,
can be encoded using $O(C)$ bits.
Counting $\pi$-avoiding permutations, together with the
unavoidable cost of the $n$ accesses, therefore yields
a worst-case lower bound of
\[
    \Omega\bigl(n(1+\log s_\pi)\bigr)
    =\Omega(n\log c_\pi),
\]
even for offline BST algorithms, for every fixed $\pi$
and all sufficiently large $n$.

Our analysis is naturally expressed in terms of the
F\"uredi--Hajnal limit $c_\pi$, a matrix extremal parameter
closely related to the Stanley--Wilf limit $s_\pi$. We adopt $c_{\pi}$ throughout the
remainder of the paper.

\subsection{Binary Search Tree}\label{sec:BST}

A \emph{binary search tree} (BST) is a rooted binary tree storing one key per node, with smaller keys in the left subtree and larger keys in the right subtree of every node. 

We use the following pointer model.
Each search starts with a single pointer at the root.
The unit-cost BST operations are moving the pointer to its
left child, right child, or parent; performing a left or right
rotation at the current node; and accessing the current key.
A rotation leaves the pointer at the same node.
A BST algorithm serves $X=(x_1,\ldots,x_m)\in[n]^m$ if it
accesses $x_t$ while processing each request $x_t$.
Its \emph{BST cost} is the total number of BST operations
performed, including at least one access operation per request.

For implementation with time complexity, we use the \emph{real-world
BST model}~\cite{BoseCFL12,iacono2013pursuit}.
The algorithm operates on a pointer machine with $O(1)$
additional words per node and $O(1)$ global working words,
each containing $O(\log n)$ bits.
Its decisions depend only on the current request, its global
state, and the data at the current node.
Its \emph{running time} counts all elementary operations,
including comparisons, updates to auxiliary information,
and computation used to choose and perform BST operations.
We require our implementation to run in time proportional
to its BST cost.

In the \emph{incremental model}, the tree starts empty and additionally supports inserting a requested key as a left or right child of the current node, at unit cost. The child position must be empty, and the insertion must preserve the key order. Insertion leaves the pointer at its current node; moving to and accessing the new leaf are separate operations. The first insertion creates the root and places the pointer there. 
We highlight that keys can only be inserted at their first request.

An \emph{online BST algorithm} is given the key universe $[n]$, but the length $m$ of future requests is not known upfront. The online algorithm receives the requests one at a time and must serve $x_t$ before receiving $x_{t+1}$. Its choices depend only on the requests received so far, its current state, and its fixed parameters. We write $X_{\leq t}\coloneqq\{x_1,\ldots,x_t\}$ for the set of revealed keys, with $X_{\leq0} \coloneq \emptyset$. 
Thus, in the incremental setting, since keys are only inserted at their first request, an incremental online BST stores exactly $X_{\leq t}$ after serving $x_t$.

We will use the following standard facts about BSTs. 

\begin{lemma}[{\cite{adelson1962algorithm}}]\label{lem:balanced-incremental-bst}
Any sequence $X\in[n]^m$ can be served by a balanced online incremental BST with cost and time $\Oh(n+m\log(n))$.
\end{lemma}

\begin{lemma}[{\cite{culik1982note}}]\label{lem:bst-transformation}
Given two BSTs on the same $q$ keys, the first can be transformed into the second with cost and time $\Oh(q)$.
\end{lemma}

A sequence $X$ is $\pi$-avoiding if no subsequence of $X$ has the same relative order as the permutation pattern $\pi$.
{Note that if the sequence $X$ contains repeated keys, since the entries of $\pi$ are distinct, an occurrence of $\pi$ in $X$ uses pairwise distinct keys, so no occurrence can use the same key twice.}

\subsection{Geometric view of BSTs}\label{sec:geometricBST}

To help us reason about BST executions, we use the geometric view of Demaine et al.~\cite{demaine2009geometry}. 
A point $p$ refers to a point in the plane $p = (p_x, p_y) \in \R^2$. 
For two points $p,q \in \R^2$, let $\rect{p}{q}$ be the closed axis-parallel rectangle with corners $p$ and $q$. 
\begin{definition}[{\cite[Definition~2.3]{demaine2009geometry}}] 
A pair of points $p,q\in P$ is \emph{arborally satisfied} with respect to $P$ if the points share a row or column, or $\rect{p}{q}$ contains a point of $P\setminus\{p,q\}$. A point set $P$ is \emph{arborally satisfied} if every pair of its points is arborally satisfied with respect to $P$.
\end{definition}
When it is clear from context, we omit specifying ``with respect to $P$''.
In our context, we write grid points in $(\mathrm{row},\mathrm{column})$ order: $(t,x)$ lies on row $t\in[m]$ and column $x\in[n]$.
We illustrate them on an $m\times n$ grid, with columns numbered from left to right and rows from bottom to top.

The \emph{arborally satisfied} problem asks to find for a given point set $P$ an arborally satisfied point superset $S \supset P$. In that case, we say that $S$ is a \emph{solution} to $P$. 
This problem was introduced in \cite{demaine2009geometry} to provide a geometric view of a BST execution in the following sense.

A row or column of a point set is said to be \emph{active} if it contains a point. A \emph{semi-permutation} has exactly one point in each active row; a \emph{permutation point set} also has exactly one point in each active column~\cite[Section~2.1]{ChalermsookCS20}. A point set avoids $\pi$ if its binary matrix, ordered by active rows and columns, avoids $P_\pi$.
Identify a sequence of requests $X=(x_1,\ldots,x_m)\in[n]^m$ with the point set $P_X\coloneqq\{(t,x_t):t\in[m]\}$.
Then in particular, $P_X$ is a semi-permutation, and $X$ is $\pi$-avoiding if and only if $P_X$ is.

The \emph{geometric trace} of a BST execution $E$ is $P(E)\coloneqq\{(t,x):x\text{ is touched while serving }x_t\}$ (see \cite[Definition~2.4]{demaine2009geometry}). Here we say that a key is \emph{touched} if the pointer visits its node, it is inserted, or it is incident to a rotated edge.
Thus, if $X$ is served by $E$ then $P_X\subseteq P(E)$.
Demaine et al.~\cite[Lemmas~2.1 and~2.2]{demaine2009geometry} showed that any BST execution serving $X$ with cost $T$ can be mapped to a solution $S_X$ to $P_X$ of size $|S_X| \leq \Oh(T)$. Moreover, this mapping is a bijection and can be computed in time $\Oh(T)$. See \cref{fig:bst_to_manhattan} for an example.
To extend this equivalence to the incremental and online setting of BSTs, we first define the equivalent notions for the arborally satisfied problem.

In the \emph{online arborally satisfied} problem, input points $(t,x_t)$ arrive one row at a time. 
After receiving $(t,x_t)$, the task is to output a set of points $S_t \subseteq \{t\} \times [n]$ on the current row such that $(t, x_t) \in S_t$ and the prefix $\bigcup_{\ell \leq t} S_{\ell}$ is arborally satisfied (see~\cite[Definition~2.5]{demaine2009geometry}). 
A column $j$ is \emph{revealed} at time $t$ if there is a request $(\ell, j)$ with $\ell \leq t$. We call the output \emph{insertion-compatible} if every point on row $t$ lies in a column revealed at time $t$, \ie $S_t\subseteq\{t\}\times X_{\leq t}$ for every $t$. This corresponds to the validity restriction of~\cite[Definitions~4 and~7]{DBLP:journals/corr/ChalermsookG0MS15a} in the insertion-only model. 
A \emph{solution} to the online arborally satisfied problem is therefore an insertion-compatible solution $S \supset X$ such that for every $t \in [m]$ the prefix $\bigcup_{\ell \leq t} S_{\ell}$ is arborally satisfied.

It is shown in \cite[Lemma~2.3]{demaine2009geometry} that the equivalence between the arborally satisfied problem and BST execution extends to the online setting. 
In \cite[Theorem C.1]{kozma2019smooth}, the equivalence was further extended to the incremental setting; however, only in the offline case. It is a simple exercise to extend the equivalence to the incremental online setting, as we show in \cref{sec:geometricBST_appendix}.

\begin{restatable}{lemma}{BSTgeom}[Adaptation of {\cite[Lemma~2.1]{demaine2009geometry}}]\label{lem:BSTgeom}
Given an online incremental BST algorithm serving $X$ with cost and time at most $T$, its geometric trace is an insertion-compatible online solution $S_X\supseteq P_X$ with $|S_X|=\Oh(T)$, computable online in time $\Oh(n+T)$.
\end{restatable}

\begin{restatable}{lemma}{geomBST}[Adaptation of {\cite[Lemma~2.2]{demaine2009geometry}}]\label{lem:geomBST}
Given an online algorithm producing an insertion-compatible solution $S_X\supseteq P_X$ of size $|S_X| \leq T$ running in time at most $T$, there exists an online incremental BST serving $X$ with cost and time $\Oh(n +T)$.
\end{restatable}

Since the two viewpoints are equivalent, we therefore use $X$ interchangeably for the sequence of requests and the set of points $P_X$. 
Combining \cref{lem:balanced-incremental-bst,lem:BSTgeom} gives the following useful fact. 

\begin{lemma}\label{lem:balancedBST_gridpoint}
Given a point set $P$ with $k\geq2$ points, there is an algorithm that constructs an arborally satisfied superset $S\supseteq P$, using only active rows and columns of $P$, of size $\Oh(k\log k)$ in time $\Oh(k\log k)$.
\end{lemma}

\subsection{\texorpdfstring{$k$}{k}-finger BST}\label{sec:preliminaries_kfingerBST}

A \emph{$k$-finger BST} is a variant of the BST model with $k$ pointers, called \emph{fingers}. The positions of the fingers persist between requests. For each operation, the algorithm chooses a finger and  performs a move, rotation, or access operation on it, at unit cost. 
Fingers may point to the same node. A rotation keeps every finger at the same node, so its key does not change. Accessing the requested key with any finger serves the request. 

Chalermsook et al.~\cite[Theorem~2]{chalermsook2018multi} show how to simulate a given $k$-finger BST execution by a standard BST with $\Oh(\log k)$ amortized overhead in cost and time per operation. The simulation processes operations one at a time and uses no future requests. In particular, given an online $k$-finger BST serving $X$ with cost and time at most $T$, there is an online BST serving $X$ with cost and time $\Oh(T\log k)$.

In the proof of \cref{thm:online-bst-bound} we exploit this simulation in the incremental model.
In that setting, the tree starts empty, and a finger may insert a new key at one of its null children at unit cost. 
The first insertion creates the root on which all fingers are then placed. As for standard BSTs, a key can only be inserted at its first request. We show in \cref{app:online_kfinger_simulation} that the $k$-finger BST simulation in \cite[Theorem~2]{chalermsook2018multi} can easily be extended to the incremental model.

\begin{restatable}{theorem}{kfingerSimulataion}[Adaptation of {\cite[Theorem~2]{chalermsook2018multi}}]\label{lem:online_kfinger_simulation}
Given an online incremental $k$-finger BST algorithm serving $X\in[n]^m$ with cost and time at most $T$, there exists an online incremental BST algorithm serving $X$ with cost and time $\Oh(T\log k)$.
\end{restatable}

\subsection{Geometric decomposition}\label{sec:geometric-decomposition}

Our proof of \cref{thm:online-bst-bound} uses a generalization of the geometric recursive framework for the arborally satisfied problem given by Chalermsook et al.~\cite{ChalermsookCS20}. The idea is to partition the active columns into consecutive strips, solve the instances inside the strips locally and the instance obtained by contracting each strip to one column, and combine the solutions by adding boundary points. We use their decomposition of the optimum to bound the total cost of the local solutions.

For a finite point set $X$, let $\opt(X)\coloneqq\min\{|S\setminus X|:S\supseteq X\text{ is arborally satisfied}\}.$
Thus, $\opt(X)$ counts only added points, as in~\cite[Section~2.1]{ChalermsookCS20}. When $X$ is a sequence of requests, we write $\opt(X)$ for $\opt(P_X)$. A solution $S\supseteq X$ is \emph{canonical} if every point of $S$ lies on an active row and an active column of $X$~\cite[Section~2.2.2]{ChalermsookCS20}.

\begin{lemma}[{\cite[Observations~2.4 and~2.5]{ChalermsookCS20}}]\label{lem:contraction-canonicalization}
Contracting consecutive active rows or columns of an arborally satisfied point set preserves arboral satisfaction. Any solution for $X$ can be transformed into a canonical solution without increasing the number of added points.
\end{lemma}

The following result bounds the combined optimum of the strip instances and the contracted instance.

\begin{lemma}[{\cite[Theorem~3.3]{ChalermsookCS20}}]\label{lem:strip-decomposition}
Let $X$ be a semi-permutation, partition its active columns into consecutive nonempty intervals $C_1,\ldots,C_q$, and let $X_i$ be its restriction to $C_i$. If $\widetilde X$ is obtained by contracting each $C_i$ to one column, then
\[
    \sum_{i=1}^q\opt(X_i)+\opt(\widetilde X)\leq\opt(X).
\]
\end{lemma}

We also use the following elementary lower bound.

\begin{lemma}\label{lem:linear-optimum}
If $X$ is a permutation point set of size $n\geq1$, then $\opt(X)\geq n-1$.
\end{lemma}
\begin{proof}
Let $S\supseteq X$ be any finite arborally satisfied set. Form a bipartite graph with one vertex for each active row and each active column of $S$, and an edge for each point of $S$. The graph is connected: otherwise, choose points $p,q$ whose edges lie in different components, minimizing $|p_x-q_x|+|p_y-q_y|$. A third point in $\rect{p}{q}$ gives a closer pair whose edges lie in different components, a contradiction. Since $X$ uses $n$ distinct rows and $n$ distinct columns, the graph has at least $2n$ vertices. Hence $|S|\geq2n-1$, so $|S\setminus X|\geq n-1$.
\end{proof}

\section{Offline Warmup}
\label{sec:offline_warmup}
\label{sec:offline-warmup}
In this section we prove an existential version of \Cref{thm:online-bst-bound} as a warmup for \Cref{sec:online-algorithm}.
We use the geometric view from \Cref{sec:geometricBST}.

\begin{proposition}[Offline version of~\cref{thm:online-bst-bound}]\label{prop:offline-warmup}
Given a $\pi$-avoiding permutation $X$ of $[n]$, there is a solution $S_X$ of $X$ with size $\Oh(n\log c_\pi)$; i.e.,
\[
\opt(X) = \Oh(n\log c_\pi).
\]
\end{proposition}

Our approach is to recursively reduce the size of the input permutation by contracting blocks of side-length $\Theta(c_\pi)$.
Thus, we fix $d=\lceil32c_\pi\rceil$ as the side-length of the blocks and set $q\coloneq\lceil n/d\rceil$ as the number of strips in each dimension.
We partition the columns into consecutive \emph{vertical strips} $V_j\coloneq\{(j-1)d+1,\ldots,jd\}\cap[n]$ for $j\in[q]$, and partition the rows in the same way into \emph{horizontal strips} $H_1,\ldots,H_q$.\footnote{As mentioned in \cref{sec:technical-overview}, a similar partition is used in the online algorithm of \cref{sec:online_algorithm}. Notably, the vertical strips are the same. However to avoid confusion, we use distinct names in the offline (vertical and horizontal strips) and in the online setting (super-columns and super-rows).
}

Naively, we would like to contract each block and recurse; however, that need not produce a permutation.
Instead, we say that a \emph{block} $H_i\times V_j$ is occupied if it contains a point of $X$.
Then, our \emph{block reduction} selects one representative point of $X$ per occupied block and renumbers the selected rows and columns to obtain a $\pi$-avoiding permutation $Z$.
We recursively solve $Z$ and lift its solution back to $X$.

The permutation $Z$ has few points and we need to add only a small number of points to lift its solution to a solution for $X$.

\begin{lemma}\label{lem:offline-block-reduction}
Let $X$ be a $\pi$-avoiding permutation of $[n]$ and let $d=\lceil32c_\pi\rceil$.
There is a $\pi$-avoiding permutation $Z$ on at most $n/16$ elements such that
every solution $S_Z$ for $Z$ admits a solution $S_X$ for $X$ with
\[
    |S_X|\leq4|S_Z|+\Oh(n\log d).
\]
\end{lemma}

Since we can handle the base cases using the balanced BST approach of \Cref{lem:balancedBST_gridpoint}, we have the ingredients to prove \Cref{prop:offline-warmup}.

\begin{proofof}{\Cref{prop:offline-warmup} assuming \Cref{lem:offline-block-reduction}}
If $|X| \leq 1$, it suffices to take $S_X=X$.
Thus, we may assume $n\geq2$.

If $2d\geq n$, solve $X$ directly using \Cref{lem:balancedBST_gridpoint}, obtaining a solution of size $\Oh(n\log n)$ which is $\Oh(n \log c_\pi)$.
Otherwise, recursively solve $Z$ and lift its solution using \Cref{lem:offline-block-reduction}.
Since $|Z|\leq n/16$, the recursion terminates, and correctness follows by induction.

Let $N_\pi(n)$ be the maximum size of the resulting solution, counting all points, over $\pi$-avoiding permutations with at most $n$ points.
The direct case has $n\leq2d=\Oh(c_\pi)$, so its size is $\Oh(n\log c_\pi)$.
In the recursive case, $d=\Oh(c_\pi)$ and \Cref{lem:offline-block-reduction} give
\[
    N_\pi(n)\leq4N_\pi(\lfloor n/16\rfloor)+\Oh(n\log c_\pi).
\]
This recurrence sums to $N_\pi(n)=\Oh(n\log c_\pi)$, proving the proposition.
\end{proofof}

It remains to prove \Cref{lem:offline-block-reduction}.
We can bound the size of $Z$ using the Marcus--Tardos theorem, so most of the work is in lifting a solution for $Z$ to a solution for $X$. Since we obtain $Z$ in two stages (contracting vertical strips, and then selected points per horizontal strip), we lift the solution of $Z$ in two stages as well, see~\Cref{fig:offline-warmup-lifting}.

\begin{figure}[H]
\centering
\begin{tikzpicture}[mcfig,x=.40cm,y=.40cm,
  warmupinput/.style={circle,fill=mcink,draw=none,inner sep=1.5pt},
  warmuprecursive/.style={rectangle,draw=warmuporange,fill=warmuporange,minimum size=4.6pt,inner sep=0pt},
  warmuphorizontal/.style={isosceles triangle,shape border rotate=90,draw=warmupteal,fill=warmupteal,minimum width=5.7pt,minimum height=5.2pt,inner sep=0pt},
  warmupvertical/.style={text=warmupblue,font=\normalsize,inner sep=0pt},
  warmupinputprojection/.style={circle,draw=mcink,fill=white,line width=.75pt,minimum size=5.5pt,inner sep=0pt},
  warmuprecursiveprojection/.style={rectangle,draw=warmuporange,fill=white,line width=.75pt,minimum size=5.5pt,inner sep=0pt},
  warmuphorizontalprojection/.style={warmuphorizontal,fill=white,line width=.75pt}]
\definecolor{warmuporange}{HTML}{C06B16}
  \definecolor{warmupteal}{HTML}{23817A}
  \definecolor{warmupblue}{HTML}{245EA6}
  \definecolor{warmupinactive}{HTML}{E6E6E6}

\def\WarmupInputs{3/1,1/2,5/3,7/4,8/5,9/6,6/7,2/8,4/9}
  \def\WarmupRepresentatives{3/1,5/3,7/4,6/7,2/8}
  \def\WarmupRecursive{2/1,5/1,5/4,6/4,5/7,5/8}
  \def\WarmupAInputs{1/1,2/1,3/2,1/3,2/3}
  \def\WarmupAAdded{2/2}
  \def\WarmupYInputs{1/1,1/2,2/3,3/4,3/5,3/6,2/7,1/8,2/9}
  \def\WarmupYHorizontal{1/3,2/8}
  \def\WarmupYInputProjections{2/1,1/7,1/9}
  \def\WarmupYRecursiveProjections{2/4,2/6}
  \def\WarmupXVertical{2/1,2/2,8/4,8/6,5/7,5/9}
  \def\WarmupXInputProjections{1/1,4/1,6/1,3/2,4/3,6/3,9/4,7/5,9/5,7/6,1/7,3/7,4/7,1/8,3/8,1/9,3/9,6/9}
  \def\WarmupXHorizontalProjections{1/3,3/3,4/8,6/8}
  \def\WarmupXRecursiveProjections{4/4,6/4,4/6,6/6}

  \newcommand{\WarmupPoints}[2]{\edef\WarmupPointList{#2}\foreach \x/\t in \WarmupPointList
      \node[#1] at ({\x-.5},{\t-.5}) {};}
  \newcommand{\WarmupGrid}[2]{\draw[mcbox] (0,0) rectangle (#1,#2);
    \pgfmathtruncatemacro{\WarmupLastColumn}{#1-1}
    \pgfmathtruncatemacro{\WarmupLastRow}{#2-1}
    \foreach \x in {1,...,\WarmupLastColumn}
      \draw[mcgrid] (\x,0)--(\x,#2);
    \foreach \t in {1,...,\WarmupLastRow}
      \draw[mcgrid] (0,\t)--(#1,\t);}

\begin{scope}
    \foreach \x in {1,4,8,9}
      \fill[warmupinactive] ({\x-1},0) rectangle (\x,9);
    \foreach \t in {2,5,6,9}
      \fill[warmupinactive] (0,{\t-1}) rectangle (9,\t);
    \WarmupGrid{9}{9}
    \foreach \x in {3,6} \draw[mcguide,draw=mcink] (\x,0)--(\x,9);
    \foreach \t in {3,6} \draw[mcguide,draw=mcink] (0,\t)--(9,\t);
    \WarmupPoints{warmuprecursive}{\WarmupRecursive}
    \WarmupPoints{warmupinput}{\WarmupRepresentatives}
    \node at (4.5,10.1) {$S_Z$};
  \end{scope}

\begin{scope}[shift={(13,3)}]
    \WarmupGrid{3}{3}
    \WarmupPoints{warmuprecursive}{\WarmupAAdded}
    \WarmupPoints{warmupinput}{\WarmupAInputs}
    \node at (1.5,7.1) {$S_A$};
  \end{scope}

\begin{scope}[shift={(22,0)}]
    \WarmupGrid{3}{9}
    \foreach \t in {3,6} \draw[mcguide,draw=mcink] (0,\t)--(3,\t);
    \WarmupPoints{warmuprecursiveprojection}{\WarmupYRecursiveProjections}
    \WarmupPoints{warmupinputprojection}{\WarmupYInputProjections}
    \WarmupPoints{warmuphorizontal}{\WarmupYHorizontal}
    \WarmupPoints{warmupinput}{\WarmupYInputs}
    \node at (1.5,10.1) {$S_Y$};
  \end{scope}

\begin{scope}[shift={(31,0)}]
    \WarmupGrid{9}{9}
    \foreach \x in {3,6} \draw[mcguide,draw=mcink] (\x,0)--(\x,9);
    \foreach \t in {3,6} \draw[mcguide,draw=mcink] (0,\t)--(9,\t);
    \WarmupPoints{warmuprecursiveprojection}{\WarmupXRecursiveProjections}
    \WarmupPoints{warmuphorizontalprojection}{\WarmupXHorizontalProjections}
    \WarmupPoints{warmupinputprojection}{\WarmupXInputProjections}
    \foreach \x/\t in \WarmupXVertical
      \node[warmupvertical] at ({\x-.5},{\t-.5}) {$\times$};
    \WarmupPoints{warmupinput}{\WarmupInputs}
    \node at (4.5,10.1) {$S_X$};
  \end{scope}

  \draw[mcarrow] (9.5,4.5)--(12.5,4.5);
  \draw[mcarrow] (16.5,4.5)--(21.5,4.5);
  \draw[mcarrow] (25.5,4.5)--(30.5,4.5);

\node[warmupinput] at (.5,-1.1) {};
  \node[anchor=west,font=\footnotesize] at (1.1,-1.1) {input};
  \node[warmuprecursive] at (9.4,-1.2) {};
  \node[anchor=west,font=\footnotesize] at (10,-1.1) {recursive};
  \node[warmuphorizontal] at (20.4,-1.2) {};
  \node[anchor=west,font=\footnotesize] at (21,-1.1) {horizontal local};
  \node[warmupvertical] at (33,-1.2) {$\times$};
  \node[anchor=west,font=\footnotesize] at (33.8,-1.1) {vertical local};
\end{tikzpicture}
\caption{The two lifts for $d=3$. 
Contracting the blocks of $S_Z$ gives $S_A$. We obtain $S_Y$ by projecting $S_A$ to the top and bottom rows of each horizontal strip (marked as hollow squares and circles) and adding the contracted horizontal local solutions (marked as triangles). We obtain $S_X$ by projecting $S_Y$ to the left and right columns of each vertical strip (marked as hollow triangles and circles) and adding the vertical local solutions (marked as crosses).
Inactive rows and columns in $S_Z$ are grey. Coincident points are drawn once, giving priority to inputs, then local additions, then projections of inputs.
}
\label{fig:offline-warmup-lifting}
\end{figure}
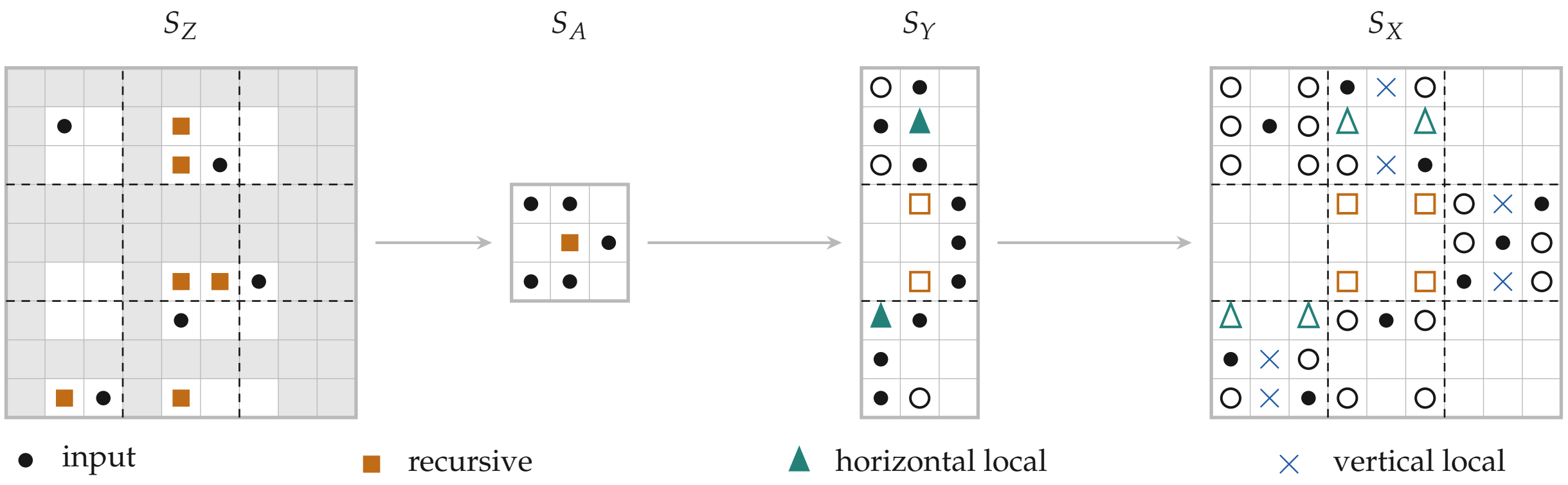 
\begin{proofof}{\Cref{lem:offline-block-reduction}}
Write $X=(x_1,\ldots,x_n)$.
Contracting the vertical strips gives $Y=(y_1,\ldots,y_n)$, where $y_t=\lceil x_t/d\rceil$.
We represent request $x_t$ by a point $(t,x_t)$ on row $t$ and column $x_t$.
Thus $P_X=\{(t,x_t):t\in[n]\}$; we define $P_Y,P_Z$ analogously,
numbering the rows of each sequence consecutively.

\textbf{Block contraction.}
If $n\leq2d$, choose any subset of at most $n/16$ points of $X$ to form $Z$ and return a solution for $X$ of size $\Oh(n\log d)$ using \Cref{lem:balancedBST_gridpoint}.
Thus, we may assume $n>2d$.
Keeping one point per occupied block and renumbering the selected coordinates gives a $\pi$-avoiding permutation.
We set $q=\lceil n/d\rceil$ and define $A$ as the $q\times q$ binary matrix whose entry $A_{ij}$ is $1$ precisely when $H_i$ contains a request to $V_j$.
This matrix has $|Z|$ ones and avoids $\pi$: we can lift an occurrence to $X$ by choosing one request in each corresponding block.
Thus, \Cref{lem:rectangular-furedi-hajnal} gives $|Z|\leq c_\pi q\leq2c_\pi n/d{}\leq n/16$.

\textbf{Solving $Y$ via $Z$.}
We first construct $S_Y$ from $S_Z$.
We set $P_A=\{(i,j):A_{ij}=1\}$.
We restore the selected row and column coordinates and contract each $V_j$ to column $j$ and each $H_i$ to row $i$, mapping $P_Z$ onto $P_A$ and, by \Cref{lem:contraction-canonicalization}, $S_Z$ to a solution $S_A$ with $|S_A|\leq|S_Z|$.
We form glue points by copying each $(i,j)\in S_A$ to $(\min H_i,j)$ and $(\max H_i,j)$, and add the column-contracted horizontal local solutions, obtaining $S_Y{}\supseteq P_Y$ with $|S_Y|\leq2|S_Z|+\Oh(n\log d)$.

We now prove that $S_Y$ is arborally satisfied.
A local point $(t,j)$ with $t\in H_i$ uses a key $j$ occurring in $H_i$, so $A_{ij}=1$ and row contraction maps it to $(i,j)\in P_A\subseteq S_A$.
For two non-aligned points in the same horizontal strip, we use their local solution if both are local points.
If one of them is a glue point, we use the other point's copy on that same boundary row.
For non-aligned points in different horizontal strips, their images under row contraction belong to $S_A$ and have a third point in their rectangle.
We use its top copy when its row corresponds to the lower endpoint's horizontal strip, its bottom copy for the upper endpoint's horizontal strip, and either copy for an intermediate horizontal strip.
In each case we obtain a third point in the original rectangle, so $S_Y$ is arborally satisfied.

\textbf{Solving $X$ via $Y$.}
To construct $S_X$ from $S_Y$, we define $S_L$ as the union of the local points in the vertical strips.
For each $(t,j)\in S_Y$, we add the two glue points $\LB(t,j)=(t,\min V_j)$ and $\RB(t,j)=(t,\max V_j)$ on the left and right boundaries of $V_j$ to a set $S_G$.
We set $S_X{}\coloneq P_X\cup{} S_L\cup S_G$.
We then have $P_X\subseteq S_X$ and $|S_X|\leq2|S_Y|+\Oh(n\log d)$.
Contracting the vertical strips maps every point of $S_X$ into $S_Y$, because each local point lies on an input row of its vertical strip.

We now prove arboral satisfaction of $S_X$ by considering non-aligned points $p=(s,a),q=(t,b)\in S_X$.
If they are in the same vertical strip, we obtain a third point in $R[p,q]$ from their local solution when both belong to $S_L$, and otherwise from the glue point $(t,a)$, assuming $p\in S_G$.
Next, consider the case when $p,q$ are in distinct vertical strips $V_i,V_j$.
Without loss of generality assume $i<j$.
Column contraction maps them to $p'=(s,i),q'=(t,j)\in S_Y$.
Since $S_Y$ is arborally satisfied, we choose a third point $r'=(\ell,k)\in S_Y\cap R[p',q']$ and use $\RB(r')$ if $k=i$, $\LB(r')$ if $k=j$, and either copy if $i<k<j$.
This is a third point in $R[p,q]$: if $k=i$ then $\ell\neq s$, and if $k=j$ then $\ell\neq t$.
Thus $S_X$ is a solution for $P_X$ with $|S_X|\leq4|S_Z|+\Oh(n\log d)$.
\end{proofof}

We end by noting that it is straightforward to turn \Cref{prop:offline-warmup} into a recursive algorithm that runs in time $\Oh(n\log c_\pi)$.
Further, by guessing the value of $c_\pi$ the algorithm can be made oblivious to $\pi$.
 
\section{Online Algorithm}
\label{sec:online_algorithm}
\label{sec:online-algorithm}
In this section we prove the following bound in terms of the F\"uredi--Hajnal limit.
\begin{theorem}\label{thm:online-cpi-bound}
    There is an online BST algorithm that, given neither $\pi$ nor $\cpi$, serves every $\pi$-avoiding semi-permutation $X \in [n]^m$ with cost and time $\Oh((n+m)\log \cpi)$.
\end{theorem}
Therefore, \cref{thm:online-bst-bound} follows from the relation $\cpi = (s_\pi)^{O(1)}$ (see~\cref{sec:convention}).
As in \cref{sec:offline-warmup}, we construct a smaller $\pi$-avoiding semi-permutation and extend an execution serving it to one serving $X$.
The following lemma gives an online version of this reduction.

\begin{lemma}\label{lem:online_existenceZ}
    Let $X$ be a $\pi$-avoiding semi-permutation on $n$ keys of length $m = |X|$, and let $d$ be an integer with $n>d>2\cpi$. There exists a $\pi$-avoiding semi-permutation $Z$ on at most $\lceil n / d \rceil$ keys of length $|Z| \leq 2 \cpi  n /d $ such that
    \begin{enumerate}
        \item The sequence $Z$ can be computed online in time $\Oh(n + m)$. 
        \item If there exists an online incremental BST algorithm serving $Z$ with cost and time $T$, then there exists an online  incremental BST algorithm serving $X$ with cost and time $\Oh(n + (m + T) \log d)$.
    \end{enumerate}
\end{lemma}

Recall that any sequence $X$ of length $m$ on $n$ keys can be served by a balanced incremental BST with cost and time $\Oh(m \log n)$. 
So the idea is to recursively apply \cref{lem:online_existenceZ} until $Z$ has at most $d$ keys, solve the base case using a balanced incremental BST and then combine each recursive call using the algorithm in \cref{lem:online_existenceZ}.  
We will show that we can achieve cost and time $\Oh((n + m) \log \cpi)$ by carefully tuning the parameter $d$. However, the optimal value $d$ depends on $\cpi$ (note that \cref{lem:online_existenceZ} requires $d > 2 \cpi$), which is not given as input. We show in \cref{lem:online_guessCpi} that this is not an issue since we can efficiently guess $\cpi$.

\begin{restatable}{lemma}{onlineCpi}
    \label{lem:online_guessCpi}
    Let $X$ be a semi-permutation on $n$ keys of length $m = |X|$. 
    Suppose that for every $c \geq 2$ there exists an online incremental BST algorithm $\Aa_c$ serving $X$ with cost and time at most $\Oh((n+m)\log c)$ whenever $X$ avoids a permutation $\pi$ with $\cpi \leq c$. 
    Then there exists an online incremental BST algorithm $\Aa$ serving $X$ with cost and time $\Oh((n+m)\log \cpi)$ for every permutation pattern $\pi$ such that $X$ is $\pi$-avoiding. 
\end{restatable}

Assuming \cref{lem:online_existenceZ,lem:online_guessCpi}, \cref{thm:online-cpi-bound} follows.

\begin{proofof}{\cref{thm:online-cpi-bound} assuming \cref{lem:online_existenceZ,lem:online_guessCpi}}
For every $c \geq 2$, we define an online BST algorithm $\Aa_c$ that serves $X$ with cost and time $\Oh((n+m)\log c)$ whenever $X$ avoids a permutation $\pi$ with $\cpi \leq c$. 
By \cref{lem:online_guessCpi}, this is sufficient to prove \cref{thm:online-cpi-bound}.

Hence, let $\pi$ be a permutation pattern such that $X$ is $\pi$-avoiding and fix $c \geq \cpi$.
Let $d > 2 c$ be a parameter to tune later.
We define $\Aa_c$ recursively. If $n \leq d$, then serve $X$ using a balanced incremental BST. Otherwise, compute $Z$ using \cref{lem:online_existenceZ}(1), recursively serve $Z$ with the same parameter $d$ and extend the execution to serve $X$ using \cref{lem:online_existenceZ}(2). Since $Z$ is a semi-permutation on at most $\lceil n / d\rceil$ keys, the recursion depth is $\Oh(\log_d n)$. Since $d > 2c \geq 2\cpi$, the correctness of $\Aa_c$ follows from \cref{lem:online_existenceZ} by induction. We analyse its cost and running time. 

Let $s = \max \{n, m\}$ and $T(s)$ be the maximum cost and running time of $\Aa_c$ over sequences with at most $n$ keys and length at most $m$. In the base case $n \leq d$, the balanced incremental BST has cost and time $T(s) \leq \Oh(n + m \log d) \leq \Oh(s \log d)$.
In the recursive case, the sequence $Z$ has at most $\lceil n/d\rceil \leq 2 s/d$ keys and length at most $ 2 \cpi n/d \leq 2 c s / d $. By \cref{lem:online_existenceZ}, the cost and time to serve $X$ are 
$$
T(s) \leq a (s + T( 2 c s / d)) \log d 
$$ 
for some constant $a$.
To obtain the claimed bound, we want to set $d$ so that $a \frac{2c}{d} \log d \leq 1/2$, so we choose $d \coloneq \lceil K c \log c\rceil$  for a sufficiently large constant $K$.
The recurrence then resolves to 
$
T(s) \leq  \Oh(2as \log d) \leq \Oh(s \log c).
$
In particular, $\Aa_c$ serves $X$ with cost and time at most $\Oh((n+m)\log c)$, as desired.
\end{proofof}

The remainder of this section is structured as follows.
\cref{sec:online_blockContraction,sec:online_glueZtoY,sec:online_glueYtoX} focus on establishing \cref{lem:online_existenceZ}. More precisely, in \cref{sec:online_blockContraction} we first define the semi-permutation $Z$ of \cref{lem:online_existenceZ}. In fact, we define two sequences $Y$ and $Z$ such that $Z \subset Y$ and we show how to construct them in linear time in an online setting. We then show in \cref{sec:online_glueZtoY} how to extend an online BST serving $Z$ to an online BST serving $Y$, before showing in \cref{sec:online_glueYtoX} how to extend an online BST serving $Y$ to an online BST serving $X$. Together, this will prove \cref{lem:online_existenceZ}. Finally, we prove \cref{lem:online_guessCpi} in \cref{sec:online_guessCpi}.

\subsection{Block Contraction}\label{sec:online_blockContraction}

Fix a semi-permutation $X=(x_1,\ldots,x_m)\in[n]^m$ and an integer $d \geq 2$. We define sequences $Y$ and $Z$ as in \cref{lem:online_existenceZ} such that they can be computed online.

\begin{figure}[t]
\centering
\begin{tikzpicture}[mcmainfig,x=.55cm,y=.36cm]
  \newcommand{\ExamplePoint}[3][mcink]{\node[mcpoint,fill=#1] at ({#3-.5},{#2-.5}) {};}
  
\draw[mcbox] (0,0) rectangle (6,8);
  \foreach \x in {1,...,5} \draw[mcgrid] (\x,0)--(\x,8);
  \foreach \y in {1,...,7} \draw[mcgrid] (0,\y)--(6,\y);
  \foreach \x in {2,4} \draw[mcguide,draw=mcink] (\x,0)--(\x,8);
\ExamplePoint{1}{1}
  \ExamplePoint{2}{2}
  \ExamplePoint{3}{5}
  \ExamplePoint{4}{4}
  \ExamplePoint{5}{3}
  \ExamplePoint{6}{6}
  \ExamplePoint{7}{5}
  \ExamplePoint{8}{5}
  \foreach \x/\j in {1/1,3/2,5/3} \node[mclabel] at (\x,8.65) {$C_\j$};
  \node[mcpanel] at (3,-.6) {$X$};
  \draw[mcarrow] (6.7,4)--(8.3,4);

\begin{scope}[shift={(10,0)}]
    \draw[mcbox] (0,0) rectangle (3,8);
    \foreach \x in {1,2} \draw[mcgrid] (\x,0)--(\x,8);
    \foreach \y in {1,...,7} \draw[mcgrid] (0,\y)--(3,\y);
    \foreach \y in {3,6} \draw[mcguide,draw=mcink] (0,\y)--(3,\y);
    \ExamplePoint[mcgray]{2}{1}
    \ExamplePoint[mcgray]{5}{2}
    \ExamplePoint[mcgray]{8}{3}
    \ExamplePoint[mcblue]{1}{1}
    \ExamplePoint[mcblue]{3}{3}
    \ExamplePoint[mcblue]{4}{2}
    \ExamplePoint[mcblue]{6}{3}
    \ExamplePoint[mcblue]{7}{3}
    \node[mcpanel] at (1.5,-.6) {$Y$};
    \foreach \y/\i in {1.5/1,4.5/2,7/3}
      \node[mclabel,anchor=west] at (-1.25,\y) {$R_\i$};
  \end{scope}

  \draw[mcarrow] (13.7,4)--(15.3,4);

\begin{scope}[shift={(16,1.5)}]
    \draw[mcbox] (0,0) rectangle (3,5);
    \foreach \x in {1,2} \draw[mcgrid] (\x,0)--(\x,5);
    \foreach \y in {1,...,4} \draw[mcgrid] (0,\y)--(3,\y);
\ExamplePoint[mcblue]{1}{1}
    \ExamplePoint[mcblue]{2}{3}
    \ExamplePoint[mcblue]{3}{2}
    \ExamplePoint[mcblue]{4}{3}
    \ExamplePoint[mcblue]{5}{3}
  \end{scope}
\node[mcpanel] at (17.5,-.6) {$Z$};
\end{tikzpicture}
\caption{Contracting the super-columns of $X$ gives $Y$.
Dark points mark the first occurrence of each super-column within each
super-row; retaining them gives $Z$, with empty rows omitted.}
\label{fig:online_block_contraction}
\end{figure}
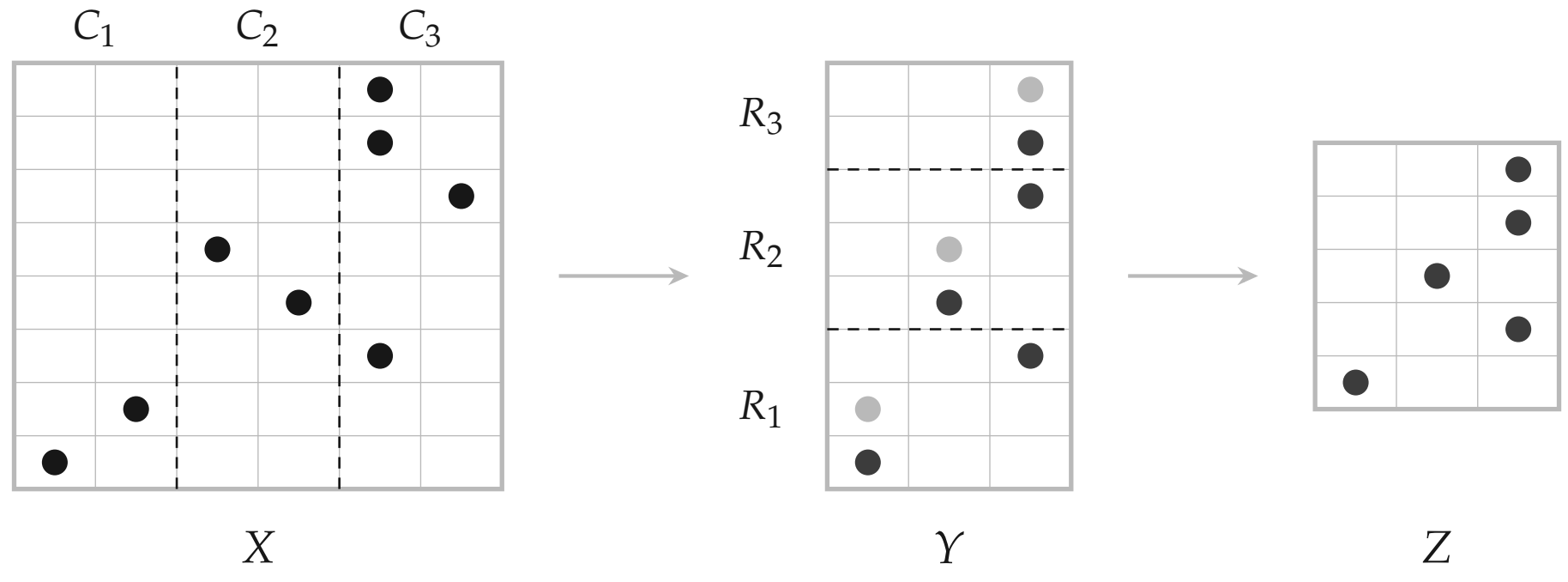

Consider an $m \times n$ grid where rows are numbered bottom-to-top and columns are numbered left-to-right. It will be helpful to represent $X$ as points on this grid, \ie for every $i \in [m]$ there is a point on the $i$-th row and $x_i$-th column. See \cref{fig:online_block_contraction} for an example of $X$ and the corresponding sequences $Y$ and $Z$.

\paragraph*{Super-columns and sequence $Y$.}
Similarly to~\cref{sec:offline-warmup}, we group consecutive columns into \emph{super-columns}, and we define $Y$ as the resulting \emph{column-contraction} of $X$.
More precisely, let $q \coloneq \lceil n/d \rceil$ be the number of super-columns so that the $j$-th super-column is the set of columns $C_j \coloneq \{(j-1)d + 1,  \dots, jd\} \cap [n]$ for $j \in [q]$, and let
$$
Y \coloneq (\lceil x_1 /d \rceil, \dots, \lceil x_m /d \rceil).
$$
The sequence $Y$ has key universe $[q]$ and length $|Y| = m$, \ie $Y \in [q]^m$. 
By abuse of notation, we refer to the keys in $[q]$ as super-columns.

\paragraph*{Super-rows and sequence $Z$.}
Next, we group consecutive rows into \emph{super-rows}, and we define $Z \subset Y$ to contain the first request to each super-column within each super-row. Since in the online setting the final length $m$ is not known in advance
we cannot do this the same way as in~\cref{sec:offline-warmup}.
Instead, we greedily construct super-rows as requests arrive so that the last row of each super-row requests the $d$-th distinct super-column within that super-row. The last super-row may contain fewer than $d$ distinct super-columns. These conditions uniquely determine super-rows $R_1, \dots, R_r$ for some $r \leq \lceil m/d\rceil$.

Within each super-row, mark the first rows that request each super-column. Let $t_1 < \dots < t_\ell$ be all marked rows. They define the sequence
$$
Z \coloneq (y_{t_1}, \dots, y_{t_\ell})
$$
with key universe $[q]$ and length $|Z| = \ell$, \ie $Z \in [q]^{\ell}$. By construction, $Z \subset Y$.

\begin{observation}\label{lem:online_defXYZ}
    Suppose that $X$ is $\pi$-avoiding for some permutation pattern $\pi$. Then $Y$ and $Z$ are also $\pi$-avoiding.
    If additionally $n>d>2\cpi$, then $|Z| \leq 2 \cpi n / d$.
\end{observation}
\begin{proof}
    The first observation holds because column-contraction and row deletion preserve pattern avoidance.
For the length bound, let $A$ be the binary matrix with $r$ rows and $q$ columns whose entry $A_{ij}$ is $1$ precisely when super-row $R_i$ contains a request to super-column $C_j$. This matrix is $\pi$-avoiding: an occurrence of $\pi$ in $A$ could be lifted to an occurrence in $X$ by choosing one request in each corresponding block $R_i\times C_j$.

Suppose that $\am{r > q}$. Then the first $q$ rows of $A$ form a $q\times q$ matrix with exactly $dq$ ones. By \cref{lem:rectangular-furedi-hajnal}, this is at most $\cpi q$, contradicting $d > 2\cpi$. Hence, $r \leq q$.
Each $1$ in $A$ corresponds to exactly one request selected in $Z$. Applying \cref{lem:rectangular-furedi-hajnal} again, we obtain
$$
|Z|=\sum_{i=1}^r |\{y_t:t\in R_i\}| \leq \cpi\max\{q,r\}=\cpi q\leq 2\cpi n/d,
$$
where the last inequality uses $q=\lceil n/d\rceil\leq 2n/d$ for $n>d$.
\end{proof}

\begin{observation}\label{lem:online_constructionXYZ}
    The sequences $Y$ and $Z$ can be constructed online in $O(n+m)$ total time.
\end{observation}
\begin{proof}
    The sequences $Y$ and $Z$ can be computed greedily. For completeness, we describe the algorithm.
Maintain the index $i$ of the current super-row, a counter $k$ of its distinct super-columns, and an array $L[1..q]$, where $L[j]$ records the index of the last super-row in which super-column $C_j$ was requested. Initialize $i=1$, $k=0$, and all array entries to $0$. This takes time $\Oh(n)$.

When $x_t$ arrives, output $y_t = \lceil x_t / d \rceil$ as the next request of $Y$. 
If $L[y_t] \neq i$, this is the first request to that super-column in the current super-row. So select $y_t$ as the next request of $Z$. Set $L[y_t] = i$ and increment $k$. Additionally, if $k = d$, this means that row $t$ closes the current super-row. So we increment $i$ and reset $k=0$. 

These operations implement exactly the definitions of $Y$, $Z$, and the super-rows. They take constant time per request and use no knowledge of future requests or of $m$. 
\end{proof}

\subsection{Solving $Y$ via $Z$: using $k$-finger BSTs}\label{sec:online_glueZtoY}

We now show how a BST serving $Z$ can be extended to a BST serving $Y$. The idea is to exploit the \emph{$k$-finger} variant of BSTs. In this model, instead of having a single pointer starting at the root of the tree at each request, there are $k$ pointers, called \emph{fingers}, that maintain their position between requests. For each operation (move, rotate, insert, access) the algorithm chooses on which finger to perform it.  
See \cref{sec:preliminaries_kfingerBST} for a formal definition. 
We also recall that a $k$-finger BST strategy can be simulated by a standard BST strategy with $\Oh(\log k)$-factor overhead in the cost and running time, see \cref{lem:online_kfinger_simulation}.

In the previous section, we defined super-rows of $Y$ so that they contain requests to at most $d$ distinct super-columns, and $Z$ selects their first occurrences. To serve $Y$, the idea is therefore to first execute the strategy serving $Z$ using a distinguished finger, which we will call the \emph{working finger}. 
Each time the working finger serves a request, we also place a finger, called a \emph{cache finger}, at the key that is accessed. This way, repetitions of a super-column in $Y$ only cost one access. Within a super-row, we therefore only need $d$ cache fingers. 
The main difficulty is to return the cache fingers to the root at the end of each super-row (\ie clearing the cache) 
without performing too many operations.

Indeed, consider a $k$-finger BST strategy $\sigma$ where all fingers start at the root. Since every operation increases the depth of every finger by at most 1, after performing $\sigma$ every finger is positioned at depth at most $\Oh(|\sigma|)$. By following parent edges, we can move all $k$ fingers back to the root in $\Oh(k \cdot |\sigma|)$ total  operations. This is too costly since the number of fingers in our application is $k = \Oh(d)$. Instead, we show in the following lemma that by carefully reversing the operations in $\sigma$, we can bring all fingers back to the root in only $\Oh(|\sigma|)$ total operations. 

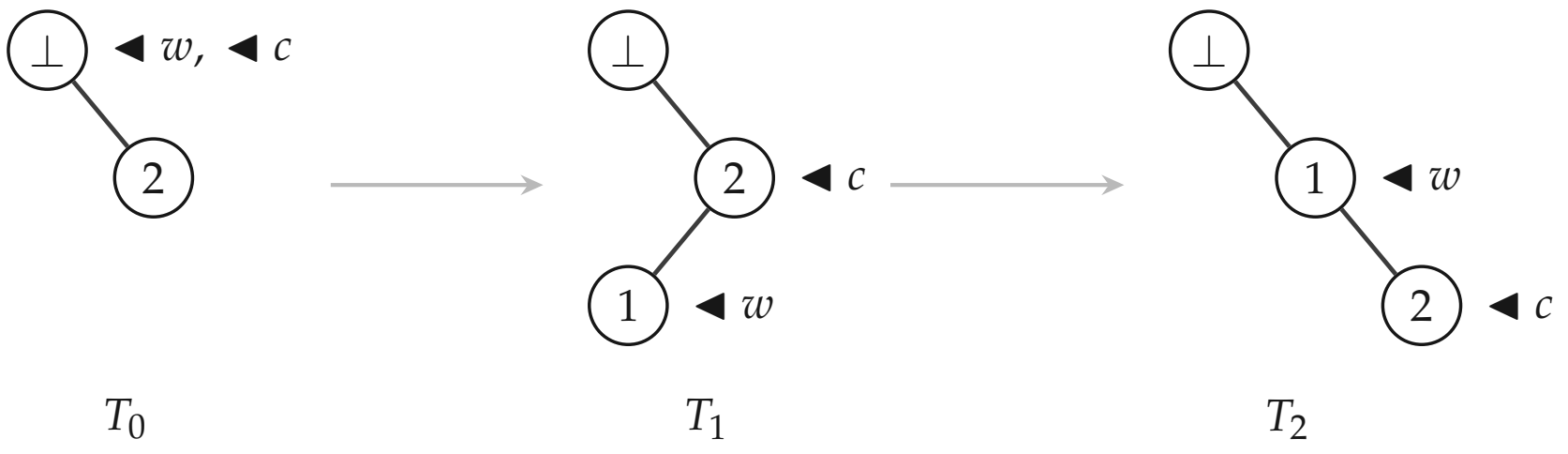
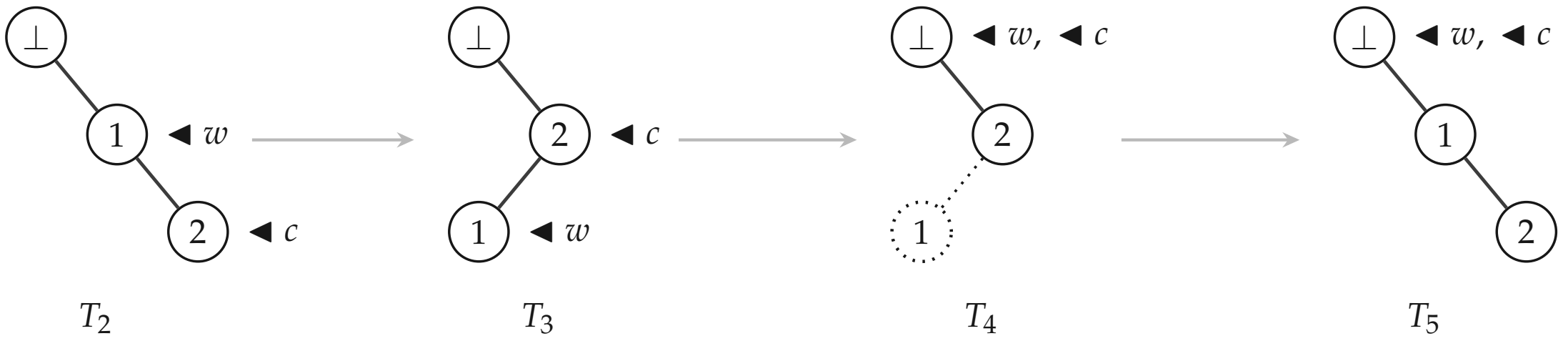
\begin{figure}[t]
\centering
\begin{subfigure}{\linewidth}
\centering
\begin{tikzpicture}[mcfig,x=1cm,y=1cm,
  key/.style={mcnode,minimum size=5.5mm},
  fingers/.style={font=\footnotesize,inner sep=1.5pt},
  operationlabel/.style={font=\footnotesize,align=center,inner sep=2pt},
  state/.style={font=\footnotesize,align=center},
  pics/initial/.style={code={
    \node[key] (-root) at (-.55,0) {$\bot$};
    \node[key] (-two) at (.2,-.9) {$2$};
    \draw[mcpath] (-root)--(-two);
  }},
  pics/before/.style={code={
    \node[key] (-root) at (-.55,0) {$\bot$};
    \node[key] (-two) at (.2,-.9) {$2$};
    \node[key] (-one) at (-.55,-1.8) {$1$};
    \draw[mcpath] (-root)--(-two)--(-one);
  }},
  pics/after/.style={code={
    \node[key] (-root) at (-.55,0) {$\bot$};
    \node[key] (-one) at (.2,-.9) {$1$};
    \node[key] (-two) at (.95,-1.8) {$2$};
    \draw[mcpath] (-root)--(-one)--(-two);
  }}]
  \path[use as bounding box] (-.15,.5) rectangle (15.05,-2.9);
\pic (a) at (.8,0) {initial};
  \node[fingers,right=3pt] at (a-root.east)
    {$\blacktriangleleft w,\;\blacktriangleleft c$};
  \node[state] at (.8,-2.6) {$T_0$};

\pic (b) at (4.9,0) {before};
  \node[fingers,right=3pt] at (b-two.east) {$\blacktriangleleft c$};
  \node[fingers,right=3pt] at (b-one.east)
    {$\blacktriangleleft w$};
  \node[state] at (4.9,-2.6) {$T_1$};

\pic (c) at (9.0,0) {after};
  \node[fingers,right=3pt] at (c-one.east) {$\blacktriangleleft w$};
  \node[fingers,right=3pt] at (c-two.east) {$\blacktriangleleft c$};
  \node[state] at (9.0,-2.6) {$T_2$};

  \draw[mcarrow] (2.25,-.95)--(3.75,-.95)
    node[midway,above=5pt,operationlabel] {};
  \draw[mcarrow] (6.2,-.95)--(7.85,-.95)
    node[midway,above=5pt,operationlabel] {};

\end{tikzpicture}
\caption{
    Execution of $\sigma$ on $T_0$. Initially both fingers are at the root $\bot$ of the tree $T_0$. On request $1$, the algorithm moves both fingers to $2$, inserts $1$ using $w$, and moves $w$ into $1$ to serve the request.
    On request $2$, the algorithm rotates the edge $(1, 2)$ using $w$, and serves the request using $c$.
}
\end{subfigure}
\begin{subfigure}{\linewidth}
\centering
\begin{tikzpicture}[mcfig,x=1cm,y=1cm,
  key/.style={mcnode,minimum size=5.5mm},
  retained/.style={key,draw=mcink,dotted,line width=.85pt},
  fingers/.style={font=\footnotesize,inner sep=1.5pt},
  operationlabel/.style={font=\footnotesize,align=center,inner sep=2pt},
  state/.style={font=\footnotesize,align=center},
  pics/before/.style={code={
    \node[key] (-root) at (-.55,0) {$\bot$};
    \node[key] (-two) at (.2,-.9) {$2$};
    \node[retained] (-one) at (-.55,-1.8) {$1$};
    \draw[mcpath] (-root)--(-two);
    \draw[draw=mcink,dotted,line width=.7pt] (-two)--(-one);
  }},
  pics/middle/.style={code={
    \node[key] (-root) at (-.55,0) {$\bot$};
    \node[key] (-two) at (.2,-.9) {$2$};
    \node[key] (-one) at (-.55,-1.8) {$1$};
    \draw[mcpath] (-root)--(-two)--(-one);
  }},
  pics/after/.style={code={
    \node[key] (-root) at (-.55,0) {$\bot$};
    \node[key] (-one) at (.2,-.9) {$1$};
    \node[key] (-two) at (.95,-1.8) {$2$};
    \draw[mcpath] (-root)--(-one)--(-two);
  }}]
  \path[use as bounding box] (-.15,.5) rectangle (15.05,-2.9);
\pic (d) at (.8,0) {after};
  \node[fingers,right=3pt] at (d-one.east) {$\blacktriangleleft w$};
  \node[fingers,right=3pt] at (d-two.east) {$\blacktriangleleft c$};
  \node[state] at (.8,-2.6) {$T_2$};

\pic (g) at (4.9,0) {middle};
  \node[fingers,right=3pt] at (g-one.east) {$\blacktriangleleft w$};
  \node[fingers,right=3pt] at (g-two.east) {$\blacktriangleleft c$};
  \node[state] at (4.9,-2.6) {$T_3$};

\pic (e) at (9.0,0) {before};
  \node[fingers,right=3pt] at (e-root.east)
    {$\blacktriangleleft w,\;\blacktriangleleft c$};
  \node[state] at (9.0,-2.6) {$T_4$};

\pic (f) at (13.1,0) {after};
  \node[fingers,right=3pt] at (f-root.east)
    {$\blacktriangleleft w,\;\blacktriangleleft c$};
  \node[state] at (13.1,-2.6) {$T_5$};

  \draw[mcarrow] (2.25,-.95)--(3.75,-.95)
    node[midway,above=5pt,operationlabel] {};
  \draw[mcarrow] (6.2,-.95)--(7.85,-.95)
    node[midway,above=5pt,operationlabel] {};
  \draw[mcarrow] (10.3,-.95)--(11.95,-.95);

\end{tikzpicture}
\caption{
    Execution of $\sigma^{-1}$ on $T_2$, followed by the replay of $\sigma$ on $w$. 
    The sequence $\sigma^{-1}$ reverts operations in $\sigma$ (skipping insertions and access operations), and correctly brings all fingers back to the root $\bot$. 
    To transform the resulting tree $T_4$ back to the tree $T_2$ obtained by executing $\sigma$, we replay the moves and rotations of $w$ (skipping insertions and access operations), and bring $w$ back to $\bot$ by following parent edges. The cache finger stays at $\bot$ during the replay.}
\end{subfigure}

\textcolor{blue}{\caption{    
    Bringing all fingers back to the root using the auxiliary cache-reset procedure in the proof of \cref{lem:online_kfinger_cache_reset} after executing an incremental $2$-finger BST strategy $\sigma$ serving requests $(1,2)$, with working finger $w$ and cache finger $c$. Only the working finger $w$ performs rotations or insertions. Note that the final tree $T_5$ agrees with the tree $T_2$ obtained by executing $\sigma$ with all fingers placed at $\bot$.
    }\label{fig:online_cache_reset}}
\end{figure}

\begin{lemma}\label{lem:online_kfinger_cache_reset}
    Let $\sigma$ be an incremental $k$-finger BST strategy where all fingers start at the root $\bot$ of a tree $T_0$. Suppose that rotations and insertions are only performed by a single designated finger, and that $\bot$ is never rotated in $\sigma$. Then, after executing $\sigma$, all fingers can be returned to $\bot$ while preserving the final tree with $\Oh(|\sigma|)$ additional operations and in time $\Oh(|\sigma|)$.
\end{lemma}
\begin{proof}
    Call the finger performing rotations and insertions the \emph{working finger} and all other fingers the \emph{cache fingers}. 
Let $T_i$ and $P_i$ be the tree and finger positions after performing the first $i$ operations in $\sigma$ starting from $T_0$. 
    Define $\sigma^{-1}$ to be the sequence of operations in $\sigma$ in reverse order where we replace insertions and access operations by null operations. See \cref{fig:online_cache_reset} for an example.
    
    We claim that $\sigma^{-1}$ is a valid $k$-finger strategy when starting from $T_{|\sigma|}$. In fact, we show that after performing the first $i$ operations in $\sigma^{-1}$, the maintained tree $T_{|\sigma| + i}$ is the tree $T_{|\sigma| - i}$ with additional finger-free subtrees attached to its leaves, and that the finger positions match $P_{|\sigma| + i} = P_{|\sigma| -i}$.
    This follows by a simple induction by observing that reverse moves are performed along existing edges and reverse rotations simply carry the additional subtrees. Moreover, accesses in $\sigma$ change neither the tree nor the finger positions and so are safe to skip. Finally, in $\sigma^{-1}$, when skipping an insertion of $\sigma$, all subsequent operations in $\sigma$ have already been reverted. Thus in the original tree, the working finger is at a leaf and no finger occupies its subtree because it didn't exist yet. 
    Keeping this subtree at the child position therefore preserves the invariant without deleting any node.

    Since at the start of $\sigma$, all fingers are positioned at the root, after performing $\sigma^{-1}$ all fingers are positioned back at the root as well. However, the resulting tree $T_{|\sigma| + |\sigma^{-1}|}$ might differ from $T_{|\sigma|}$. To transform $T_{|\sigma| + |\sigma^{-1}|}$ back to $T_{|\sigma|}$, it suffices to perform every move and rotation operation of the working finger in $\sigma$, and then move the finger back to the root by following parent edges. Since $T_{|\sigma| + |\sigma^{-1}|}$ agrees with $T_0$ with additional subtrees attached to the leaves, all operations are valid, and since rotations are only performed by the working finger, the resulting tree agrees with $T_{|\sigma|}$. This is illustrated in \cref{fig:online_cache_reset}. Furthermore, every operation in $\sigma$ increases the depth of the working finger by at most 1, so after executing (a subsequence of) $\sigma$, the working finger is at depth at most $\Oh(|\sigma|)$. Thus following parent edges to the root takes $\Oh(|\sigma|)$ additional move operations. Finally, since the root is never rotated, all cache fingers remain at the root during this transformation.
\end{proof}

To serve $Y$, we can simulate the BST algorithm serving $Z$ with a working finger, keep cache fingers at every key served within a super-row, and use the above strategy at the end of each super-row to bring all fingers back to the root.

\begin{lemma}\label{lem:online_glueZtoY}
    Given an online incremental BST algorithm serving $Z$ with cost and time at most $T$, there exists an online incremental BST algorithm serving $Y$ with cost and time $\Oh(n + (m + T) \log d)$.
\end{lemma}
\begin{proof}
    By \cref{lem:online_kfinger_simulation} a $k$-finger BST strategy can be simulated by a standard BST with $\Oh(\log k)$-factor overhead. Hence, it suffices to show an online incremental $k$-finger BST algorithm serving $Y$ with cost and time $\Oh(m + T)$, where $k = \Oh(d)$. 

    Initialize a root node $\bot$ and place $d + 1$ fingers at $\bot$. Distinguish one finger and call it the \emph{working finger}. All other fingers are called \emph{cache fingers}. Let $\Aa$ be the online incremental BST algorithm serving $Z$.
Consider a request $y \in Y$. If it is the first request of the key $y$ within the current super-row, then $y \in Z$ and there is a sequence of operations $\sigma_y$ used by $\Aa$ to serve $y$.  
    Using the working finger and a cache finger placed at $\bot$, perform all operations of $\sigma_y$ on the current tree, until both fingers reach the key $y$. Serve $y$ by accessing the key with the working finger. Then move the working finger back to the root $\bot$ by following parent edges, and leave the cache finger at the key $y$. 
    If the request $y \in Y$ is a repetition of a super-column within the current super-row, then there is a cache finger placed on the key $y$ in the current tree. So serve $y$ by accessing the key using that cache-finger. 
Additionally, if the request $y \in Y$ is the last request of the current super-row, after serving $y$ according to the above rule, we bring all cache fingers back to the root $\bot$ by applying \cref{lem:online_kfinger_cache_reset} with the total sequence of operations $\sigma$ performed to serve the current super-row.
    
    We argue correctness of the described $k$-finger BST strategy.
    First, observe that the right subtree of $\bot$ agrees with the tree maintained by $\Aa$. This holds at initialization and follows by induction on the subsequent key requests: the working finger is clearly placed at $\bot$ after each request is served, so it can legally simulate the strategy $\sigma_y$ of $\Aa$ (after moving from $\bot$ to the right subtree), and rotations and insertions are only performed by the working finger during that step. 

    Moreover, at the beginning of each super-row, all $d$ cache fingers are placed at $\bot$.
    This can be shown again by induction since we initialized the cache fingers at $\bot$, and, at the end of each super-row, we placed them back at $\bot$ using \cref{lem:online_kfinger_cache_reset}. 
    Within each super-row, there are at most $d$ distinct super-columns requested in $Y$. Hence, whenever a super-column is requested for the first time within a super-row, there is always an available cache finger placed at $\bot$.

    Finally, it remains to analyse the cost and time of the $k$-finger BST strategy. Observe that the working finger only performs operations dictated by $\Aa$ followed by move operations along parent edges to bring it back to $\bot$. Since every operation increases the depth of the working finger by at most 1, the total number of operations performed by the working finger is at most $\Oh(T)$.
    Moreover, cache fingers perform three types of operations: either they simulate operations dictated by $\Aa$ (in order to place them at respective keys); or they access the key they are placed at (when the requested super-column is a repetition within the current super-row); or they perform operations dictated by \cref{lem:online_kfinger_cache_reset} (in order to place them back at $\bot$). In total, all $d$ cache fingers perform $\Oh(m + T)$ operations. So the $k$-finger BST algorithm has cost and time $\Oh(m +T)$ as required. It works in the online incremental model.
\end{proof}

\subsection{Solving $X$ via $Y$: defining glue points}\label{sec:online_glueYtoX}
We now show how a BST serving $Y$ can be extended to a BST serving $X$.
The idea is similar to what is done in the offline setting: we use a balanced BST to serve each super-column of $X$ separately and use the strategy serving $Y$ to \emph{glue} the points between super-columns. The main difference is that the glue has to be online, \ie it can only visit keys already requested by $X$.
In the geometric view of BSTs defined in \cref{sec:geometricBST}, this corresponds to \emph{insertion-compatibility}.

\begin{lemma}\label{lem:online_glueYtoX}
Given an online incremental BST algorithm serving $Y$ with cost and time at most $T$, there exists an online incremental BST algorithm serving $X$ with cost and time $\Oh(n+m\log d+T)$.
\end{lemma}
\begin{proof}
    For simplicity, we present the proof in the geometric view of BSTs defined in \cref{sec:geometricBST}, as we now summarize.
    Recall that $X$ has key universe $[n]$ and length $m$, and $Y$ has key universe $[q]$ and length $m$, with $q = \lceil n / d \rceil$. Let $P_X$ be the point set on an $m \times n$ grid corresponding to $X$, and $P_Y$ be the point set on an $m \times q$ grid corresponding to $Y$. 
    See \cref{fig:bst_to_manhattan} for an example of such correspondence. 
    An online algorithm is given a point set $P$ row by row. We therefore say that row $t$ is \emph{revealed at time $t$}.
    We also say that a column $j$ is \emph{revealed at time $t$} if $P$ contains a point $(t', j)$ with $t' \leq t$. 
    To avoid confusions, we will refer to columns in the $m \times n$ grid as \emph{columns} and columns in the $m \times q$ grid as \emph{super-columns}. 

    Let $S_Y$ be the set of points on the $m \times q$ grid corresponding to the keys visited by the BST serving $Y$. In particular, we have that $|S_Y| \leq T$, and $S_Y$ is arborally satisfied and insertion-compatible for $P_Y$. 
    We describe a set of points $S_X$ on the $m \times n$ grid such that $S_X$ is arborally satisfied and insertion-compatible for $P_X$, and $S_X$ can be computed online in time $\Oh(|S_X|)$ given $P_X$, $P_Y$ and $S_Y$. 
    By \cref{lem:geomBST}, this can be translated to an online incremental BST algorithm serving $X$ with cost and time $\Oh(|S_X|)$. 
    We therefore aim at a solution $S_X$ of size $|S_X| \leq \Oh(n + m \log d + T)$.

    We first describe a set of \emph{local points} $S_L$ on the $m \times n$ grid that makes every pair of points in $X$ within the same super-column arborally satisfied. Then, to connect points across different super-columns, we first add \emph{neighbouring points} to the set $S_Y$ and obtain $S_Y' \supset S_Y$, before converting the points in $S_Y'$, that lie on the $m \times q$ grid, to \emph{glue points} $S_G$, that lie on the $m \times n$ grid. In the end, we let $S_X = S_L \cup S_G$ be the union of local and glue points and will show that $P_X \cup S_X$ is arborally satisfied and insertion-compatible. The construction is illustrated in \cref{fig:online_SLYG}.

    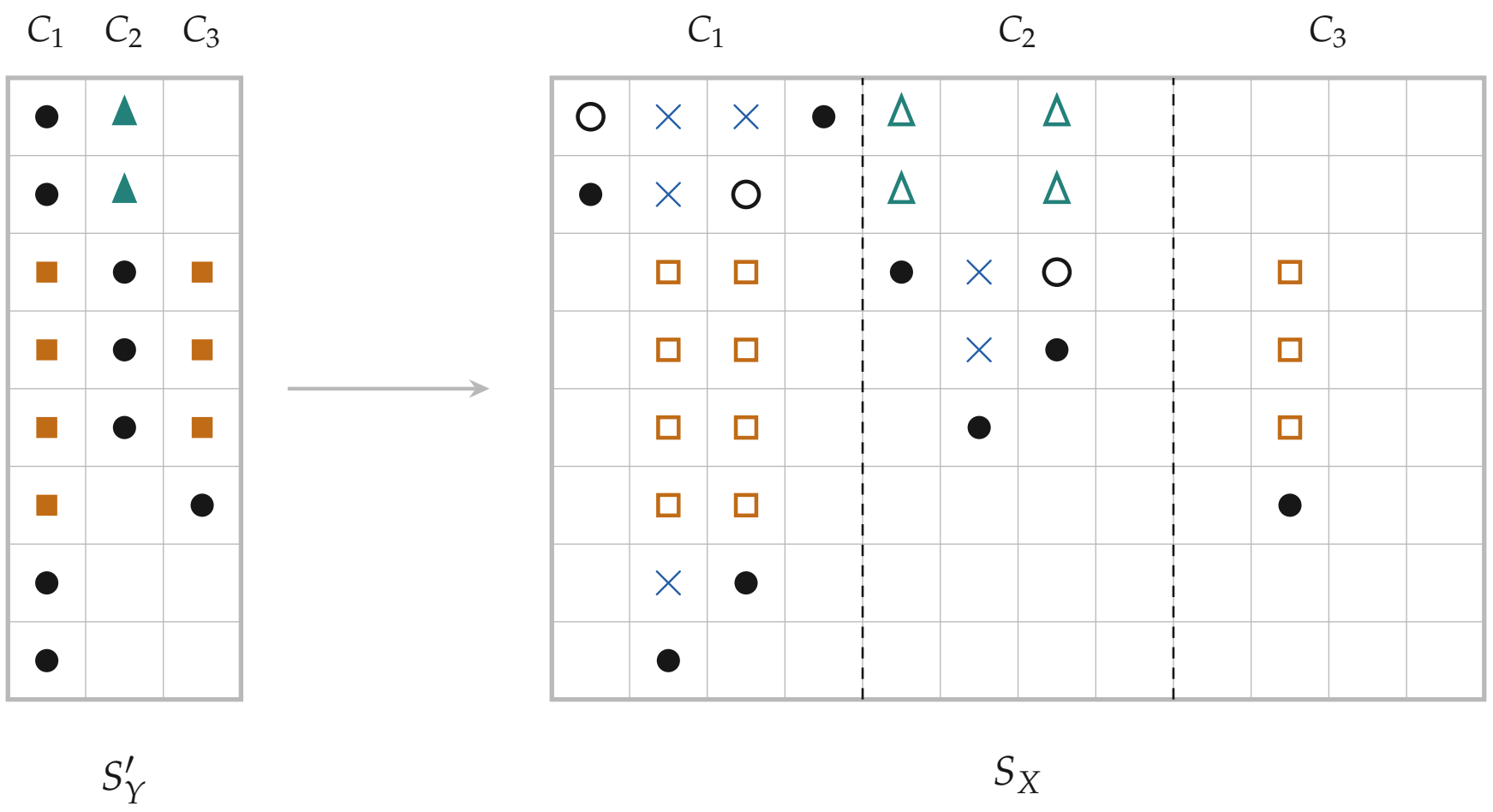
\begin{figure}[t]
\centering
\usetikzlibrary{shapes.geometric}
\begin{tikzpicture}[mcfig,x=.60cm,y=.60cm,
  localpoint/.style={text=liftlocal,font=\normalsize,inner sep=0pt},
  sourceadded/.style={rectangle,draw=lifttransfer,fill=lifttransfer,minimum size=4.2pt,inner sep=0pt},
  sourceneighbour/.style={isosceles triangle,shape border rotate=90,draw=liftglue,fill=liftglue,minimum width=5pt,minimum height=4.5pt,inner sep=0pt},
  inputboundary/.style={circle,draw=mcink,fill=white,line width=.85pt,inner sep=2pt},
  addedboundary/.style={rectangle,draw=lifttransfer,fill=white,line width=.85pt,minimum size=4.5pt,inner sep=0pt},
  neighbourboundary/.style={isosceles triangle,shape border rotate=90,draw=liftglue,fill=white,line width=.85pt,minimum width=5pt,minimum height=4.5pt,inner sep=0pt}]
  \definecolor{liftlocal}{HTML}{245EA6}
  \definecolor{lifttransfer}{HTML}{C06B16}
  \definecolor{liftglue}{HTML}{23817A}
\draw[mcbox] (0,0) rectangle (3,8);
  \draw[mcbox] (7,0) rectangle (19,8);
  \foreach \t in {1,...,7} {
    \draw[mcgrid] (0,\t)--(3,\t);
    \draw[mcgrid] (7,\t)--(19,\t);
  }
  \foreach \x in {1,2} \draw[mcgrid] (\x,0)--(\x,8);
  \foreach \x in {8,...,18} \draw[mcgrid] (\x,0)--(\x,8);
  \foreach \x in {11,15} \draw[mcguide,draw=mcink] (\x,0)--(\x,8);
  \foreach \j in {1,2,3} {
    \node[mclabel] at ({\j-.5},8.6) {$C_\j$};
    \node[mclabel] at ({7+4*\j-2},8.6) {$C_\j$};
  }
  \draw[mcarrow] (3.6,4)--(6.2,4) node[midway,above,mclabel] {};
  \node[mcpoint] at (0.5,0.5) {};
  \node[mcpoint] at (8.5,0.5) {};
  \node[mcpoint] at (0.5,1.5) {};
  \node[localpoint] at (8.5,1.5) {$\times$};
  \node[mcpoint] at (9.5,1.5) {};
  \node[sourceadded] at (0.5,2.5) {};
  \node[mcpoint] at (2.5,2.5) {};
  \node[addedboundary] at (8.5,2.5) {};
  \node[addedboundary] at (9.5,2.5) {};
  \node[mcpoint] at (16.5,2.5) {};
  \node[sourceadded] at (0.5,3.5) {};
  \node[sourceadded] at (2.5,3.5) {};
  \node[mcpoint] at (1.5,3.5) {};
  \node[addedboundary] at (8.5,3.5) {};
  \node[addedboundary] at (9.5,3.5) {};
  \node[addedboundary] at (16.5,3.5) {};
  \node[mcpoint] at (12.5,3.5) {};
  \node[sourceadded] at (0.5,4.5) {};
  \node[sourceadded] at (2.5,4.5) {};
  \node[mcpoint] at (1.5,4.5) {};
  \node[addedboundary] at (8.5,4.5) {};
  \node[addedboundary] at (9.5,4.5) {};
  \node[localpoint] at (12.5,4.5) {$\times$};
  \node[addedboundary] at (16.5,4.5) {};
  \node[mcpoint] at (13.5,4.5) {};
  \node[sourceadded] at (0.5,5.5) {};
  \node[sourceadded] at (2.5,5.5) {};
  \node[mcpoint] at (1.5,5.5) {};
  \node[addedboundary] at (8.5,5.5) {};
  \node[addedboundary] at (9.5,5.5) {};
  \node[inputboundary] at (13.5,5.5) {};
  \node[addedboundary] at (16.5,5.5) {};
  \node[localpoint] at (12.5,5.5) {$\times$};
  \node[mcpoint] at (11.5,5.5) {};
  \node[sourceneighbour] at (1.5,6.5) {};
  \node[mcpoint] at (0.5,6.5) {};
  \node[inputboundary] at (9.5,6.5) {};
  \node[neighbourboundary] at (11.5,6.5) {};
  \node[neighbourboundary] at (13.5,6.5) {};
  \node[localpoint] at (8.5,6.5) {$\times$};
  \node[mcpoint] at (7.5,6.5) {};
  \node[sourceneighbour] at (1.5,7.5) {};
  \node[mcpoint] at (0.5,7.5) {};
  \node[inputboundary] at (7.5,7.5) {};
  \node[neighbourboundary] at (11.5,7.5) {};
  \node[neighbourboundary] at (13.5,7.5) {};
  \node[localpoint] at (8.5,7.5) {$\times$};
  \node[localpoint] at (9.5,7.5) {$\times$};
  \node[mcpoint] at (10.5,7.5) {};
  \node[mcpanel] at (1.5,-.6) {$S_Y'$};
  \node[mcpanel] at (13,-.6) {$S_X$};
\end{tikzpicture}
\caption{Lifting $S_Y'$ to $S_X$ in~\cref{lem:online_glueYtoX}.
Input points are marked with black circles.
Left: squares mark $S_Y\setminus P_Y$ and triangles mark neighbours.
Right: crosses mark local additions; hollow symbols mark glue points
obtained by projecting $S_Y'$ to the currently revealed extreme columns
of the corresponding super-columns.}
\label{fig:online_SLYG}
\end{figure}
 
    \paragraph*{Local points.}
        Let $P_j \coloneq P_X \cap ([m] \times C_j)$ be the points of $P_X$ in the super-column $C_j \subset [n]$, and let $X_j \subset X$ be the corresponding requests in $X$. 
        Run a balanced incremental BST on $X_j$ and apply \cref{lem:BSTgeom} to obtain a set of points $S_j$ on the $m \times n$ grid. In particular, $S_j$ contains points on row $t$ for every key visited when serving the request at time $t$.
Let $S_L \coloneq \bigcup_j S_j$ be the resulting set of points. 
        We claim that $S_L$ is arborally satisfied within each super-column and has sufficiently small size.
    \begin{claim}\label{lem:online_SL}
        The set of points $S_L$ has size $|S_L| \leq \Oh(n + m \log d)$ and can be computed online in time $\Oh(n + m \log d)$ given $P_X$.
Moreover, $P_X \cup S_L$ is insertion-compatible, and each pair of its points in the same super-column is arborally satisfied.
    \end{claim}
    \begin{claimproof}    
        Since $X_j$ requests at most $|C_j| \leq d$ distinct keys, the balanced incremental BST on $X_j$ has cost and time at most $\Oh(|X_j| \log d)$. Thus, by \cref{lem:BSTgeom}, $|S_j| \leq \Oh(d + |X_j| \log d)$ for every $j \in [q]$, and therefore $|S_L| \leq \Oh(n + m \log d)$. By running the balanced incremental BSTs in parallel, we can compute $S_L$ online in the total time $\Oh(n + m \log d)$.
        The second claim follows directly from the definition of $S_L$. 
    \end{claimproof}

    \paragraph*{Neighbouring points.}
    We now add points to $S_Y$ and obtain $S_Y' \supset S_Y$ such that $S_Y'$ is still arborally satisfied and insertion-compatible for $P_Y$.
    Initially, let $S_Y' = S_Y$. 
    For each point $(t, j) \in S_Y$, let $\LN(t, j) = (t, L)$ and $\RN(t, j) = (t, R)$ be the \emph{left} and \emph{right neighbouring} points to $(t, j)$, where $L$ and $R$ are the nearest super-columns strictly to the left and right of $j$ revealed at time $t$ in $S_Y$. 
    Add those points to $S_Y'$ (if defined). 
We claim that the added points preserve arboral satisfaction and insertion-compatibility of $S_Y'$. 
    \begin{claim}\label{lem:online_SYprime}
        The set of points $S_Y'$ has size $|S_Y'| \leq \Oh(m + T)$ and can be computed online in time $\Oh(m + T)$ given $S_Y$.
        Moreover, $S_Y'$ is arborally satisfied and insertion-compatible for $P_Y$. 
    \end{claim}

    The proof is a straightforward case analysis, similar to those in~\cite{ChalermsookCS20}. We recommend skipping it on a first reading.

    \begin{figure}[t]
\centering
\begin{tikzpicture}[x=.85cm,y=.80cm,font=\small,
  label distance=.5pt,every label/.style={font=\small,inner sep=1pt},
  source/.style={circle,fill=black,inner sep=2pt},
  copy/.style={rectangle,draw=teal!80!black,fill=white,line width=.8pt,inner sep=2.1pt},
  witnesscopy/.style={rectangle,draw=teal!80!black,fill=teal!80!black,inner sep=2.1pt},
  sourcebox/.style={draw=black!45,line width=.7pt},
  targetbox/.style={draw=blue!65!black,dashed,line width=.8pt},
  copying/.style={-{Stealth[length=2mm]},draw=teal!80!black,line width=.8pt}]
\foreach \offset/\side in {0/left,8.5/right} {
    \begin{scope}[shift={(\offset,0)}]
\fill[black!5] (.7,.7) rectangle (2,4.2);
      \fill[black!5] (5,.7) rectangle (6.3,4.2);
      \draw[sourcebox] (.7,.7) rectangle (6.3,4.2);
      \draw[targetbox] (2,.7) rectangle (5,4.2);
      \node[source,label=below left:{$p' = (s, \alpha)$}] at (.7,.7) {};
      \node[source,label=above right:{$q'= (t, \beta)$}] at (6.3,4.2) {};
      \node[copy,label=below right:{$p = (s, a)$}] at (2,.7) {};
      \node[copy,label=above left:{$q = (t, b)$}] at (5,4.2) {};
      \draw[copying] (.85,.7)--(1.8,.7);
      \draw[copying] (6.15,4.2)--(5.2,4.2);
    \end{scope}
  }
\node[source,label=left:{$r$}] at (.7,2.5) {};
  \node[copy,label=right:{$\RN(r)$}] at (2,2.5) {};
  \draw[copying] (.85,2.5)--(1.8,2.5);
\begin{scope}[shift={(8.5,0)}]
    \node[source,label=right:{$r$}] at (6.3,2.5) {};
    \node[copy,label=left:{$\LN(r)$}] at (5,2.5) {};
    \draw[copying] (6.15,2.5)--(5.2,2.5);
  \end{scope}
\end{tikzpicture}
\caption{The final case considered in the proof of \cref{lem:online_SYprime}. Black circles are points in $S_Y$; squares are their neighbouring points in $S_Y'$. 
The interior of rectangles $R[p',(t, a)]$ and $R[(s, b),q']$ is empty. Thus, when a third point $r \in S_Y$ lies on the boundary, its neighbouring point lies inside $R[p, q]$.
}
\label{fig:online_SYprime_lastcase}
\end{figure}
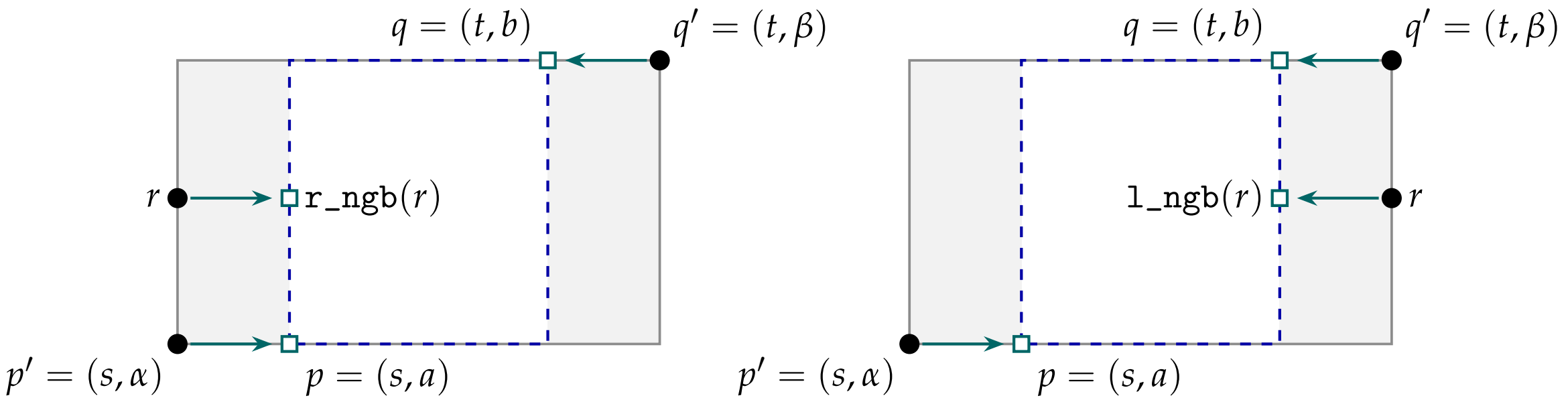

    \begin{claimproof}
        Each point in $S_Y$ contributes to at most two points in $S_Y'$, so we bound $|S_Y'| \leq 3 |S_Y| \leq \Oh(m + T)$. Since we only added points on revealed super-columns on the current row, $S_Y'$ can be computed online in the same total time and $S_Y'$ is insertion-compatible for $P_Y$. We verify that it is arborally satisfied.

For the sake of contradiction, assume that there are two non-aligned points $p = (s, a)$ and $q = (t, b)$ in $S_Y'$ such that the rectangle $R[p, q]$ contains no third point of $S_Y'$.
        Without loss of generality, assume that $s < t$ and $a < b$.

        \proofcase{Case 1: $p$ is a left neighbouring point.}
        Suppose $p = \LN(s, \alpha)$ for some $\alpha > a$. By emptiness of $R[p, q]$, we have $a < b < \alpha$. Let $v= (\ell, b)$ be the first point in $P_Y$ that reveals super-column $b$, \ie $s < \ell \leq t$. Then the left neighbouring point $\LN(v) = (\ell, c)$ is on super-column $a \leq c < b$, and is thus inside $R[p, q]$, contradiction.
        Hence, let $p' = (s, \alpha) \in S_Y$ be such that either $p = p'$ or $p = \RN(p')$. In particular, $\alpha \leq a$.

        \proofcase{Case 2: $q$ is a right neighbouring point.}
        Suppose $q = \RN(t, \beta)$ for some $\beta < b$. Since super-column $a$ is revealed at time $t$, we get $a \leq \beta < b$, and thus the point $(t, \beta) \in S_Y$ is inside $R[p, q]$, contradiction.
        Hence, let $q' = (t, \beta) \in S_Y$ be such that either $q = q'$ or $q = \LN(q')$. In particular, $b \leq \beta$.

        \proofcase{Case 3: column $b$ is not revealed at time $s$.}
        In that case, there exists a point $v = (\ell, b) \in P_Y$ with $s < \ell \leq t$, and its left neighbouring point is contained in $R[p, q]$, contradiction.

        \proofcase{Case 4: remaining situation.}
        Since $S_Y$ is arborally satisfied, there is a point $r \in S_Y\setminus\{p',q'\}$ inside the rectangle $R[p', q']$. Since $\alpha \leq a < b \leq \beta$, the column of $r$ lies in $[a,b]$, in $[\alpha, a)$, or in $(b, \beta]$ (see \cref{fig:online_SYprime_lastcase}).

        \emph{(a) Column of $r$ in $[a,b]$.} Then $r \in R[p,q]$, so $r = p$ or $r = q$ by emptiness.
        If $r = p$, then $S_Y'$ contains the right neighbour $\RN(p)$ of the point $p \in S_Y$. Since column $b$ is revealed at time $s$, $\RN(p)$ is inside $R[p, q]$, contradiction.
        If $r = q$, then $S_Y'$ contains the left neighbour $\LN(q)$ of point $q$. Since column $a$ is revealed at time $t$, $\LN(q)$ is inside $R[p, q]$, contradiction.

        \emph{(b) Column of $r$ in $[\alpha, a)$, \ie $r \in R[p', (t, a-1)]$.}
        Let $v = (\ell, k)$ be the first point revealing a column $k \in (\alpha, a)$ with $s < \ell \leq t$.
        If none exist, then $r$ lies on the column $\alpha$ and $\RN(r)$ is on column $a$ and thus inside $R[p, q]$, contradiction.
        Otherwise, $\RN(v)$ is on column $a$, and thus inside $R[p, q]$, contradiction.

        \emph{(c) Column of $r$ in $(b, \beta]$, \ie $r \in R[(s, b+1), q']$.}
        By definition of $q'$, all columns $c \in (b, \beta)$ are unrevealed until time $t$, so $r$ is on column $\beta$. Since column $b$ is revealed at time $s$, the left neighbour of $r$ is in $R[p, q]$, contradiction.
    \end{claimproof}

    \paragraph*{Glue points.}
    Finally, we convert the points in $S_Y'$ on the $m \times q$ grid to \emph{glue points} $S_G$ on the $m \times n$ grid. 
    Initially, let $S_G = \emptyset$. 
For each point $(t, j) \in P_Y \cup S_Y'$, let $\LB(t, j) = (t, L)$ and $\RB(t, j) = (t, R)$ be the \emph{left} and \emph{right boundary} points to $(t, j)$, where $L$ and $R$ are the leftmost and rightmost columns in the super-column $C_j$ revealed at time $t$ in $P_X$.
    Add those points to $S_G$.
In the end, let $S_X \coloneq P_X \cup S_L \cup S_G$ be the union of input, local and glue points. \cref{fig:online_SLYG} summarizes the construction. 
    We claim that $S_X$ is arborally satisfied and insertion-compatible for $P_X$.

    \begin{claim}\label{lem:online_SX}
        The set of points $S_X$ has size $|S_X| \leq \Oh(n + m \log d + T)$ and can be computed online in time $\Oh(n + m \log d + T)$ given $P_X$ and $S_Y$.
        Moreover, $S_X$ is arborally satisfied and insertion-compatible for $P_X$.
    \end{claim}

    The proof is a straightforward case analysis, similar to those in~\cite{ChalermsookCS20}. We recommend skipping it on a first reading.

    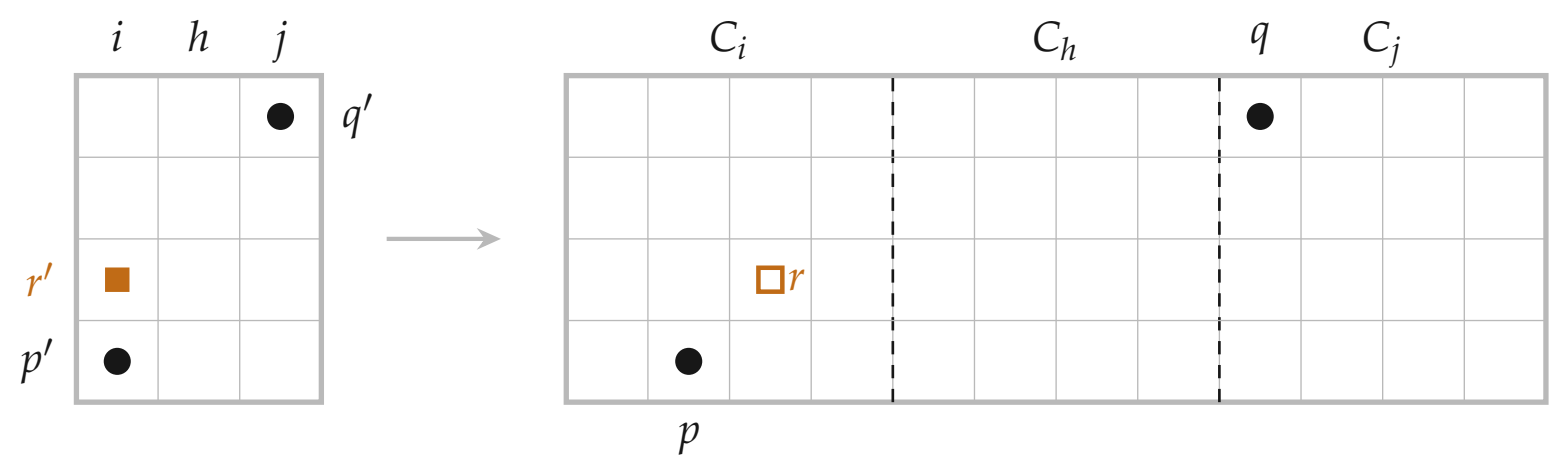
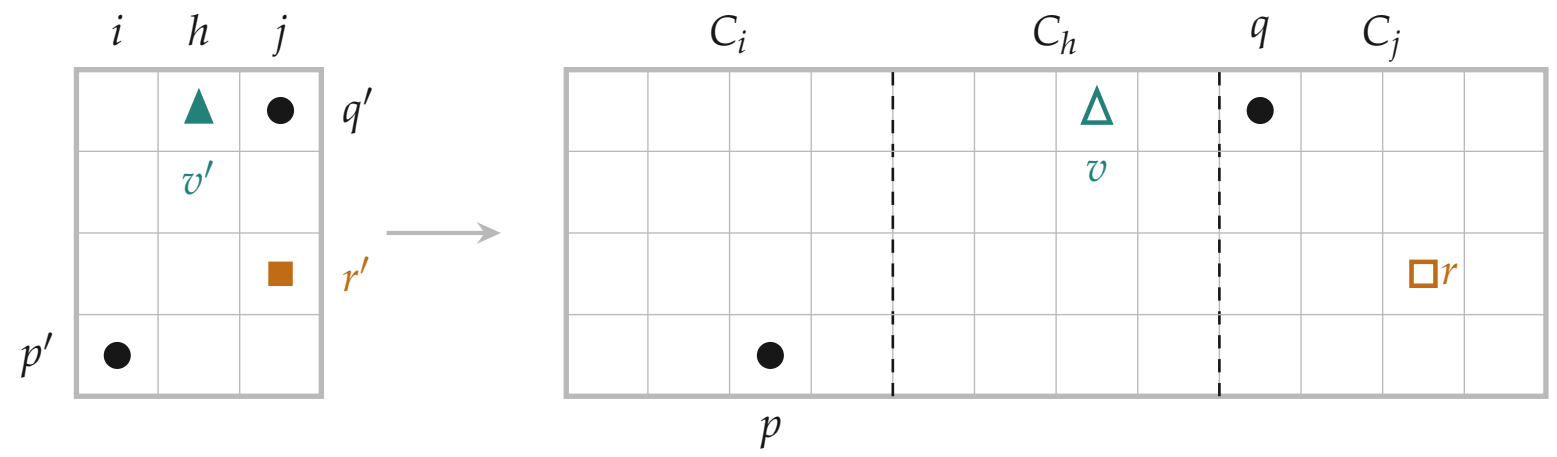
\begin{figure}[t]
\usetikzlibrary{shapes.geometric}
\centering
\begin{subfigure}{\textwidth}
\centering
\begin{tikzpicture}[mcfig,x=1.08cm,y=1.08cm,
  label distance=.5pt,
  mclabel/.append style={fill=white,inner sep=.6pt},
  source/.style={rectangle,draw=caseorange,fill=caseorange,minimum size=4.2pt,inner sep=0pt},
  neighbour/.style={isosceles triangle,shape border rotate=90,draw=casegreen,fill=casegreen,minimum width=5pt,minimum height=4.5pt,inner sep=0pt},
  neighbourboundary/.style={isosceles triangle,shape border rotate=90,draw=casegreen,fill=white,line width=.85pt,minimum width=5pt,minimum height=4.5pt,inner sep=0pt},
  boundary/.style={rectangle,draw=caseorange,fill=white,line width=.85pt,minimum size=4.5pt,inner sep=0pt},
  rect/.style={draw=caseblue,dash pattern=on 3pt off 2pt,line width=.7pt},
  copyarrow/.style={mcarrow,draw=casegreen},
  liftarrow/.style={mcarrow,draw=caseorange}]
  \definecolor{caseblue}{HTML}{245EA6}
  \definecolor{caseorange}{HTML}{C06B16}
  \definecolor{casegreen}{HTML}{23817A}
  \begin{scope}[xscale=.5]
  \draw[mcbox] (0,0) rectangle (3,2);
  \draw[mcgrid] (1,0)--(1,2) (2,0)--(2,2);
  \node[mclabel,anchor=base] at (0.5,2.15) {$i$};
  \node[mclabel,anchor=base] at (1.5,2.15) {$h$};
  \node[mclabel,anchor=base] at (2.5,2.15) {$j$};
  \draw[mcgrid] (0,0.5)--(3,0.5);
  \draw[mcgrid] (0,1)--(3,1);
  \draw[mcgrid] (0,1.5)--(3,1.5);
  \end{scope}
  \draw[mcarrow] (1.9,1)--(2.6,1);
  \begin{scope}[shift={(3,0)}]
  \draw[mcbox] (0,0) rectangle (6,2);
  \draw[mcgrid] (0.5,0)--(0.5,2);
  \draw[mcgrid] (1,0)--(1,2);
  \draw[mcgrid] (1.5,0)--(1.5,2);
  \draw[mcgrid] (2,0)--(2,2);
  \draw[mcgrid] (2.5,0)--(2.5,2);
  \draw[mcgrid] (3,0)--(3,2);
  \draw[mcgrid] (3.5,0)--(3.5,2);
  \draw[mcgrid] (4,0)--(4,2);
  \draw[mcgrid] (4.5,0)--(4.5,2);
  \draw[mcgrid] (5,0)--(5,2);
  \draw[mcgrid] (5.5,0)--(5.5,2);
  \draw[mcguide,draw=mcink] (2,0)--(2,2);
  \draw[mcguide,draw=mcink] (4,0)--(4,2);
  \node[mclabel,anchor=base] at (1,2.15) {$C_i$};
  \node[mclabel,anchor=base] at (3,2.15) {$C_h$};
  \node[mclabel,anchor=base] at (5,2.15) {$C_j$};
  \draw[mcgrid] (0,0.5)--(6,0.5);
  \draw[mcgrid] (0,1)--(6,1);
  \draw[mcgrid] (0,1.5)--(6,1.5);
  \node[mcpoint] at (0.75,0.25) {};
  \node[mclabel,anchor=north] at (0.75,-.1) {$p$};
  \node[mcpoint] at (4.25,1.75) {};
  \node[mclabel,anchor=south] at (4.25,2.1) {$q$};
  \end{scope}
  \begin{scope}[xscale=.5]
  \node[mcpoint] at (0.5,0.25) {};
  \node[mclabel,anchor=east] at (-.2,0.25) {$p'$};
  \node[mcpoint] at (2.5,1.75) {};
  \node[mclabel,anchor=west] at (3.2,1.75) {$q'$};
  \node[source] at (0.5,0.75) {};
  \node[mclabel,text=caseorange,anchor=east] at (-.2,0.75) {$r'$};
  \end{scope}
  \begin{scope}[shift={(3,0)}]
  \node[boundary] (case-a-r) at (1.25,0.75) {};
  \node[mclabel,text=caseorange,anchor=west,inner sep=.35pt] (case-a-r-label) at (1.34,0.75) {$r$};
  \end{scope}
\end{tikzpicture}
\caption{If $i = k$, then there exists a point $r' \in S_Y'$ in $R[p', q']$. Its right boundary $r=\RB(r')$ is inside $R[p, q]$.}\label{fig:online_SX_kleft}
\end{subfigure}

\medskip
\begin{subfigure}{\textwidth}
\centering
\begin{tikzpicture}[mcfig,x=1.08cm,y=1.08cm,
  label distance=.5pt,
  mclabel/.append style={fill=white,inner sep=.6pt},
  source/.style={rectangle,draw=caseorange,fill=caseorange,minimum size=4.2pt,inner sep=0pt},
  neighbour/.style={isosceles triangle,shape border rotate=90,draw=casegreen,fill=casegreen,minimum width=5pt,minimum height=4.5pt,inner sep=0pt},
  neighbourboundary/.style={isosceles triangle,shape border rotate=90,draw=casegreen,fill=white,line width=.85pt,minimum width=5pt,minimum height=4.5pt,inner sep=0pt},
  boundary/.style={rectangle,draw=caseorange,fill=white,line width=.85pt,minimum size=4.5pt,inner sep=0pt},
  rect/.style={draw=caseblue,dash pattern=on 3pt off 2pt,line width=.7pt},
  copyarrow/.style={mcarrow,draw=casegreen},
  liftarrow/.style={mcarrow,draw=caseorange}]
  \definecolor{caseblue}{HTML}{245EA6}
  \definecolor{caseorange}{HTML}{C06B16}
  \definecolor{casegreen}{HTML}{23817A}
  \begin{scope}[xscale=.5]
  \draw[mcbox] (0,0) rectangle (3,2);
  \draw[mcgrid] (1,0)--(1,2) (2,0)--(2,2);
  \node[mclabel,anchor=base] at (0.5,2.15) {$i$};
  \node[mclabel,anchor=base] at (1.5,2.15) {$h$};
  \node[mclabel,anchor=base] at (2.5,2.15) {$j$};
  \draw[mcgrid] (0,0.5)--(3,0.5);
  \draw[mcgrid] (0,1)--(3,1);
  \draw[mcgrid] (0,1.5)--(3,1.5);
  \end{scope}
  \draw[mcarrow] (1.9,1)--(2.6,1);
  \begin{scope}[shift={(3,0)}]
  \draw[mcbox] (0,0) rectangle (6,2);
  \draw[mcgrid] (0.5,0)--(0.5,2);
  \draw[mcgrid] (1,0)--(1,2);
  \draw[mcgrid] (1.5,0)--(1.5,2);
  \draw[mcgrid] (2,0)--(2,2);
  \draw[mcgrid] (2.5,0)--(2.5,2);
  \draw[mcgrid] (3,0)--(3,2);
  \draw[mcgrid] (3.5,0)--(3.5,2);
  \draw[mcgrid] (4,0)--(4,2);
  \draw[mcgrid] (4.5,0)--(4.5,2);
  \draw[mcgrid] (5,0)--(5,2);
  \draw[mcgrid] (5.5,0)--(5.5,2);
  \draw[mcguide,draw=mcink] (2,0)--(2,2);
  \draw[mcguide,draw=mcink] (4,0)--(4,2);
  \node[mclabel,anchor=base] at (1,2.15) {$C_i$};
  \node[mclabel,anchor=base] at (3,2.15) {$C_h$};
  \node[mclabel,anchor=base] at (5,2.15) {$C_j$};
  \draw[mcgrid] (0,0.5)--(6,0.5);
  \draw[mcgrid] (0,1)--(6,1);
  \draw[mcgrid] (0,1.5)--(6,1.5);
  \node[mcpoint] at (1.25,0.25) {};
  \node[mclabel,anchor=north] at (1.25,-.1) {$p$};
  \node[mcpoint] at (4.25,1.75) {};
  \node[mclabel,anchor=south] at (4.25,2.1) {$q$};
  \end{scope}
  \begin{scope}[xscale=.5]
  \node[mcpoint] at (0.5,0.25) {};
  \node[mclabel,anchor=east] at (-.2,0.25) {$p'$};
  \node[mcpoint] at (2.5,1.75) {};
  \node[mclabel,anchor=west] at (3.2,1.75) {$q'$};
  \node[source] at (2.5,0.75) {};
  \node[mclabel,text=caseorange,anchor=west] at (3.2,0.75) {$r'$};
  \node[neighbour,yshift=-.6pt] (case-b-vprime) at (1.5,1.75) {};
  \node[mclabel,text=casegreen,anchor=north] (case-b-vprime-label) at (1.5,1.47) {$v'$};
  \end{scope}
  \begin{scope}[shift={(3,0)}]
  \node[boundary] (case-b-r) at (5.25,0.75) {};
  \node[mclabel,text=caseorange,anchor=west,inner sep=.35pt] (case-b-r-label) at (5.34,0.75) {$r$};
  \node[neighbourboundary,yshift=-.6pt] (case-b-v) at (3.25,1.75) {};
  \node[mclabel,text=casegreen,anchor=north] (case-b-v-label) at (3.25,1.47) {$v$};
  \end{scope}
\end{tikzpicture}
\caption{If $k=j$ and the left boundary $r=\LB(r')$ is on a column $c>b$, then column $b$ is first revealed after the row of $r'$. Its first input point either lies in $R[p,q]$ or equals $q$. In the latter case, illustrated here, $S_Y'$ contains the left neighbouring point $v'=\LN(q')$, whose right boundary $v=\RB(v')$ lies inside $R[p,q]$.}\label{fig:online_SX_krigh}
\end{subfigure}
\caption{Two cases analysed in \cref{lem:online_SX}. On the left, we represent the points in $S_Y'$; on the right we represent the corresponding points in $S_X$.}
\label{fig:online_distinct_supercolumns}
\end{figure}
 
    \begin{claimproof}
        Each point in $S_Y'$ contributes to at most two boundary points in $S_G$, so we bound $|S_G| \leq 3 |S_Y'| \leq \Oh(m + T)$ by \cref{lem:online_SYprime}.
        Since we only add points in $S_G$ to revealed columns on the current row, $S_G$ can be computed online in the same total time and $S_G$ is insertion-compatible. 
        By \cref{lem:online_SL}, we deduce that $|S_X| \leq \Oh(n + m \log d + T)$, $S_X$ can be computed online in time $\Oh(n + m \log d + T)$ and $S_X$ is insertion-compatible for $P_X$. We verify that it is also arborally satisfied. 
        
        For the sake of contradiction, suppose there are non-aligned points $p = (s,a)$ and $q = (t,b)$ in $S_X$ such that no third point of $S_X$ lies in the rectangle spanned by $p$ and $q$, which we denote $R[p, q]$.
        Without loss of generality, let $s < t$ and $a < b$; the case $a > b$ is symmetric.
        We use the following fact: if $(t', c) \in S_X$ with $c \in C_j$, then $(t', j) \in P_Y \cup S_Y'$. Indeed, points of $P_X \cup S_L$ on row $t'$ lie in the super-column of $x_{t'}$, and a point of $S_G$ on row $t'$ in $C_j$ is a boundary point of $(t', j) \in P_Y \cup S_Y'$.
        We distinguish whether $p$ and $q$ lie in the same super-column (Case~1) or (Case~2).

        \proofcase{Case 1:  $a, b \in C_j$ for some $j$.}

        \emph{Case 1.1: $p, q \in P_X \cup S_L$.} Then by \cref{lem:online_SL} points $p$ and $q$ are arborally satisfied in $P_X \cup S_L$ and thus in $S_X$, contradiction.
        
        Otherwise, $p$ or $q$ is a boundary point in $S_G$, and we may assume that $p$ is a boundary point. Indeed, suppose that $q$ is a boundary point. Then, since column $a < b$ is revealed at time $t$, $q$ is a right boundary point, \ie $q = \RB(t, j)$. Consider the right boundary point $r = \RB(s, j)$, which exists by the fact above. The column $c$ of $r$ satisfies $a \leq c \leq b$. Since we assume that $R[p, q]$ is empty, we deduce that $p = r = \RB(s, j)$.

        \emph{Case 1.2: $p = \LB(s, j)$.}
        Suppose $p = \LB(s, j)$.  If $\LB(t, j) = (t, a)$, then that point is inside $R[p, q]$, contradiction. Hence, let $v = (\ell, c)$ be the first point in $P_X$ that reveals column $c < a$ and $c \in C_j$.
        This means that the balanced BST serving $X_j$ needs to insert the key $c$ before accessing the key. Before insertion, the smallest key in the BST is $a$, and thus inserting $c$ requires visiting the key $a$, \ie the corresponding set of points $S_j$ contains the point $(\ell, a)$ (see \cref{lem:BSTgeom}).  Since $s < \ell \leq t$, point $(\ell, a) \in S_L$ is in $R[p, q]$, contradiction. 

        \emph{Case 1.3: $p = \RB(s, j)$.}
        Suppose $p = \RB(s, j)$. Let $v = (\ell, c)$ be the first point in $P_X$ that reveals column $c > a$ with $c \in C_j$. We have $s < \ell \leq t$ and $a < c$. Using the same reasoning as above, we conclude that the balanced BST serving $X_j$ visits the key $a$ at time $\ell$, and thus $S_L$ contains the point $(\ell, a)$. This is inside $R[p, q]$, contradiction.

        \proofcase{Case 2: $a \in C_i$ and $b \in C_j$ for some $i < j$.}
        By the fact above, the projected points $p' = (s, i)$ and $q' = (t, j)$ are contained in $P_Y \cup S_Y'$. Since $P_Y \cup S_Y'$ is arborally satisfied, there exists a point $r'= (\ell, k)$ in $P_Y \cup S_Y'$ in the rectangle spanned by $p'$ and $q'$ on the $m \times q$ grid. In particular, $i \leq k \leq j$. We distinguish three cases (see \cref{fig:online_distinct_supercolumns}).

        \emph{Case 2.1: $i < k < j$.}
        Suppose $i < k < j$, \ie the super-column $C_k$ is contained strictly between super-columns $C_i$ and $C_j$. By definition, $S_G$ contains boundary points of $r'$, \eg the point $r=\LB(r')$, which is inside $R[p, q]$, contradiction.

        \emph{Case 2.2: $k = i$.}
        Suppose $i = k$, \ie $p'$ and $r'$ are in the same super-column.
Then $s < \ell$, as otherwise $p' = r'$. By definition, $S_G$ contains the right boundary point of $r'$, the point $r = \RB(r') = (\ell, c)$. Observe that $a\leq c$, since column $a$ is revealed at time $s$ (by the point $p$), and thus is also revealed at time $\ell > s$. On the other hand, $c \in C_i$ and thus $c < b$. So the point $r$ is inside $R[p, q]$, contradiction.

        \emph{Case 2.3: $k = j$.}
        Finally, suppose that $j = k$. Then $\ell < t$, as otherwise $q' = r'$. 
        By definition, $S_G$ contains the left boundary point of $r'$, the point $r = \LB(r')$. Let $c$ be the column of $r$.  
        Since $c \in C_j$, we have $a < c$. Now if $c \leq b$, then we conclude as in the previous case that the point $r$ is inside $R[p, q]$, contradiction.
        Assume therefore that $c > b$. 
This means that the column $b$ has not yet been revealed at time $\ell$. Let $u = (t_u, b)$ be the first point in $P_X$ on column $b$. We have $\ell < t_u$. If $t_u < t$, then the point $u$ is inside $R[p, q]$, contradiction. If $t_u = t$, then $u = q$, and thus $q \in P_X$. In particular, $q' \in P_Y$, and thus $S_Y'$ contains the left neighbouring point $v' = \LN(q')$ to $q'$. 
        Let $i_v$ be the super-column of $v'$. 
        Since the super-column $i$ is revealed at time $t$ in $P_Y$, we have $i \leq i_v < j$. The right boundary point $v = \RB(v')$ is therefore inside $R[p, q]$, contradiction.
    \end{claimproof}

    As already argued above, by \cref{lem:geomBST}, we can convert the online algorithm of \cref{lem:online_SX} into an online incremental BST algorithm serving $X$ with cost and time $\Oh(n + |S_X|) \leq \Oh(n + m \log d + T)$, as required. 
\end{proof}

Finally, we can prove \cref{lem:online_existenceZ}.

\begin{proof}[Proof of \cref{lem:online_existenceZ}]
    Combine \cref{lem:online_constructionXYZ,lem:online_defXYZ,lem:online_glueYtoX,lem:online_glueZtoY}.
\end{proof}

\subsection{Guessing $c_\pi$}\label{sec:online_guessCpi}

We prove that the online algorithm serving $X$ does not need to be given the constant $\cpi$. The idea in \cref{lem:online_guessCpi} is to guess a value $c$ for $\cpi$, run $\Aa_c$ until we witness that $\cpi \geq c$, then increase the guess $c$ and repeat with the new guess. The difference with the offline setting is that we cannot discard the past operations. Hence, at each new guess, we additionally perform a sequence of operations to simulate a restart of the requests. 

\onlineCpi*

\begin{proof}
    Fix a permutation pattern $\pi$ such that $X$ avoids $\pi$.
    Let $K \in \N$ be a sufficiently large constant, such that, for any $c \geq 2$, $\Aa_c$ serves $X$ with cost and time at most $K (n+m) \log c$ whenever $\cpi \leq c$. 
We call $B(c, t) \coloneq K (n + t) \log c$ the \emph{budget} allocated to the algorithm $\Aa_c$.\footnote{Notice that the budget depends on the current row $t$ and not $m$, since the total length of $X$ is not known in advance.} While running $\Aa_c$ on the first $t$ requests of $X$, if the cost or time exceeds the budget $B(c, t)$, then we know that $c < \cpi$.
    The idea is therefore to guess the value $\cpi$ in phases: for each guessed value $c$, run the algorithm $\Aa_c$ until the budget is exceeded, or $X$ is successfully served. If the budget is exceeded, then start a new phase by increasing $c$ to $c'$ and using the algorithm $\Aa_{c'}$.
    However, in order to run $\Aa_{c'}$ for subsequent requests, we first need to modify the maintained BST so that it agrees with the BST constructed by $\Aa_{c'}$ if run on the previous requests. We therefore start every phase by a \emph{restart} procedure that transforms the maintained BST into the correct shape.

More precisely, let $i$ be the index of the current phase. Initially, $i = 0$.
    In phase $i$, serve the arriving request $x_t$ using the algorithm $\Aa_{c_i}$ for $c_i \coloneq 2^{2^i}$. As soon as the total cost or time spent exceeds the budget $B(c_i, t)$, abort and serve $x_t$ by a direct search (and insertion, if needed). This finishes phase $i$.
    Start phase $i+1$ with the following \emph{restart} procedure:
    simulate $\Aa_{c_{i+1}}$ on requests $(x_1, \dots, x_t)$ starting from an empty tree; let $T$ be the resulting tree, and transform the maintained BST so that it agrees with $T$. 
Continue running phase $i+1$ with the algorithm $\Aa_{c_{i+1}}$. If the cost or time of the \emph{restart} procedure exceeds the new budget $B(c_{i+1}, t)$, start phase $i+2$. 

    We argue that each phase $i$ costs at most $\Oh((n + m)2^i)$. Indeed, in the restart procedure of phase $i$ after request $x_t$ is served, we can compute the tree $T$ in time at most $B(c_i, t) \leq \Oh((n+t)\log c_i) \leq \Oh((n+m)2^i)$. Since $\Aa_{c_i}$ is an incremental algorithm, $T$ contains the same set of keys as the maintained BST. The number of keys is at most $n$. Thus, transforming the maintained BST into $T$ can be done in $\Oh(n)$ operations. Finally, the direct search that finishes the phase can also be done in $\Oh(n)$ operations. Since we abort the algorithm $\Aa_{c_i}$ as soon as the cost or time exceeds the budget, phase $i$ costs in total at most $\Oh((n + m)2^i)$.

    Suppose that we run phases $i=0, 1, \dots, r$ and complete serving $X$ in phase $r$. Then the total cost is at most
    $
    \sum_{i=0}^r \Oh((n+m)2^i) \leq \Oh((n+m)2^r). 
    $
    If $c_r < \cpi$, then in particular $2^r < \log \cpi$ and we obtain the claimed running time. 
    If $c_r \geq \cpi$, then we claim that $r$ is the smallest index $i$ with $c_i \geq \cpi$. 
    Indeed, for every $i$ with $c_i \geq \cpi$, by assumption on algorithms $\Aa_c$, simulating and running $\Aa_{c_i}$ has cost and running time at most $\Oh((n+m) \log c_i)$. By choosing $K$ sufficiently large, this means that phase $i$ never exceeds the budget. In particular, since we complete serving $X$ in phase $r$, we deduce that $r$ is the smallest index with $c_r \geq \cpi$, and thus $2^r \leq \Oh(\log \cpi)$.
    In any case, the total cost and time are at most $\Oh((n+m)2^r)\leq \Oh((n+m)\log \cpi)$. 
\end{proof}

\section{FPT Approximation}
\label{sec:fpt_approximation}
In this section, we prove our FPT approximation result. 
\begin{theorem}\label{thm:fpt-approximation}
    Given a $\pi$-avoiding permutation $X$ of $[n]$ with $n\geq2$ and
    $\varepsilon\in(0,1]$, an algorithm computes, in time
    $2^{\widetilde{\Oh}(c_\pi/\varepsilon)}n$, an arborally satisfied set $S_X$ with
        $|S_X|\leq(\am{4}+\varepsilon)\opt(X)$.
    The algorithm does not receive $c_\pi$.
\end{theorem}

Taking $\varepsilon=1$, \cref{thm:fpt-approximation} implies~\cref{thm:fpt-approx}, since $c_\pi=s_\pi^{\Oh(1)}$ and arborally satisfied sets correspond to BST executions within a constant factor in cost~\cite{demaine2009geometry}. The case $n=1$ is immediate.

The construction is similar to the recursive ideas from \Cref{sec:offline-warmup}: solve a version with contracted blocks $Z$ and
extend its solution through contracted columns $Y$ to the original setting $X$.
The key difference is that we replace the local BST search of \Cref{lem:balancedBST_gridpoint} by exact solutions given by the following theorem.

\begin{theorem}[{\cite[Section~6.6.3]{ChalermsookCS20}}]\label{thm:fpt-exact}
For a nonempty semi-permutation with $m$ points
and $d$ active columns, an exact algorithm computes an optimal solution
in $m\,d^{\Oh(d)}$ time.
\end{theorem}

\begin{proofof}{\Cref{thm:fpt-approximation}}

As in \Cref{sec:offline-warmup}, we define the point sets
$P_X=\{(t,x_t):t\in[n]\}$, and $P_Y,P_Z$ similarly.
We iterate \am{$d=\min\{2^k,n\}$ for $k=1,\ldots,\lceil\log_2 n\rceil$}.
If we ever reach $d=n$, we return an exact solution for $P_X$ computed by
\Cref{thm:fpt-exact}\am{, of size at most $3\opt(P_X)$ by \cref{lem:linear-optimum}}. Otherwise,
the construction in \Cref{sec:online_blockContraction,lem:online_constructionXYZ} gives $Y,Z$, super-columns $C_1,\ldots,C_q$,
and super-rows $R_1,\ldots,R_r$.
The resulting requests form a $\pi$-avoiding
permutation, so \Cref{thm:online-bst-bound} (to apply it we renumber keys
selected for $Z$ as $1,\ldots,|Z|$) and \Cref{lem:BSTgeom}
give a geometric solution for these requests.
Contracting each $C_j$ gives a
solution $S_Z$ for $P_Z$ by \Cref{lem:contraction-canonicalization}.  The main
difference with \Cref{sec:offline-warmup} is that we stop at the first $d$ for
which $|S_Z|\leq\varepsilon n/10$.

For this $d$, let $X_j$ be the restriction of $P_X$ to super-column $C_j$, and
let $Y_i$ be the restriction of $P_Y$ to super-row $R_i$.  We compute their
local solutions with an exact algorithm from~\cref{thm:fpt-exact} (notice that
in~\cref{sec:offline-warmup} we instead use balanced BSTs).  We map each row of
$S_Z$ to row $i$ if its selected input point lies in $R_i$, obtaining $S_A$.

We glue the local solutions in two steps.
First, we construct $S_Y$ by copying each $(i,j)\in S_A$ to the two boundary
rows of $R_i$, at $(\min R_i,j)$ and $(\max R_i,j)$, and adding the local
solutions for $Y_i$.  We then construct the output $S_X$ by copying each
$(t,j)\in S_Y$ to the two boundary columns of $C_j$, at $(t,\min C_j)$ and
$(t,\max C_j)$, and adding the local solutions for $X_j$. This concludes the
construction.

To prove correctness, we show that $S_Y$ and then $S_X$ are arborally
satisfied by finding a third point in $R[p,q]$ for every non-aligned pair
$p,q$.  We follow the gluing argument in \Cref{sec:offline-warmup}.

First, $S_A$ is arborally satisfied by
\Cref{lem:contraction-canonicalization}.  Contracting the super-rows maps
every point of $S_Y$ into $S_A$: each column used by a local solution
occurs in its super-row and hence has a selected representative in $Z$.
Let $p,q\in S_Y$ be non-aligned and first suppose they lie in the same
super-row.  If both are local points, its local solution contains a third
point in $R[p,q]$.  If one is a boundary point, assume it is $p$.
The contraction of $q$ belongs to $S_A$, so $q$ has a copy on this
boundary row.  Non-alignment makes this copy a third point in $R[p,q]$.
If $p,q$ lie in distinct super-rows $R_a,R_b$, with
$a<b$, their contractions $p',q'\in S_A$ have a third point
$r'=(k,j)\in S_A\cap R[p',q']$.
Use $(\max R_a,j)$ if $k=a$, $(\min R_b,j)$ if $k=b$, and either boundary
copy of $r'$ if $a<k<b$.  This point lies in $S_Y\cap R[p,q]$ and differs
from $p,q$, since it contracts to $r'\notin\{p',q'\}$.
Thus $S_Y$ is arborally satisfied.

Now, we prove that $S_X$ is arborally satisfied. Let $p,q\in S_X$ be
non-aligned. Contracting the super-columns maps every point of $S_X$ into
$S_Y$, since each local point lies on an input row of its super-column.  Suppose
first that $p,q$ lie in the same super-column.  If both are local points, its
local solution contains a third point in $R[p,q]$. If one is a boundary point,
assume it is $p$.  The contraction of $q$ belongs to $S_Y$, so $q$ has a copy on
this boundary column.  Non-alignment makes this copy a third point in $R[p,q]$.
If $p,q$ lie in distinct super-columns $C_a,C_b$, with $a<b$, their contractions
$p',q'\in S_Y$ have a third point $r'=(t,k)\in S_Y\cap R[p',q']$, since $S_Y$ is
arborally satisfied.  Use $(t,\max C_a)$ if $k=a$, $(t,\min C_b)$ if $k=b$, and
either boundary copy if $a<k<b$.  Again this gives a third point in $S_X\cap
R[p,q]$, so $S_X$ is arborally satisfied.

For the running time, fix a pattern $\pi$ avoided by $X$.
Each trial computes $S_Z$ in $\Oh(n\log c_\pi)$ time with
$|S_Z|=\Oh(|Z|\log c_\pi)$.  For $d>2c_\pi$,
\Cref{lem:online_defXYZ} gives
$|S_Z|=\Oh(c_\pi n\log(c_\pi)/d)$.
Thus, for $d<n$, a sufficiently large constant multiple of
$c_\pi\log(c_\pi)/\varepsilon$ ensures the stopping test succeeds.
Since consecutive scales differ by at most a factor of two, the last scale
is $d=\Oh(c_\pi\log(c_\pi)/\varepsilon)$, including the exact algorithm for the case $d=n$.
Each $X_j$ and $Y_i$ has at most $d$ active columns, with
$\sum_j|X_j|=\sum_i|Y_i|=n$, so \Cref{thm:fpt-exact} bounds the exact calls
by $n\,d^{\Oh(d)}$ time. In total, the running time is $2^{\widetilde{\Oh}(c_\pi/\varepsilon)}n$.

The important difference with~\Cref{sec:offline-warmup} is the cost analysis.
The strip decomposition in \cref{lem:strip-decomposition} gives
$\sum_j\opt(X_j)+\opt(P_Y)\leq\opt(P_X)$.
Partitioning an optimal solution for $P_Y$ by super-row and applying
\Cref{lem:contraction-canonicalization} to each part gives
$\sum_i\opt(Y_i)\leq\opt(P_Y)$.  Hence
\[
    \sum_j\opt(X_j)+\sum_i\opt(Y_i)\leq\opt(P_X).
\]
Every $X_j$ is a nonempty permutation point set, so
\Cref{lem:linear-optimum} gives $\sum_j\opt(X_j)\geq n-q$.
Also $q\leq|Z|\leq|S_Z|$, since every super-column occurs in $Z$.

During gluing we copy each point at most twice, so it gives
$|S_Y\setminus P_Y|\leq\sum_i\opt(Y_i)+2|S_Z|$ and
$|S_X\setminus P_X|\leq\sum_j\opt(X_j)+2|S_Y|$.
\am{Thus, $|S_X|\leq n+\sum_j\opt(X_j)+2|S_Y|$, and therefore}
\begin{align*}
    \am{|S_X|}
    &\leq\sum_j\opt(X_j)+2\sum_i\opt(Y_i)+\am{3}|P_Y|+4|S_Z|\\
    &\leq2\opt(P_X)-\sum_j\opt(X_j)+\am{3}|P_Y|+4|S_Z|\\
    &\leq2\opt(P_X)+\am{2}|P_Y|+q+4|S_Z|\\
    &\leq2\opt(P_X)+\am{2}|P_Y|+5|S_Z|\\
    &\leq(\am{4}+\varepsilon)\opt(P_X).
\end{align*}
The last inequality uses
$|S_Z|\leq\varepsilon n/10$, $|P_Y|=n$ and $n-|S_Z| \le \opt(P_X)$.

\end{proofof}

The analysis in fact gives a solution of cost at most
$\am{3}\cdot\opt(X)+(1+\varepsilon)n$, but we decided to state it as a single
parameter for simplicity.

\phantomsection
\addcontentsline{toc}{section}{AI Acknowledgment}
\section*{AI Acknowledgment}
Large language models were used for minor editing and figure preparation. 
Most ideas were developed by the human authors.
LLMs suggested cleaner constants in~\cref{sec:offline_warmup} and the use of $k$-finger BSTs in~\cref{sec:online_glueZtoY}, which simplified our earlier ideas and which we decided to present here. This work has received access to OpenAI models through the ChatGPT for Academic Researchers program.

\begingroup
\raggedright
\newcommand{\etalchar}[1]{$^{#1}$}

\endgroup

\clearpage
\appendix
\section{Extension to incremental online setting}
\label{sec:incremental_extension}
In this section we extend results from previous work to the incremental online setting. 

\subsection{Geometric view of BSTs}\label{sec:geometricBST_appendix}

We prove the extension of \cite[Lemmas 2.1 and 2.2]{demaine2009geometry} to the online and incremental setting. 
Note that \cite[Theorem C.1]{kozma2019smooth} already provided an extension to the incremental setting. However, this was only shown in the offline model, which is not sufficient for our purpose. 

\BSTgeom*
\begin{proof}
Output a point $(t,a)$ for every key $a$ touched while serving $x_t$. For two trace points whose keys are present after the earlier row, the argument of~\cite[Lemma~2.1]{demaine2009geometry} applies: the touched keys form a subtree containing the root, and inserting a leaf preserves the ancestor relations between existing keys.

It remains to consider $p=(s,a)$ and $q=(t,b)$ with $s<t$ when $b$ is inserted on a row $u\in(s,t]$. Let $c$ be the predecessor of $b$ at insertion time if $a<b$, and its successor otherwise. This key exists because $a$ is already present. The root-to-insertion path contains $c$, so $c$ is touched on row $u$. Since $c$ lies between $a$ and $b$ and differs from $b$, the point $(u,c)$ belongs to $\rect{p}{q}\setminus\{p,q\}$. Thus every trace prefix is arborally satisfied.

Only revealed keys are touched, so the output is insertion-compatible and contains $P_X$. Each operation touches constantly many keys, giving $|S_X|=\Oh(T)$. An array recording the last output row for each key suppresses duplicate points in constant time per touch, after $\Oh(n)$ initialization. The conversion therefore takes $\Oh(n+T)$ time and uses no future requests.
\end{proof}

\geomBST*

\begin{proof}
We use the online conversion and split trees of~\cite[Lemmas~2.3 and~2.4]{demaine2009geometry}, starting with an empty BST. A key $a$ is selected on row $t$ if $(t,a)\in S_X$. Before processing row $t$, we maintain that the BST stores exactly $X_{\leq t-1}$. Insertion-compatibility ensures that every selected key belongs to $X_{\leq t-1}\cup\{x_t\}$, while $P_X\subseteq S_X$ ensures that $x_t$ is selected. Thus $x_t$ is the only key that may need to be inserted. If it is already present, use the usual row update.

Suppose that $x_t$ is new. For each defined neighbour $a$ of $x_t$ among the previously revealed keys, insertion-compatibility implies that the rectangle between its last point and $(t,x_t)$ can contain a third point only at $(t,a)$. Arboral satisfaction therefore forces $a$ to be selected on row $t$. Removing $(t,x_t)$ leaves an arborally satisfied prefix, since any rectangle spanned by old points that used it as a witness also contains one of these selected neighbours. Run the splitting phase of the usual update on the selected old keys. Their traversal exposes the empty gap for $x_t$. Insert $x_t$ there as a leaf and include it when building the new root supernode. For the first request, create a singleton tree. After the update, the BST stores exactly $X_{\leq t}$, preserving the invariant.

The cited split-tree analysis gives $\Oh(|S_X|)$ total BST operations and additional time, including the insertion traversals, which use only selected keys. Initializing column records takes $\Oh(n)$ time. Since $|S_X|\leq\Oh(T)$, the total time, including the given point algorithm, is $\Oh(n+T)$. Every update uses only the current prefix, and a key is inserted exactly at its first request. Thus the resulting BST algorithm is online and incremental.
\end{proof}

\subsection{Online incremental simulation of $k$-finger BST}
\label{app:online_kfinger_simulation}

We prove \cref{lem:online_kfinger_simulation} by extending the $k$-finger BST simulation of \cite[Theorem~2]{chalermsook2018multi} to the incremental model. The only added difference is that we need to handle insertion operations. Since the overall idea of the proof remains the same as the one presented in \cite[Section~2 and Appendix~A]{chalermsook2018multi}, we only describe how to simulate insertion operations.

\kfingerSimulataion*

\begin{proof}[Proof sketch of \cref{lem:online_kfinger_simulation}]
      Let $\Aa$ be an online incremental $k$-finger BST algorithm serving $X$ with cost and time at most $T$, and let $\Tt_{\Aa}$ be the tree maintained by $\Aa$. We construct an online incremental BST algorithm $\Bb$  maintaining a tree $\Tt_{\Bb}$ on the same keys. The idea of \cite[Appendix~A]{chalermsook2018multi} is to keep the following invariant after every operation performed by $\Aa$ is simulated by $\Bb$.
      \begin{quote}
            Every key on which a finger is positioned in $\Tt_{\Aa}$ is at depth $\Oh(\log k)$ in $\Tt_{\Bb}$.
      \end{quote}
      Whenever $\Aa$ uses a finger to perform an operation, the algorithm $\Bb$ first visits the corresponding key from the root of $\Tt_{\Bb}$ with $\Oh(\log k)$ operations, and then simulates the operation in a way that preserves this invariant.

      The main effort in the proof of \cite[Theorem 2]{chalermsook2018multi} is therefore to show that, after each operation of $\Aa$ is simulated by $\Bb$, the tree $\Tt_{\Bb}$ can be updated to 
satisfy the invariant in amortized cost and time $\Oh(\log k)$. 
      We will use the same update algorithm as in \cite[Theorem 2]{chalermsook2018multi} whenever the simulated operation is a move, a rotation, or key access. The only new operations we consider are \emph{insertion} operations, which are specific to the incremental model. We therefore only describe the update for the insertion case.
We do so using the same terminology of the \emph{extended hand} as in \cite[Appendix A]{chalermsook2018multi}, which we now recall. 

      \paragraph*{The extended hand of \cite[Appendix A]{chalermsook2018multi}.}
Add a finger that stays at the root of $\Tt_{\Aa}$. 
Let $S$ be the smallest subtree of $\Tt_{\Aa}$ containing all fingers. A finger or a degree-three vertex of $S$ is called a \emph{pseudofinger}. A \emph{tendon} is the internal part of a path between two pseudofingers with no intermediate pseudofinger. Split each tendon into the keys smaller and larger than its lower endpoint, which we call its two \emph{half-tendons}. The components of $\Tt_{\Aa} \setminus S$ are called \emph{knuckles} and contain no finger.

      There are $\Oh(k)$ pseudofingers and half-tendons, and their key intervals are disjoint. Maintain a balanced BST on the key intervals. A half-tendon is represented by a BST that stores its minimum and maximum at the top and its remaining keys in a deque, while each knuckle is attached in its inorder gap. This way, every pseudofinger and every half-tendon extremum has depth $\Oh(\log k)$. The deque supports insertion and removal at either end in constant amortized time. A finger move or rotation changes only constantly many hand entries and deque ends, so it can be simulated with $\Oh(\log k)$ amortized operations and time.

      \paragraph*{Insertion operation.}

      Suppose that $\Aa$ inserts a new key $x$ at a null right child of a finger $f$. 
      {We claim that each defined neighbour of $x$ has depth $\Oh(\log k)$ in $\Tt_{\Bb}$. Then, searching for $x$  in $\Tt_{\Bb}$ reaches an appropriate null child of one of its neighbours in $\Oh(\log k)$ cost and time.}
      Suppose without loss of generality that $f$ is the predecessor of $x$ (the left-child insertion is symmetric). Then by the invariant, $f$ is at depth $\Oh(\log k)$. Now the successor of $x$, if defined, is the lowest ancestor $a>f$.
      Since we keep a finger at the {root of $\Tt_{\Aa}$}, every ancestor of $f$ belongs to $S$. If $a$ is not a pseudofinger, then it belongs to a tendon whose lower endpoint is an ancestor of $f$. The keys in the larger half of this tendon decrease down the path, and $a$ is the last ancestor larger than $f$, so $a$ is the minimum of that half-tendon. {This proves that each defined neighbour of $x$ has depth $\Oh(\log k)$ in $\Tt_{\Bb}$.}

      To insert $x$ in $\Tt_{\Bb}$, we proceed as follows. {Search for $x$ in $\Tt_{\Bb}$ and insert it at the null child reached by this search. The parent is one of the defined neighbours, so this takes $\Oh(\log k)$ cost and time.}
      Insertion leaves the fingers at their current nodes in $\Tt_{\Aa}$, so $S$ and all pseudofingers and half-tendons remain unchanged. The new leaf is a knuckle in the previously empty inorder gap, and the leaf inserted in $\Tt_{\Bb}$ occupies exactly that gap. Hence, the extended hand representation is preserved. 

      \paragraph*{Overall simulation.}
      Using the same simulation procedure for move, rotation, and key access operations as in \cite{chalermsook2018multi}, every operation of $\Aa$ is simulated with $\Oh(\log k)$ amortized cost and time while preserving the extended-hand representation. Running $\Aa$ itself takes at most $T$ time, so $\Bb$ serves $X$ with cost and time $\Oh(T\log k)$. Each request is fully simulated before reading the next request, so the algorithm works in the online incremental model.
\end{proof}


\begin{thebibliography}{CGK{\etalchar{+}}15b}

\bibitem[Art07]{arthur2007fast}
David Arthur.
\newblock Fast sorting and pattern-avoiding permutations.
\newblock In {\em Proceedings of the Fourth Workshop on Analytic Algorithmics
  and Combinatorics (ANALCO 2007)}, pages 169--174. SIAM, 2007.
\newblock \href {https://doi.org/10.1137/1.9781611972979.1}
  {\path{doi:10.1137/1.9781611972979.1}}.

\bibitem[AVL62]{adelson1962algorithm}
Georgii~M. Adel'son-Vel'skii and Evgenii~M. Landis.
\newblock An algorithm for organization of information.
\newblock {\em Doklady Akademii Nauk SSSR}, 146(2):263--266, 1962.
\newblock URL: \url{https://www.mathnet.ru/eng/dan26964}.

\bibitem[Bay72]{Bay72}
Rudolf Bayer.
\newblock Symmetric binary {B}-trees: Data structure and maintenance
  algorithms.
\newblock {\em Acta Informatica}, 1(4):290--306, 1972.
\newblock \href {https://doi.org/10.1007/BF00289509}
  {\path{doi:10.1007/BF00289509}}.

\bibitem[BCFL12]{BoseCFL12}
Prosenjit Bose, S{\'e}bastien Collette, Rolf Fagerberg, and Stefan Langerman.
\newblock De-amortizing binary search trees.
\newblock In {\em Automata, Languages, and Programming (ICALP 2012), Part I},
  volume 7391 of {\em Lecture Notes in Computer Science}, pages 121--132.
  Springer, 2012.
\newblock \href {https://doi.org/10.1007/978-3-642-31594-7_11}
  {\path{doi:10.1007/978-3-642-31594-7_11}}.

\bibitem[BKO24]{berendsohn2024optimization}
Benjamin~Aram Berendsohn, L{\'a}szl{\'o} Kozma, and Michal Opler.
\newblock Optimization with pattern-avoiding input.
\newblock In {\em Proceedings of the 56th Annual ACM Symposium on Theory of
  Computing (STOC 2024)}, pages 671--682. ACM, 2024.
\newblock \href {https://doi.org/10.1145/3618260.3649631}
  {\path{doi:10.1145/3618260.3649631}}.

\bibitem[CCS23]{ChalermsookCS20}
Parinya Chalermsook, Julia Chuzhoy, and Thatchaphol Saranurak.
\newblock Pinning down the strong {Wilber-1} bound for binary search trees.
\newblock {\em Theory of Computing}, 19(8):1--71, 2023.
\newblock \href {https://doi.org/10.4086/toc.2023.v019a008}
  {\path{doi:10.4086/toc.2023.v019a008}}.

\bibitem[CGJ{\etalchar{+}}23]{chalermsook2023improved}
Parinya Chalermsook, Manoj Gupta, Wanchote Jiamjitrak, Nidia~Obscura Acosta,
  Akash Pareek, and Sorrachai Yingchareonthawornchai.
\newblock Improved pattern-avoidance bounds for {Greedy BSTs} via matrix
  decomposition.
\newblock In {\em Proceedings of the Thirty-Fourth Annual ACM-SIAM Symposium on
  Discrete Algorithms (SODA 2023)}, pages 509--534. SIAM, 2023.
\newblock URL: \url{https://arxiv.org/abs/2211.04112}, \href
  {https://doi.org/10.1137/1.9781611977554.ch22}
  {\path{doi:10.1137/1.9781611977554.ch22}}.

\bibitem[CGJ{\etalchar{+}}24]{chalermsook2024group}
Parinya Chalermsook, Manoj Gupta, Wanchote Jiamjitrak, Akash Pareek, and
  Sorrachai Yingchareonthawornchai.
\newblock The group access bounds for binary search trees.
\newblock In {\em 51st International Colloquium on Automata, Languages, and
  Programming (ICALP 2024)}, volume 297 of {\em Leibniz International
  Proceedings in Informatics (LIPIcs)}, pages 38:1--38:18. Schloss Dagstuhl --
  Leibniz-Zentrum f{\"u}r Informatik, 2024.
\newblock \href {https://doi.org/10.4230/LIPIcs.ICALP.2024.38}
  {\path{doi:10.4230/LIPIcs.ICALP.2024.38}}.

\bibitem[CGK{\etalchar{+}}15a]{DBLP:journals/corr/ChalermsookG0MS15a}
Parinya Chalermsook, Mayank Goswami, L{\'{a}}szl{\'{o}} Kozma, Kurt Mehlhorn,
  and Thatchaphol Saranurak.
\newblock Greedy is an almost optimal deque.
\newblock In {\em Algorithms and Data Structures (WADS 2015)}, volume 9214 of
  {\em Lecture Notes in Computer Science}, pages 152--165. Springer, 2015.
\newblock \href {https://doi.org/10.1007/978-3-319-21840-3_13}
  {\path{doi:10.1007/978-3-319-21840-3_13}}.

\bibitem[CGK{\etalchar{+}}15b]{chalermsook2015pattern}
Parinya Chalermsook, Mayank Goswami, L{\'a}szl{\'o} Kozma, Kurt Mehlhorn, and
  Thatchaphol Saranurak.
\newblock Pattern-avoiding access in binary search trees.
\newblock In {\em 56th Annual IEEE Symposium on Foundations of Computer Science
  (FOCS 2015)}, pages 410--423. IEEE, 2015.
\newblock \href {https://doi.org/10.1109/FOCS.2015.32}
  {\path{doi:10.1109/FOCS.2015.32}}.

\bibitem[CGK{\etalchar{+}}18]{chalermsook2018multi}
Parinya Chalermsook, Mayank Goswami, L{\'a}szl{\'o} Kozma, Kurt Mehlhorn, and
  Thatchaphol Saranurak.
\newblock Multi-finger binary search trees.
\newblock In {\em 29th International Symposium on Algorithms and Computation
  (ISAAC 2018)}, volume 123 of {\em Leibniz International Proceedings in
  Informatics (LIPIcs)}, pages 55:1--55:26. Schloss Dagstuhl -- Leibniz-Zentrum
  f{\"u}r Informatik, 2018.
\newblock \href {https://doi.org/10.4230/LIPIcs.ISAAC.2018.55}
  {\path{doi:10.4230/LIPIcs.ISAAC.2018.55}}.

\bibitem[CHH{\etalchar{+}}26]{chmel2026splay}
Petr Chmel, Bernhard Haeupler, Richard Hlad{\'i}k, Michal Kouck{\'y}, Antti
  Roeyskoe, V{\'a}clav Rozho{\v{n}}, Ond{\v{r}}ej Sladk{\'y}, and Robert~E.
  Tarjan.
\newblock Splay trees are almost dynamically optimal, 2026.
\newblock Preprint.
\newblock \href {https://arxiv.org/abs/2607.18498} {\path{arXiv:2607.18498}}.

\bibitem[Cib09]{Cibulka09}
Josef Cibulka.
\newblock On constants in the {F{\"{u}}redi--Hajnal} and the {Stanley--Wilf}
  conjecture.
\newblock {\em Journal of Combinatorial Theory, Series A}, 116(2):290--302,
  2009.
\newblock \href {https://doi.org/10.1016/j.jcta.2008.06.003}
  {\path{doi:10.1016/j.jcta.2008.06.003}}.

\bibitem[CPY24]{chalermsook2024sorting}
Parinya Chalermsook, Seth Pettie, and Sorrachai Yingchareonthawornchai.
\newblock Sorting pattern-avoiding permutations via 0--1 matrices forbidding
  product patterns.
\newblock In {\em Proceedings of the Thirty-Fifth Annual ACM-SIAM Symposium on
  Discrete Algorithms (SODA 2024)}, pages 133--149. SIAM, 2024.
\newblock \href {https://doi.org/10.1137/1.9781611977912.7}
  {\path{doi:10.1137/1.9781611977912.7}}.

\bibitem[CW82]{culik1982note}
Karel {Culik II} and Derick Wood.
\newblock A note on some tree similarity measures.
\newblock {\em Information Processing Letters}, 15(1):39--42, 1982.
\newblock \href {https://doi.org/10.1016/0020-0190(82)90083-7}
  {\path{doi:10.1016/0020-0190(82)90083-7}}.

\bibitem[DHI{\etalchar{+}}09]{demaine2009geometry}
Erik~D. Demaine, Dion Harmon, John Iacono, Daniel Kane, and Mihai
  P{\u{a}}tra{\c{s}}cu.
\newblock The geometry of binary search trees.
\newblock In {\em Proceedings of the Twentieth Annual ACM-SIAM Symposium on
  Discrete Algorithms (SODA 2009)}, pages 496--505. SIAM, 2009.
\newblock \href {https://doi.org/10.1137/1.9781611973068.55}
  {\path{doi:10.1137/1.9781611973068.55}}.

\bibitem[DHIP07]{demaine2007dynamic}
Erik~D. Demaine, Dion Harmon, John Iacono, and Mihai P{\u{a}}tra{\c{s}}cu.
\newblock Dynamic optimality—almost.
\newblock {\em SIAM Journal on Computing}, 37(1):240--251, 2007.
\newblock \href {https://doi.org/10.1137/S0097539705447347}
  {\path{doi:10.1137/S0097539705447347}}.

\bibitem[FH92]{FurediH92}
Zolt{\'{a}}n F{\"{u}}redi and P{\'{e}}ter Hajnal.
\newblock {Davenport--Schinzel} theory of matrices.
\newblock {\em Discrete Mathematics}, 103(3):233--251, 1992.
\newblock \href {https://doi.org/10.1016/0012-365X(92)90316-8}
  {\path{doi:10.1016/0012-365X(92)90316-8}}.

\bibitem[Fox13]{fox2013stanley}
Jacob Fox.
\newblock {Stanley--Wilf} limits are typically exponential, 2013.
\newblock To appear in Advances in Mathematics.
\newblock \href {https://arxiv.org/abs/1310.8378} {\path{arXiv:1310.8378}}.

\bibitem[Fre76]{fredman1976information}
Michael~L. Fredman.
\newblock How good is the information theory bound in sorting?
\newblock {\em Theoretical Computer Science}, 1(4):355--361, 1976.
\newblock \href {https://doi.org/10.1016/0304-3975(76)90078-5}
  {\path{doi:10.1016/0304-3975(76)90078-5}}.

\bibitem[GG19]{goyal2019better}
Navin Goyal and Manoj Gupta.
\newblock Better analysis of {Greedy} binary search tree on decomposable
  sequences.
\newblock {\em Theoretical Computer Science}, 776:19--42, 2019.
\newblock \href {https://doi.org/10.1016/j.tcs.2018.12.021}
  {\path{doi:10.1016/j.tcs.2018.12.021}}.

\bibitem[GM14]{GM14}
Sylvain Guillemot and D{\'a}niel Marx.
\newblock Finding small patterns in permutations in linear time.
\newblock In {\em Proceedings of the Twenty-Fifth Annual ACM-SIAM Symposium on
  Discrete Algorithms (SODA 2014)}, pages 82--101. SIAM, 2014.
\newblock \href {https://doi.org/10.1137/1.9781611973402.7}
  {\path{doi:10.1137/1.9781611973402.7}}.

\bibitem[GPSW26]{guo2026greedy}
Yuhao Guo, Seth Pettie, Daniel Skora, and Chengzhang Wan.
\newblock The {Greedy} binary search tree is non-trivially competitive, 2026.
\newblock Preprint.
\newblock \href {https://arxiv.org/abs/2608.04410} {\path{arXiv:2608.04410}}.

\bibitem[Iac13]{iacono2013pursuit}
John Iacono.
\newblock In pursuit of the dynamic optimality conjecture.
\newblock In {\em Space-Efficient Data Structures, Streams, and Algorithms},
  volume 8066 of {\em Lecture Notes in Computer Science}, pages 236--250.
  Springer, 2013.
\newblock \href {https://doi.org/10.1007/978-3-642-40273-9_16}
  {\path{doi:10.1007/978-3-642-40273-9_16}}.

\bibitem[JLWY24]{jiang2024hardness}
Shunhua Jiang, Victor Lecomte, Omri Weinstein, and Sorrachai
  Yingchareonthawornchai.
\newblock Hardness amplification for dynamic binary search trees.
\newblock In {\em 35th International Symposium on Algorithms and Computation
  (ISAAC 2024)}, volume 322 of {\em Leibniz International Proceedings in
  Informatics (LIPIcs)}, pages 42:1--42:19. Schloss Dagstuhl -- Leibniz-Zentrum
  f{\"u}r Informatik, 2024.
\newblock \href {https://doi.org/10.4230/LIPIcs.ISAAC.2024.42}
  {\path{doi:10.4230/LIPIcs.ISAAC.2024.42}}.

\bibitem[Kla00]{Klazar00}
Martin Klazar.
\newblock The {F{\"u}redi--Hajnal} conjecture implies the {Stanley--Wilf}
  conjecture.
\newblock In {\em Formal Power Series and Algebraic Combinatorics}, pages
  250--255. Springer, 2000.
\newblock \href {https://doi.org/10.1007/978-3-662-04166-6_22}
  {\path{doi:10.1007/978-3-662-04166-6_22}}.

\bibitem[Knu68]{Knuth68}
Donald~E. Knuth.
\newblock {\em The Art of Computer Programming, Volume 1: Fundamental
  Algorithms}.
\newblock Addison-Wesley, 1 edition, 1968.
\newblock Third edition published in 1997.

\bibitem[Koz16]{Kozma16}
L{\'a}szl{\'o} Kozma.
\newblock {\em Binary Search Trees, Rectangles and Patterns}.
\newblock PhD thesis, Saarland University, 2016.
\newblock URL:
  \url{https://publikationen.sulb.uni-saarland.de/handle/20.500.11880/26727}.

\bibitem[KS20]{kozma2019smooth}
L{\'a}szl{\'o} Kozma and Thatchaphol Saranurak.
\newblock Smooth heaps and a dual view of self-adjusting data structures.
\newblock {\em SIAM Journal on Computing}, 49(5):STOC18--45--STOC18--93, 2020.
\newblock \href {https://doi.org/10.1137/18M1195188}
  {\path{doi:10.1137/18M1195188}}.

\bibitem[Luc88]{lucas1988canonical}
Joan~Marie Lucas.
\newblock Canonical forms for competitive binary search tree algorithms.
\newblock Technical Report DCS-TR-250, Rutgers University, Department of
  Computer Science, 1988.

\bibitem[LW20]{lecomte2020settling}
Victor Lecomte and Omri Weinstein.
\newblock Settling the relationship between {Wilber}'s bounds for dynamic
  optimality.
\newblock In {\em 28th Annual European Symposium on Algorithms (ESA 2020)},
  volume 173 of {\em Leibniz International Proceedings in Informatics
  (LIPIcs)}, pages 68:1--68:21. Schloss Dagstuhl -- Leibniz-Zentrum f{\"u}r
  Informatik, 2020.
\newblock \href {https://doi.org/10.4230/LIPIcs.ESA.2020.68}
  {\path{doi:10.4230/LIPIcs.ESA.2020.68}}.

\bibitem[MT04]{marcus2004excluded}
Adam Marcus and G{\'a}bor Tardos.
\newblock Excluded permutation matrices and the {Stanley--Wilf} conjecture.
\newblock {\em Journal of Combinatorial Theory, Series A}, 107(1):153--160,
  2004.
\newblock \href {https://doi.org/10.1016/j.jcta.2004.04.002}
  {\path{doi:10.1016/j.jcta.2004.04.002}}.

\bibitem[Mun00]{munro2000competitiveness}
J.~Ian Munro.
\newblock On the competitiveness of linear search.
\newblock In {\em Algorithms -- ESA 2000}, volume 1879 of {\em Lecture Notes in
  Computer Science}, pages 338--345. Springer, 2000.
\newblock \href {https://doi.org/10.1007/3-540-45253-2_31}
  {\path{doi:10.1007/3-540-45253-2_31}}.

\bibitem[Opl24]{opler2024sorting}
Michal Opler.
\newblock An optimal algorithm for sorting pattern-avoiding sequences.
\newblock In {\em 65th Annual IEEE Symposium on Foundations of Computer Science
  (FOCS 2024)}, pages 689--699. IEEE, 2024.
\newblock \href {https://doi.org/10.1109/FOCS61266.2024.00049}
  {\path{doi:10.1109/FOCS61266.2024.00049}}.

\bibitem[Pet10]{pettie2010applications}
Seth Pettie.
\newblock Applications of forbidden 0--1 matrices to search tree and path
  compression-based data structures.
\newblock In {\em Proceedings of the Twenty-First Annual ACM-SIAM Symposium on
  Discrete Algorithms (SODA 2010)}, pages 1457--1467. SIAM, 2010.
\newblock \href {https://doi.org/10.1137/1.9781611973075.118}
  {\path{doi:10.1137/1.9781611973075.118}}.

\bibitem[Pet15]{pettie2015three}
Seth Pettie.
\newblock Three generalizations of {Davenport--Schinzel} sequences.
\newblock {\em SIAM Journal on Discrete Mathematics}, 29(4):2189--2238, 2015.
\newblock \href {https://doi.org/10.1137/140968574}
  {\path{doi:10.1137/140968574}}.

\bibitem[ST85]{sleator1985self}
Daniel~D. Sleator and Robert~E. Tarjan.
\newblock Self-adjusting binary search trees.
\newblock {\em Journal of the ACM}, 32(3):652--686, 1985.
\newblock \href {https://doi.org/10.1145/3828.3835}
  {\path{doi:10.1145/3828.3835}}.

\bibitem[Tar85]{Tarjan85}
Robert~E. Tarjan.
\newblock Sequential access in splay trees takes linear time.
\newblock {\em Combinatorica}, 5(4):367--378, 1985.
\newblock \href {https://doi.org/10.1007/BF02579253}
  {\path{doi:10.1007/BF02579253}}.

\bibitem[WDS06]{wang2006competitive}
Chengwen~Chris Wang, Jonathan Derryberry, and Daniel~D. Sleator.
\newblock {$O(\log\log n)$}-competitive dynamic binary search trees.
\newblock In {\em Proceedings of the Seventeenth Annual ACM-SIAM Symposium on
  Discrete Algorithms (SODA 2006)}, pages 374--383. SIAM, 2006.
\newblock URL: \url{https://www.cs.cmu.edu/~jonderry/multi-splay.pdf}.

\bibitem[Wil89]{wilber1989lower}
Robert Wilber.
\newblock Lower bounds for accessing binary search trees with rotations.
\newblock {\em SIAM Journal on Computing}, 18(1):56--67, 1989.
\newblock \href {https://doi.org/10.1137/0218004} {\path{doi:10.1137/0218004}}.

\end{thebibliography}
\end{document}